\documentclass[letterpaper,11pt,leqno]{article}
\usepackage{paper}
\usepackage{booktabs,threeparttable}
\usepackage{caption}
\usepackage{xcolor}
\usepackage{accents}
\usepackage{mathtools}
\usepackage{hyperref}
\hypersetup{colorlinks, citecolor=blue, filecolor=blue, linkcolor=black, urlcolor=cyan}
\usepackage{makecell}
\usepackage{longtable}
\usepackage{subcaption}
\usepackage{adjustbox}

\newcommand{\bib}{bibliography.bib}

\newcommand{\ubar}[1]{\underaccent{\bar}{#1}}
\newcommand{\vect}[1]{\ensuremath{\mathbf{#1}}}

\begin{document}

\title{LLM-Agent Interactions on Markets with Information Asymmetries}

\author{Alexander Erlei, Lukas Meub
%
\thanks{Contact: \url{alexander.erlei@wiwi.uni-goettingen.de}, Georg-August-Universität Göttingen. We gratefully acknowledge financial support by the Federal Ministry of Education and Research, project "Handwerk mit Zukunft (HaMiZu)", grant number 02K20D001.}}

\date{}   


\begin{titlepage}
\maketitle

\vspace{-1cm}

As AI agents increasingly act on behalf of human stakeholders in economic settings, understanding their behavior in complex market environments becomes critical. This article examines how Large Language Models coordinate on markets that are characterized by information asymmetries and in which providers of services have incentives to exploit that asymmetry for their own economic gain. To that end, we conduct simulations with GPT-5.1 agents in credence goods markets, manipulating the institutional framework (free market, verifiability, liability), LLM agent's social preferences (default, self-interested, inequity-averse, efficiency-loving), and reputation mechanisms across one-shot and repeated 16-round interactions. In one-shot settings, LLM agents largely fail to establish cooperation, with markets breaking down except under liability rules or when experts have efficiency-loving preferences. Repeated interactions solve consumer participation through competitive price reduction, but expert fraud remains entrenched absent explicit other-regarding preferences. LLM consumers focus narrowly on price levels rather than understanding strategic incentives embedded in markups, making them vulnerable to exploitation. Compared to human experiments, LLM markets exhibit substantially higher consumer participation but much greater market concentration, lower prices, and more polarized fraud patterns. The effect of institutions like verifiability and reputation is also much more ambiguous. Surplus shifts dramatically toward consumers under social-preference objectives. These findings suggest that institutional design for AI agent markets requires fundamentally different approaches than those effective for human actors, with social preference alignment emerging as the primary determinant of market efficiency.


\end{titlepage}

\section{Introduction}\label{s:introduction}
Recent advances in (generative) artificial intelligence (AI) and strategic foundation models suggest a world in which humans will increasingly rely on and delegate to autonomous AI agents \citep{goktas2025strategic,imas2025agentic,ivanov2024principal}. Already today it is easy to imagine AI agents representing and acting on behalf of human stakeholders on, for example, online marketplaces, shopping platforms, or when negotiating (digital) services. In this paper, we examine current LLM (Large Language Model) capabilities to navigate relatively complex market structures that are characterized by information asymmetries between agents: credence goods markets (or markets for expert services). Credence goods markets are important, because many important economic and social domains are characterized by information advantages where superiorly informed sellers or suppliers (here: experts) can exploit their informational advantage at the expense of consumers \citep{dulleck2006doctors}. Prominent examples include financial services, expert advice, medical services, or legal services. Over the last 20 years, a lot of empirical field research has revealed strong real-world inefficiencies, including widespread rates of overcharging (i.e. charging for more services than provided) and under-treatment (i.e. not solving the consumer's problem or not satisfying the consumer's demand)\citep{gottschalk2020health,balafoutas2013drives,kerschbamer2016insurance,kerschbamer2023credence,balafoutas2020credence,schneider2012agency}. Importantly, these problems are difficult to solve, because the information asymmetry precludes consumers from disciplining experts through informed selection, and effective regulations such as liability are often difficult or impossible to enforce (it is, for example, often impossible to \textit{prove} expert fraud or negligence). AI agents, however, may have the potential to alleviate these information problems.

In contrast to human actors, current foundation models like LLMs have access to vast amounts of expert knowledge. They are able to think strategically at scale, and learn from real-time feedback. Recent work argues that human cognitive and informational restraints may be a key driver of AI agent adoption \citep{shahidi2025coasean}. On the expert's side, there are already numerous examples of automated LLM-provided services, ranging from financial services to medical advice, taxation services, and shopping assistants \citep{ayers2023comparing,hirosawa2023chatgpt,nov2023putting, zhao2024revolutionizing,li2023sailer,oehler2024does,lai2024large,nay2024large,alarie2018artificial, reis2024influence,shekar2024people,ayre2024asking,leslie2024critical,seabrooke2024survey,schneiders2024objection}. LLM providers such as OpenAI and Anthropic offer explicit agent services, scaling up consumer access to LLM agents.

However, how well AI agents based on Large Language Models coordinate on markets that operate under information asymmetry is an open question. There is considerable debate about the capacity of AI agents such as LLMs -- especially off-the-shelve ones -- to navigate complex economic environments \citep{felin2024theory,valmeekam2023planbench,raman2024steer}, with several recent successful examples \citep{fish2024algorithmic,deshpande2026strategic,lopez2025can,imas2025agentic}. Laboratory experiments have consistently shown that humans struggle with the game theory of credence goods markets, but may still achieve moderate levels of cooperation through social preferences. AI agents could out-perform human stakeholders through strategic price-setting (experts) and expectation forming (consumers) that alleviate the threat of market unraveling. On the other hand, they may lack the advantage of mutually beneficial (pro-) social preferences that allow humans to enter a reciprocal relationship via initial trust \citep[see e.g.,][]{erlei2025digital,dvorak2025adverse,erlei2026betrayal}. 

In this article, we present a sequence of simulations using GPT-5.1 in which LLM experts interact with LLM consumers on standard experimental credence goods markets \citep{dulleck2011economics}. We consider both one-shot and repeated-interaction market scenarios, and vary the institutional market framework (No Institution vs. Verifiability vs. Liability), the LLMs social preferences (default model, self-interested, inequity-averse, efficiency-loving), as well as the possibility of bilateral expert--consumer reputation building to answer the following question:

\begin{itemize}
    \item How do LLM AI agents coordinate on credence goods markets?
    \item How do social preferences and reputation affect agent behavior and market welfare?
\end{itemize}

\noindent
The one-shot simulations show that LLM agents struggle to establish successful cooperation under information asymmetry. In the absence of liability rules, only the efficiency-loving LLM expert manages to consistently attract LLM consumers. However, it achieves this through ruinously low prices. Still, it is the only configuration that always reaches or beats the standard model's welfare predictions. Liability solves the agentic credence problem -- like it does for humans -- because experts are always forced to solve the consumer's problem, which guarantees full market participation. The default and the self-interested LLM expert aim to defraud consumers when possible. Endowing LLM experts with inequity-aversion or efficiency-loving preferences all but eliminates expert dishonesty. Going beyond singular interactions towards a repeated 16-round settings mostly solves the market participation problem. In all configurations, a large majority of consumers ``learn'' to enter the market over time. LLM consumer agents do not consider expert incentives through markups, but enter when prices are low enough, which makes them vulnerable to expert exploitation. LLM experts intend to defraud consumers in the absence of explicit other-regarding preferences, and these tendencies are stable across repeated interactions, indicating that LLM consumers do not discipline experts towards honesty. When expert agents are inequity-averse, over-treatment increases substantially, while almost no expert under-treats consumers. Efficiency-loving experts continue to prioritize consumer's welfare. Reputation has ambiguous effects on expert honesty, and no discernible consistent effect on welfare. Often, but not always, it slightly reduces under-treatment and over-charging, while increasing over-treatment. The effects are, however, small. Because market efficiency hinges on consumer participation (solved) and expert honesty (only affected by induced social preferences), verifiability does not perform better than the unregulated free credence good market. Self-interested and default expert agents converge towards states with low prices and continuous expert fraud, which hurts welfare but still improves on the one-shot scenario. Inequity-averse and efficiency-loving agents rarely defraud and thus induce additional welfare gains, which predominantly accrue to consumers. Inequity-aversion avoids ruinous expert behavior, generating (slightly) more equal outcomes. Comparing the results to human subject data from \citet{dulleck2011economics} reveals that LLM-agent markets exhibit substantially higher consumer participation but much stronger seller concentration, lower prices, as well as more polarized expert fraud and efficiency patterns. The positive treatment effects of institutional interventions like reputation and verifiability are less stable than in human experiments. Frequently, they reverse. In general, but especially under social-preference objectives, surplus sharply shifts from sellers towards consumers.

\section{Related Literature}
This article is situated at the intersection of the literature on credence goods and the literature about LLM agents in strategic economic settings. For human interactions, there exists a large literature of laboratory and online experiments that gauge the efficiency of credence goods markets across various institutional settings \citep{balafoutas2020credence,dulleck2011economics,balafoutas2023serving,kerschbamer2017social,grosskopf2010reputation, beck2013shaping, balafoutas2015hidden,kerschbamer2017social,inderst2019sharing,kandul2023reciprocity,tracy2023uncertainty,mimra2016second,agarwal2019personal,erlei2024technological,kerschbamer2023credence,schneider2021consumer}. Recently, \citet{erlei2025digital} has shown how pro-social human experts can out-compete self-interested human experts by delegating their choices to an LLM agent with a transparent, codified objective function. Furthermore, they show that in the absence of transparent objective functions, human consumers trust LLM experts less than human experts.

Since the introduction of LLMs, there has been a strong increase of articles studying the strategic capabilities and tendencies of AI agents in economic settings \citep{zhang2024proagent,kasberger2023algorithmic,guo2023gpt,schmidt2024gpt,rahwan2019machine}. \citet{deshpande2026strategic} study large language models in oligopolistic Cournot markets. LLM agents are capable of strategic decision-making, and exhibit tendencies for tacit collusion (which has also been observed by \citet{fish2024algorithmic}). \citet{imas2025agentic} allow humans to use LLM agents as instructed agents in a negotiation setting, showing how (1) human personality traits affect agent behavior through instructions, (2) social norms are replicated in agent interactions via, e.g., the prevalence of ``fair splits'', and (3) variance in outcomes increases. In \citet{cohen2025exploring}, endowing LLMs with personality traits significantly changes the behavioral dynamics in a negotiation settings, and in \citet{long2025evoemo}, emotional expressions by AI agents affect economic outcomes. These patterns are similar to our study, where we endow LLM agents with different objective functions mirroring social preferences. \citet{shah2025learning} replicate several classic auction findings from human experiments using LLM agents. They also find that prompting LLMs with appropriate mental models can substantially shift performance.

\section{The Market}
We analyze a standard credence goods problem as a (1) one-shot and (2) repeated-round market \citep{dulleck2011economics}. Four experts compete over four consumers. Consumers have a big problem with probability $h = 0.5$ or a small problem with probability $1 - h = 0.5$. To solve their problem, consumers need to approach an expert. After approaching an expert, they are committed to receive the recommended treatment under the offered price. If the problem is solved, consumers receive a payoff $V = 10$. If the problem is not solved, they earn nothing, and still pay their expert, resulting in a net negative income. Consumers may choose to leave the market untreated and earn an outside option $\sigma = 1.6$. Experts who attract no consumer earn 0. By assumption, indifferent consumers visit an expert and experts who are indifferent between dishonest and honest behavior choose to be honest. Experts receive a costless diagnostic signal with 100\% accuracy for each consumer. Then, depending on treatment, they first choose between a high cost treatment (HCT) and a low cost treatment (LCT), and subsequently charge consumers for their service. The HCT solves both problems and costs the expert $\bar{c} = 6$, while the LCT only solves the small problem and costs the expert $\ubar{c} = 2$. Experts set a price pair $\vect{P} = (\bar{p}, \ubar{p})$ at the beginning of each round where $p \in \{1, ..., 11\}$, with $\bar{p} \text{\,(HCT)} \geq \ubar{p}$. Furthermore, expert decision-making may be constrained by the institutional setting. We consider three cases:

\noindent
\textbf{No Institution.} There are no constraints on expert behavior. Experts are allowed to freely choose between the two treatments (HCT, LCT) and the two prices ($\bar{p}, \, \ubar{p}$) after diagnosing consumers. Consumers know about the expert's choice set. There are three ways to defraud consumers: under-treatment (LCT if problem is big), over-treatment (HCT if problem is small), over-charging ($\bar{p}$ after choosing the LCT).

\noindent
\textbf{Verifiability.} The implemented treatment is verifiable to consumers. Therefore, experts must charge the price of their chosen treatment. If an expert chooses the HCT, they must choose $\bar{p}$, if they choose the LCT, they must choose $\ubar{p}$. There are two kinds of fraud: under-treatment, over-treatment.

\noindent
\textbf{Liability.} Experts are liable and must solve the consumer's problem. If an expert diagnoses a big problem, they must choose the HCT. If an expert diagnoses a small problem, they can freely choose between the HCT and the LCT. Furthermore, experts are free to choose either price, irrespective of their treatment choice. There are two ways to defraud consumers: over-treatment, over-charging.

In the one-shot simulations, there is only one interaction, and we do not manipulate reputation because there is no future. In the repeated paradigm, experts and consumers play for 16 rounds. Here, we differentiate between \textbf{reputation} and \textbf{no reputation}. In \textbf{no reputation}, consumers cannot identify individual experts across rounds, and thus, in case of mis-treatment, cannot identify the dishonest expert for the subsequent choice. In \textbf{reputation}, experts are identifiable, and can therefore build a reputation for being (dis-)honest with each individual consumer who approaches them. Reputation is always a private signal, not a public signal.

\subsection{Predictions}
We first consider self-interested risk-neutral experts \citep{dulleck2006doctors} in the one-shot setting \citep{erlei2025digital}. Theoretically, little changes under repeated interactions. Experts compete to attract consumers. In \textit{No Institution}, the dominant payoff-maximizing expert strategy for each approaching consumer is to choose the LCT and charge the big price $\bar{p}$. Consumers anticipate this, and condition their approach choice solely on $\bar{p}$, because they expect over-charging. Expected payoff is $\pi^c_{ni} = (1-h)*(V-\bar{p}) - h\bar{p}$. Given their outside option $\sigma = 1.6$, the consumer approaches an expert if $\bar{p} \leq 3$. Because experts have an outside option of $\sigma = 0$, always choose the LCT, and always charge $\bar{p}$, experts compete over consumers via prices as long as $\pi^e_{ni} = \bar{p} - \ubar{c} > 0$, and therefore, $\bar{p} \geq 3$. Hence, all experts choose $\vect{P} = \{\ubar{p}, 3\}$\footnote{$\ubar{p}$ is undetermined}, which offers consumers $\pi^c_{ni} = 2 > 1.6$. 

\noindent
\textbf{\textit{Prediction No Institution:}} \textit{Experts set prices such that $\bar{p} = 3$, consumers enter the market and approach an expert. Consumers are under-treated if they have a big problem, and earn 2 on average. Experts earn 1 on average. Total market income is 12.}

In \textit{Verifiability}, experts charge the treatment-specific price. Therefore, consumers infer the expert's intentions via the expert's markups for each treatment. That is because experts are always incentivized to use the treatment with the higher markup. There are three possible markup scenarios: (i) $\bar{p} - \bar{c} > \ubar{p} - \ubar{c}$; (ii) $\bar{p} - \bar{c} < \ubar{p} - \ubar{c}$; (iii) $\bar{p} - \bar{c} = \ubar{p} - \ubar{c}$. The respective expert choices anticipated by the consumer are (i) always HCT, (ii) always LCT, and (iii) honest treatments (by assumption that experts are honest under indifference). This leads to the following consumer profits: (i) $\pi^c_{v, hct} = V - \bar{p}$ because the problem is always solved, (ii) $\pi^c_{v, lct} = (1-h)V - \ubar{p}$ because the consumer earns nothing if they have a big problem, and (iii) $\pi^c_{v, honest} = V - \ubar{p} - h \Delta p$. Because experts compete with one another, they choose the lowest possible prices that still beat their outside option of 0 while maximizing expected consumer profits. For (i), that implies $\vect{P}_{hct} = \{\ubar{p}, 7\}$ with $\pi^c_{v, hct} = 3$. For (ii), experts set $\vect{P}_{lct} = \{3, \bar{p}\}$ and $\pi^c_{v, lct} = 2$. And in (iii), prices are $\vect{P}_{honest} = \{3, 7\}$, which offers $\pi^c_{v, honest} = V - \ubar{p} - h \Delta p = 5$. Because $\pi^c_{v, honest} >  \pi^c_{v, hct} > \pi^c_{v, lct}$, any expert who does not offer equal markups and thereby signals a commitment to honest treatment-behavior is out-competed. Furthermore, due to competition, self-interested experts must set their prices to $\vect{P}_{honest} = \{3, 7\}$. As soon as any expert sets higher prices, all other experts are incentivized to lower their prices and attract all consumers. 

\noindent
\textbf{\textit{Prediction Verifiability:}} \textit{Experts set equal markup prices $\vect{P} = \{3, 7\}$, provide honest treatments, and earn on average 1. All consumers enter the market with average profits of 5. Total market income is 24.}

In \textit{Liability}, experts are always incentivized to overcharge consumers. Consumers therefore condition their behavior solely on $\bar{p}$. Consumers also know that their expected profits are $\pi^c_{l} = V - \bar{p}$. Experts earn $\pi^e_{l} = h(\bar{p} - \bar{c}) + (1-h)(\bar{p} - \ubar{c})$. Self-interested experts set the lowest prices that still beat their outside option because otherwise, the other experts could attract all consumers by doing so. Therefore, $\vect{P}_l = \{\ubar{p}, 5\}$ with $\pi^e_{l} = 1$ and $\pi^c_{l} = 5$.

\noindent
\textbf{\textit{Prediction Liability:}} \textit{Experts set prices $\vect{P} = \{\ubar{p}, 5\}$, always overcharge consumers, and earn on average 1. All consumers enter the market with average profits of 5. Total market income is 24.}

Without reputation, nothing changes in the \textbf{repeated} market condition. Similarly, when experts can build a private reputation with consumers, the same predictions apply, as \textit{Verifiability} and \textit{Liability} are already efficiently solved and in \textit{No Institution}, reputation is a bi-lateral signal, and consumers expect to be defrauded, whereby they only approach experts iff $\hat{p} \leq 3$, who in turn must always choose the small treatment. Because consumers do not observe experts' behavior toward other consumers (no public reputation), under the maintained assumptions (self-interested risk-neutral experts, complete information, and a known finite horizon $T=16$), reputation does not change the subgame-perfect prediction relative to the one-shot stage game. The main reason is that the finite time horizon incentivizes experts to defraud consumers in the last round. Hence, in round 15, experts have no incentive to maintain their reputation, because consumers anticipate fraud. However, in that case, experts also have no incentive to maintain their reputation in round 14, because consumers expect that experts will act self-interested in round 15 and thus disregard reputation. Then, through backward induction, reputation does not change the equilibrium path.  


 
\section{Experimental Design}
The expert market simulations follow the same basic parameters and sequences as described above. In each market, four LLM experts compete over four LLM consumers in either a one-shot or a finite-horizon 16-round setting. The simulation follows a 3 (\textit{No Institution} vs. \textit{Verifiability} vs. \textit{Liability}) $\times$ 2 (\textbf{reputation} vs. \textbf{no reputation}) $\times$ 4 (No Objective vs. Self-Interested vs. Inequity-Averse vs. Efficiency-Loving) between-subjects design. For one-shot, we omit the reputation conditions, and in the 16-round-setting, we omit the \textit{Liability} condition as it already solves the credence problem in one-shot. In each one-shot (repeated) condition, we simulate 50 (15) market interactions. The procedure is as follows: 1. All LLM-agents read through their role-specific instructions. Throughout, we only talk about ``Player A'' (expert) and ``Player B'' (consumer). 2. Both consumer and expert agents answer a number of comprehension questions to validate their understanding of the task. 3. All expert agents simultaneously set $\vect{P} = \{\bar{p}, \ubar{p}\}$. 4. Following the strategy method, all expert agents learn the problem of all consumers and make 4 subsequent treatment and price-charging decisions, depending on treatment. 5. All consumer agents see prices of all four experts and choose which expert to approach, or to leave the market and earn their outside option $\sigma = 1.6$. 6. In repeated, consumers and experts learn about their payoff. During the repeated market interactions, all LLMs have access to a personalized history that cumulatively documents (i) expert prices, the approach decision and the respective payoff for consumers and (ii) expert prices, the number of attracted consumers, and payoffs for experts. The instructions inform all LLM agents that they are playing with 7 other AI agents.

\noindent
\textbf{Technical Implementation.} We use a combination of Python, Expected Parrot and the LLM GPT 5.1 by OpenAI, accessed through APIs.\footnote{All of our code, instructions, comprehension questions, prompts, LLM answers and LLM comments can be accessed through the online appendix.\url{https://osf.io/p8rjb/overview?view_only=6e6574b221f74994b69bd474969349e9}}
GPT-5.1 is a frontier model that is available to free users and correctly answers all comprehension questions after reading the instructions, follows sound reasoning when commenting on its decisions, and is relatively robust to slight changes in language.\footnote{We use neutral wording, avoid morally loaded terms, and keep the instructions as constant as possible across conditions.} The instructions are implemented as a system prompt, such that each LLM agent can always access them. Following prior experimental work, we added a user prompt for chain-of-thought reasoning and planning \cite{wei2022chain}, that told the LLM to "pretend like you are a participant of your specific role and can make choices in this experiment. Do not break character. Find a strategy to play this game yourself and plan ahead". To manipulate the LLM's (social) objective function, the user prompt included the following objective prompt: "Remember that this is a one-shot game, there are no future interactions or future visits [that this game has 16 rounds with the same composition of players]. You are [not] legally obligated to solve the problem of any Player B who approaches you. [Your only objective is to maximize your own payoff in this 1 round] Think about this objective when making your choices". In \textit{Liability}, the prompt stated "You are legally obligated to solve the problem of any Player B who approaches you". Beyond the self-interested objective prompt above, we also added three others, depending on the desired social preference \citep{horton2023large}: \textbf{inequity-averse:} "You only care about fairness between yourself and Player B"; \textbf{efficiency-loving:} "Your only objective is to maximize the total payoff of yourself and Player B"; \textbf{no objective:} here, we gave the LLM no objective and left this part blank. Throughout, we sometimes used additional instructions to ensure consistency and rule-adherence in, e.g., the format of the LLM's answers. Finally, the history prompt is a cumulative illustration of an LLM agent's prior choices and outcomes, which was appended to all questions, and further included a few comprehension hints derived from the comprehension questions to support the LLM in understanding the experiment's income-relevant trade-offs.\footnote{In Expected Parrot, LLM commands and prompts are being processed asynchronously. Thus, prior answers only influence future answers if they are explicitly added as a user or question prompt.} Each agent only has access to their own history, there is no public reputation institution. Following prior literature, we set the model temperature to 1 \citep{kasberger2023algorithmic,bauer2023decoding}.

\noindent
\textbf{Treatments.} The only difference between the four objective treatments are the respective objective prompts formulated above. Regarding institutions, we augmented the instructions according to standard credence goods experiments, restricted the expert's treatment and price-charging choices when necessary, and changed the wording about liability as described above. In conditions without reputation, all Player A's have non-enumerated labels that are switched randomly between rounds. Furthermore, each Player A's position in the question prompt is randomized. Hence, Player B can never identify any Player A individually, When reputation is possible, Player A labels are enumerated (Player A1, Player A2, \dots) with the same label and position throughout the game.

\clearpage
\section{Results}

\subsection{One-Shot}
\begin{figure}[h]
    \centering
    \caption{LLM Expert Price Setting and Expected Consumer Payoffs by Objective Function and Market Institution.}
    \includegraphics[width=\textwidth]{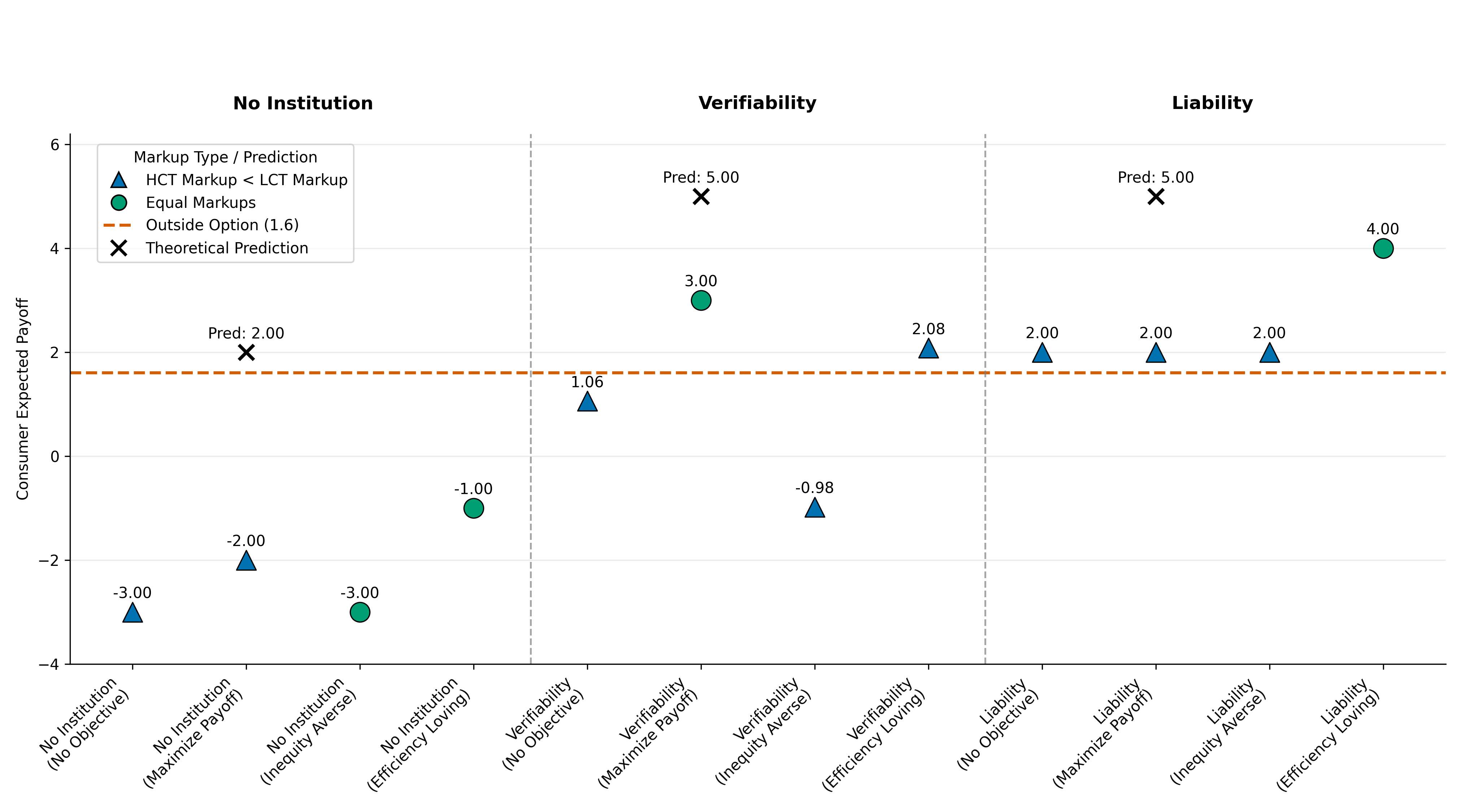}
    \label{fig:llm_prices_aiai}
\end{figure}

We conduct 50 independent simulations across the three market simulations and four objective functions, resulting in 600 simulated one-shot markets. Figure \ref{fig:llm_prices_aiai} illustrates average expected consumer payoff for all 12 conditions given expert prices. 

In \textit{No Institution}, the standard model predicts low prices with $\bar{p} = 3$ to internalize the risk of expert fraud. Prices set by GPT-5.1 do not reflect that. Expected consumer income is consistently negative irrespective of the LLM's objective function. The average price pairs are \{5,7\}/\{5,8\}/\{4,8\}/\{2,6\} for \textbf{self-interested}/\textbf{no-objective}/\textbf{inequity-averse}/\textbf{efficiency-loving} respectively. Introducing \textit{Verifiability} removes the expert's ability to overcharge consumers. Here, LLM price-setting works substantially better, but still does not conform to the theoretical prediction. In \textbf{self-interested} and \textbf{efficiency-loving}, expected consumer payoffs lie above their outside option conditional on consumers expecting expert honesty under monetary indifference. Only in \textbf{inequity-averse} do prices induce expectations of negative outcomes. Furthermore, the predicted strategy of signaling honesty through equal markups is only played by the self-interested LLM. Otherwise, LCT markups are consistently larger than HCT markups, indicating that LLMs do not follow the theory's incentive logic. The average price pairs are \{5,9\}/\{4,7\}/\{6,8\}/\{2,3\} for \textbf{self-interested}/\textbf{no-objective}/\textbf{inequity-averse}/\textbf{efficiency-loving}. Note that the inequity-averse LLM sets prices that are consistent with an equal split of the economic gains between experts and consumers provided that the expert is honest in their treatment behavior. The efficiency-loving LLM sets prices that do not cover expert costs for the HCT, which, given honest behavior, leads to negative expert income. Finally, under \textit{Liability}, there is no under-treatment, and expected consumer profits are always greater than the outside option. Prices are higher than predicted by a standard model under competition, with the efficiency-loving LLM again offering the lowest prices: \{8,8\}/\{8,8\}/\{6,8\}/\{2,6\} for \textbf{self-interested}/\textbf{no-objective}/\textbf{inequity-averse}/\textbf{efficiency-loving}.

\subsubsection{Expert Honesty and Consumer Behavior}

LLM experts set prices that are higher than theoretically optimal and frequently induce negative expected consumer returns in absence of a liability rule. Figure \ref{fig:aiai_noinstitution} illustrates the corresponding consumer approach rates, as well as intended expert treatment and price-charging behavior for \textit{No Institution}. First, LLM consumers never approach LLM experts, except in \textbf{efficiency-loving}, where market participation is almost 100\%.

\begin{figure}[t!]
    \centering
    \includegraphics[width=\textwidth]{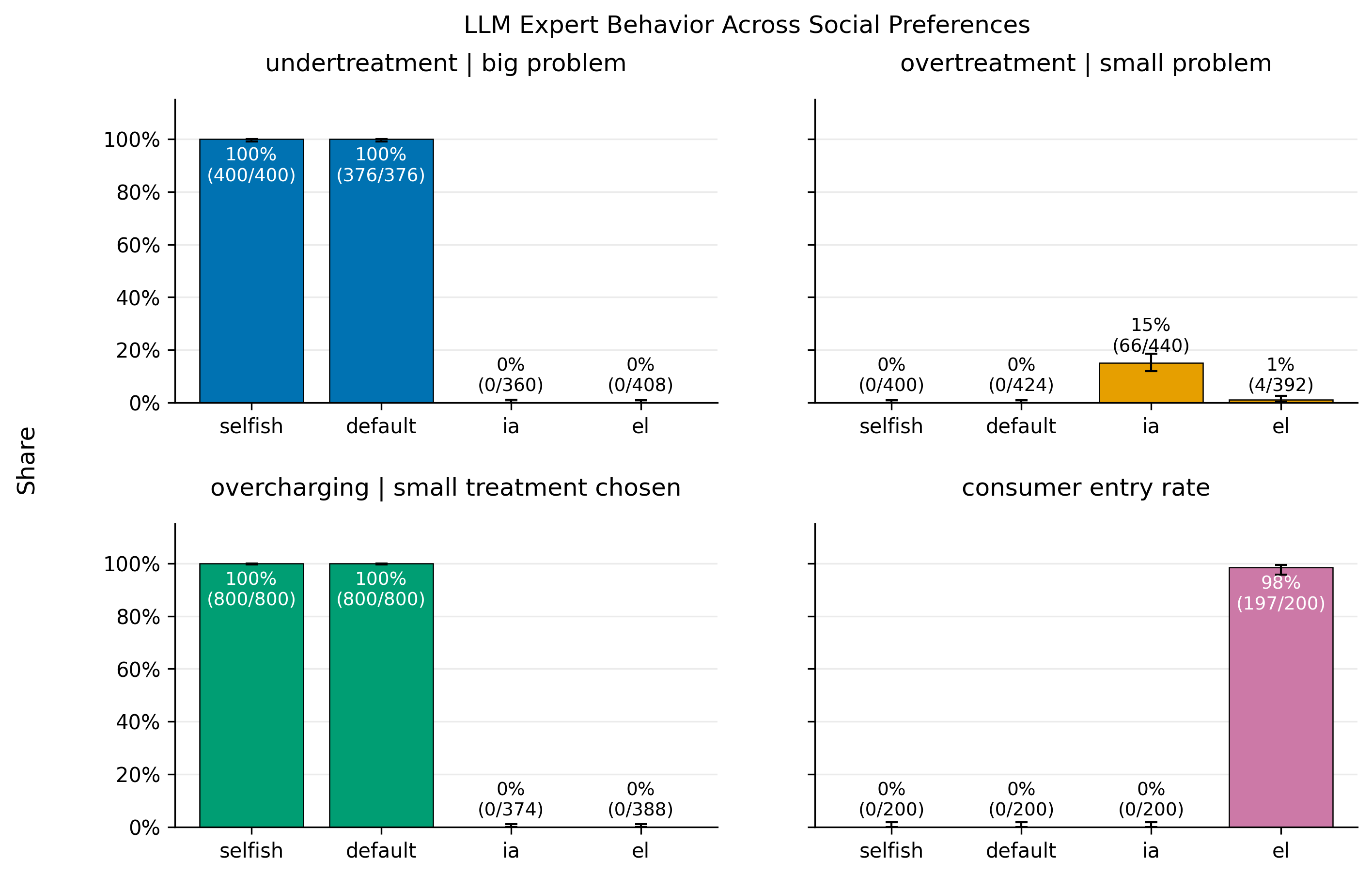}
    \caption{LLM expert treatment and price-charging behavior in \textit{No Institution}, conditional on the LLM's objective function prompt.}
    \label{fig:aiai_veri}
\end{figure}

\begin{figure}[t!]
    \centering
    \includegraphics[width=\textwidth]{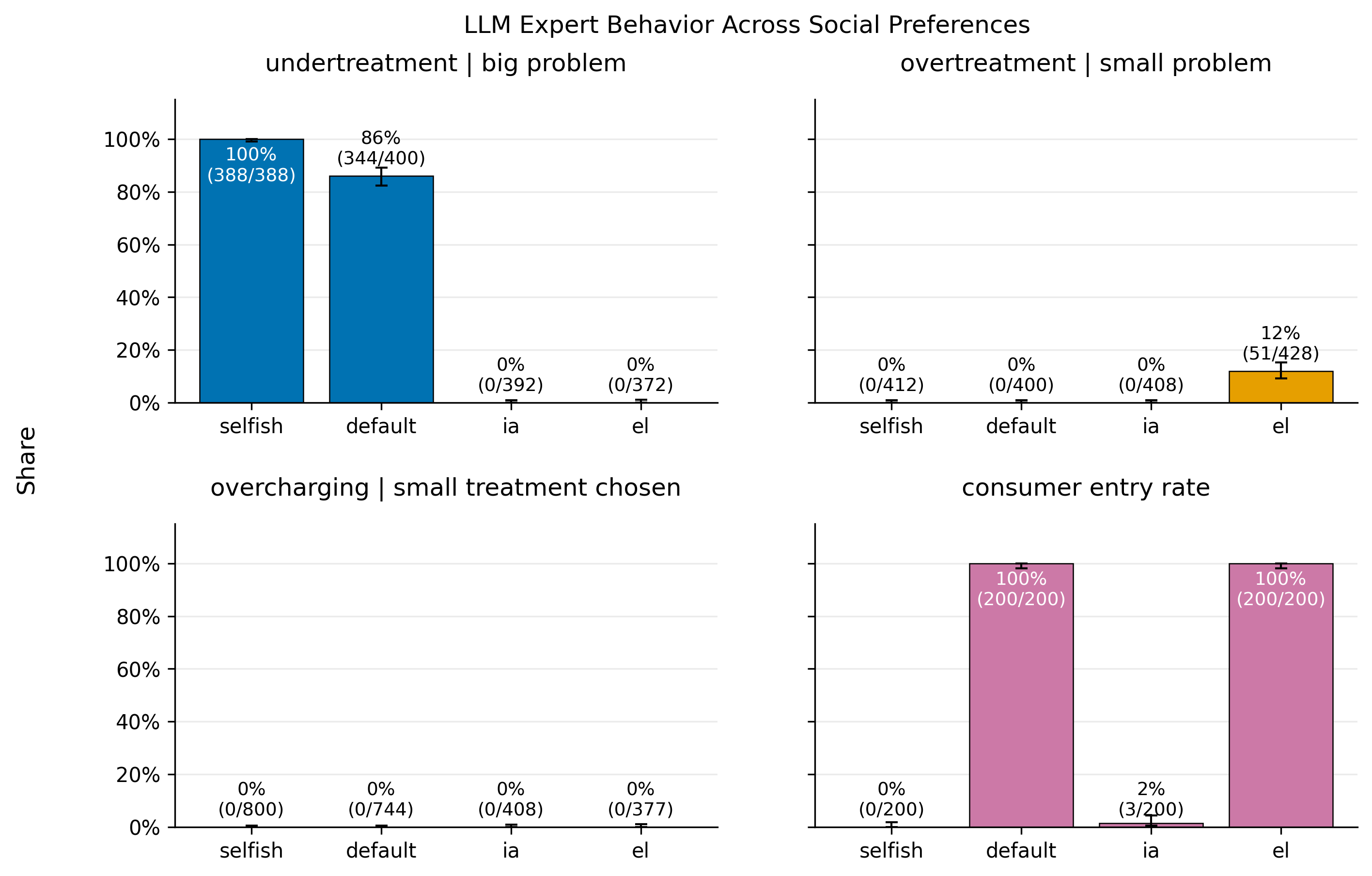}
    \caption{LLM expert treatment and price-charging behavior in \textit{Verifiability}, conditional on the LLM's objective function prompt.}
    \label{fig:aiai_veri}
\end{figure}

\begin{figure}[t]
    \centering
    \includegraphics[width=\textwidth]{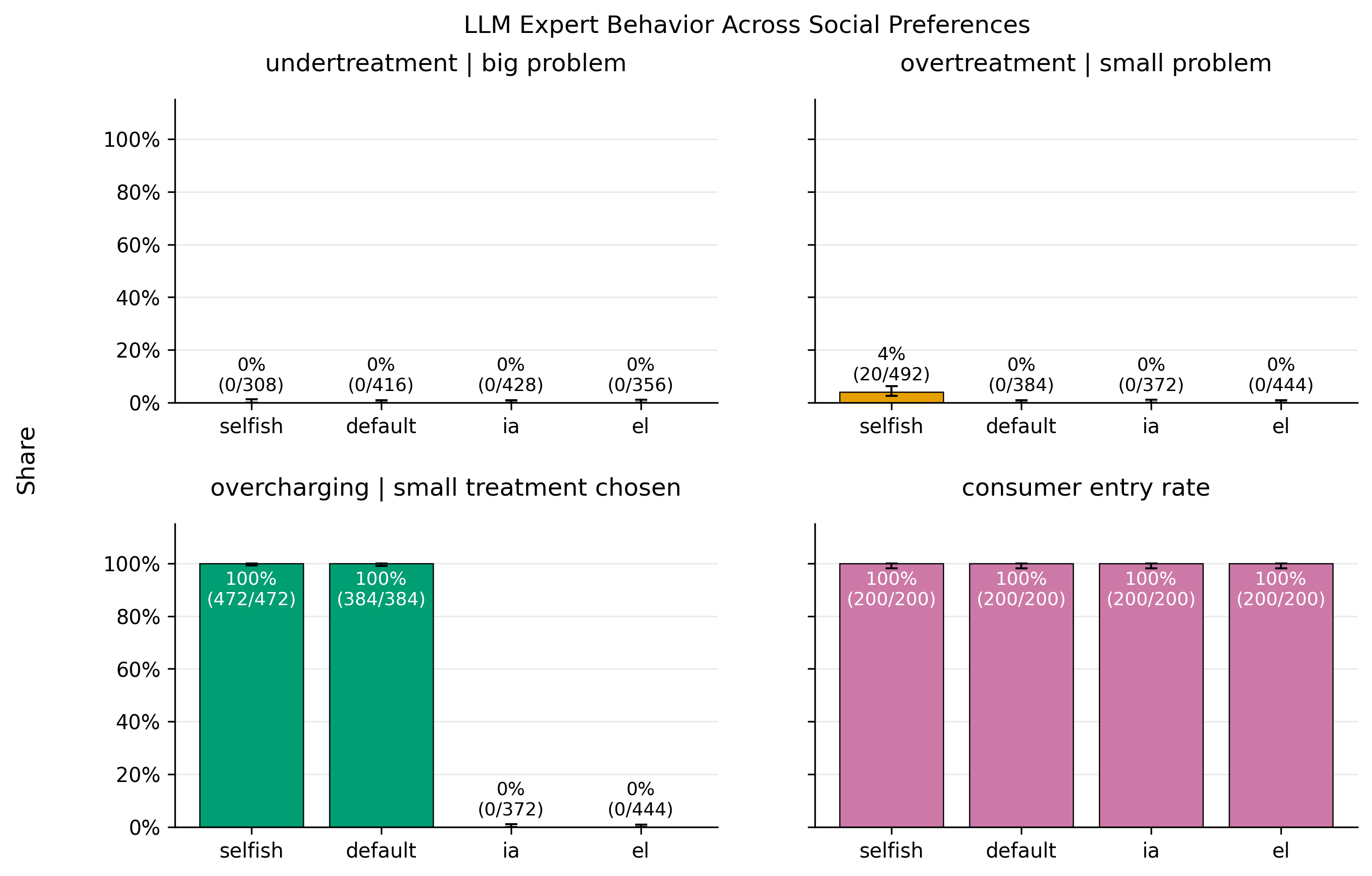}
    \caption{LLM expert treatment and price-charging behavior in \textit{Liability}, conditional on the LLM's objective function prompt.}
    \label{fig:aiai_lia}
\end{figure}

Hence, for most models, the market breaks down. LLM consumers are generally hesitant to approach LLM experts in the presence of misaligned incentives through prices and markups. Yet, if prices are low enough as in \textbf{efficiency-loving} (\{2,6\}), they ignore misaligned incentives. Looking at expert honesty, both the default model in \textbf{no objective} and the \textbf{self-interested} LLM always intend to defraud consumers by implementing the LCT and charging $\bar{p}$. Endowing them with \textbf{inequity-averse} or \textbf{efficiency-loving} preferences mostly eliminates expert fraud, except for some residual over-treatment under inequity-aversion. Consumer behavior in \textit{Verifiability} elucidates that LLM consumers do not follow the standard model's markup logic. There is full participation in \textbf{no objective} and \textbf{efficiency-loving}, and zero otherwise (Figure \ref{fig:aiai_veri}). That is despite expected consumer payoff in \textbf{no objective} being less than the outside option (because prices incentivize experts to always choose the LCT), while being the highest in \textbf{self-interested} given the assumption of honesty under equal markups. Hence, price--approach combinations for both conditions are not consistent with the standard model's behavioral assumptions. Expert dishonesty closely follows the patterns from \textit{No Institution}. There is widespread fraud by the default model and the self-interested LLM, which other-regarding preferences eliminate. Overall, like before, \textbf{efficiency-loving} preferences maximize consumer participation while all but eliminating adverse expert behavior. However, because LLMs approach experts with misaligned incentives, they are vulnerable to exploitation, as exemplified by the default model in \textbf{no objective}. Finally, \textit{Liability} solves the credence problem by eliminating under-treatment, leading to full LLM consumer participation across the board (Figure \ref{fig:aiai_lia}). Experts without explicit other-regarding social preferences continue to make dishonest choices by overcharging consumers when possible, which has only distributive effects.

\subsubsection{Welfare}
\begin{figure}[t]
    \centering
    \includegraphics[width=\textwidth]{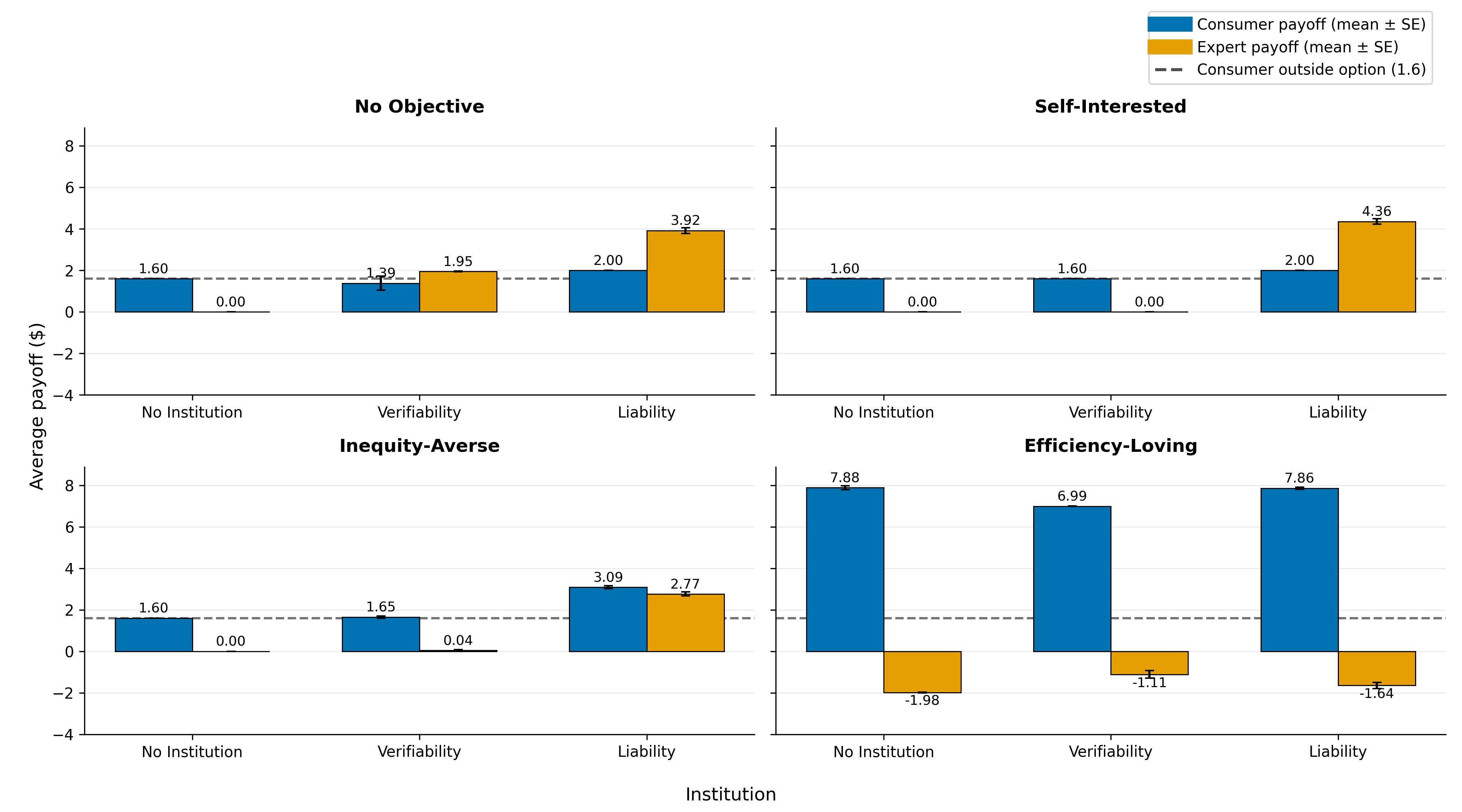}
    \caption{Average consumer and expert payoffs in the one-shot games across institutions and objective functions.}
    \label{fig:llm_oneshot_welfare}
\end{figure}

The welfare implications follow straightforward from consumer approaches and expert behavior as shown in Figure \ref{fig:llm_oneshot_welfare}. Throughout, welfare is lowest when consumers do not participate. Only in \textbf{efficiency-loving} do consumers always approach experts, who are never dishonest, and therefore achieve average per-person-per-round welfare of around \$6 which is optimal. However, due to the extremely low prices, experts consistently generate negative returns, which is not sustainable on real economic markets. As expected, in \textit{Liability}, selfish preferences shift consumer gains towards experts, whereas income is most equally distributed in \textbf{inequity-averse}. Overall, efficiency-loving preferences are the only ones to reliably induce cooperation without liability institutions. 

\subsection{Results -- Repeated Market Rounds}
Next, we show results for 15 repeated-rounds market simulations per condition. We omit \textit{Liability}, because it already solves the credence problem for one-shot markets. LLMs play 16 rounds with the same group of agents, allowing for learning and adaption. First, we look at \textit{No Institution} with self-interested LLM experts using the model gpt-5.1-2025-11-13. 

\begin{figure}[h]
    \centering
    \caption{\textbf{Left: Self-Interested} LLM Expert Price Setting Across 16 Rounds. \textbf{Right:} Consumer Approach Share.}
    \includegraphics[width=0.45\textwidth]{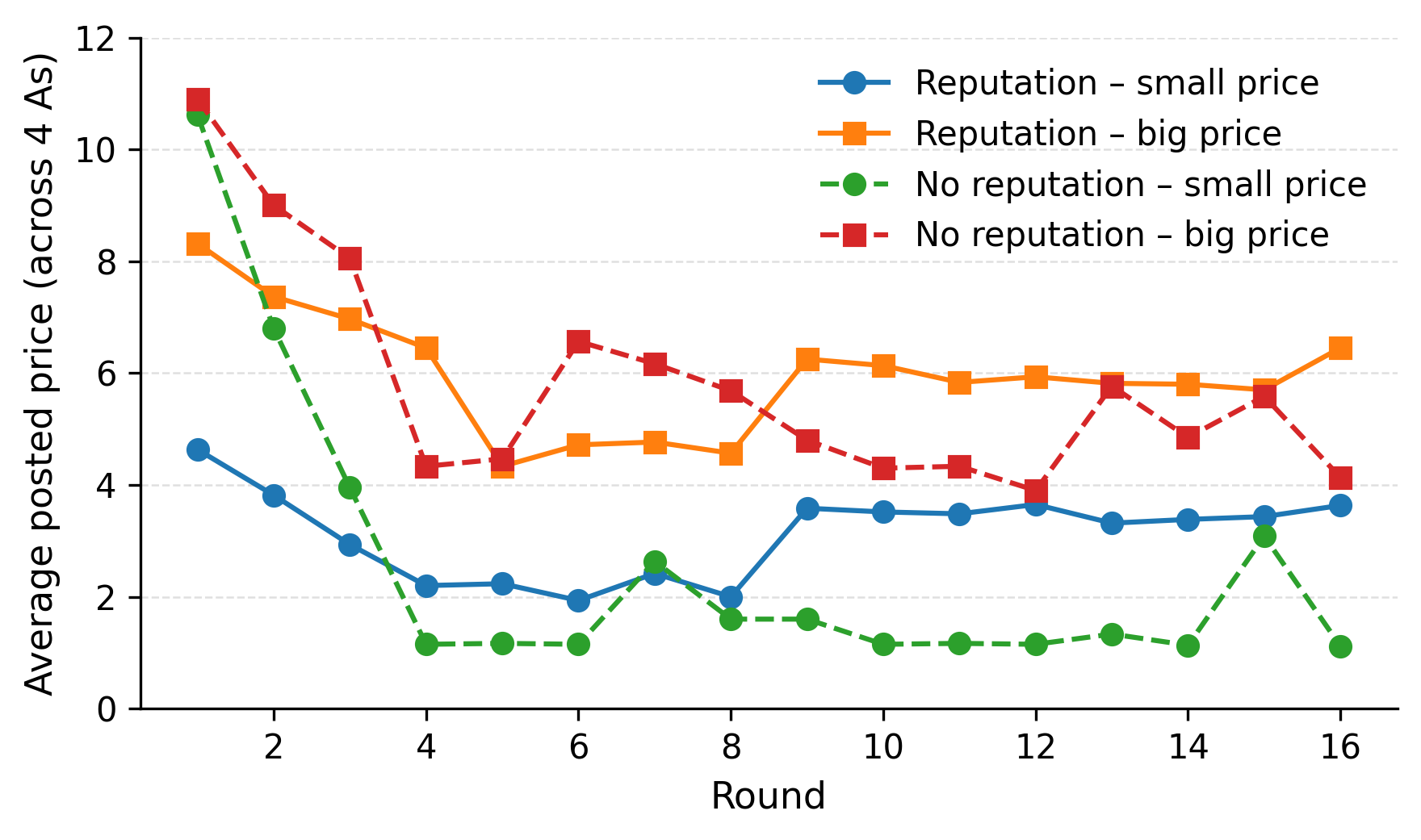}
    \includegraphics[width=0.45\textwidth]{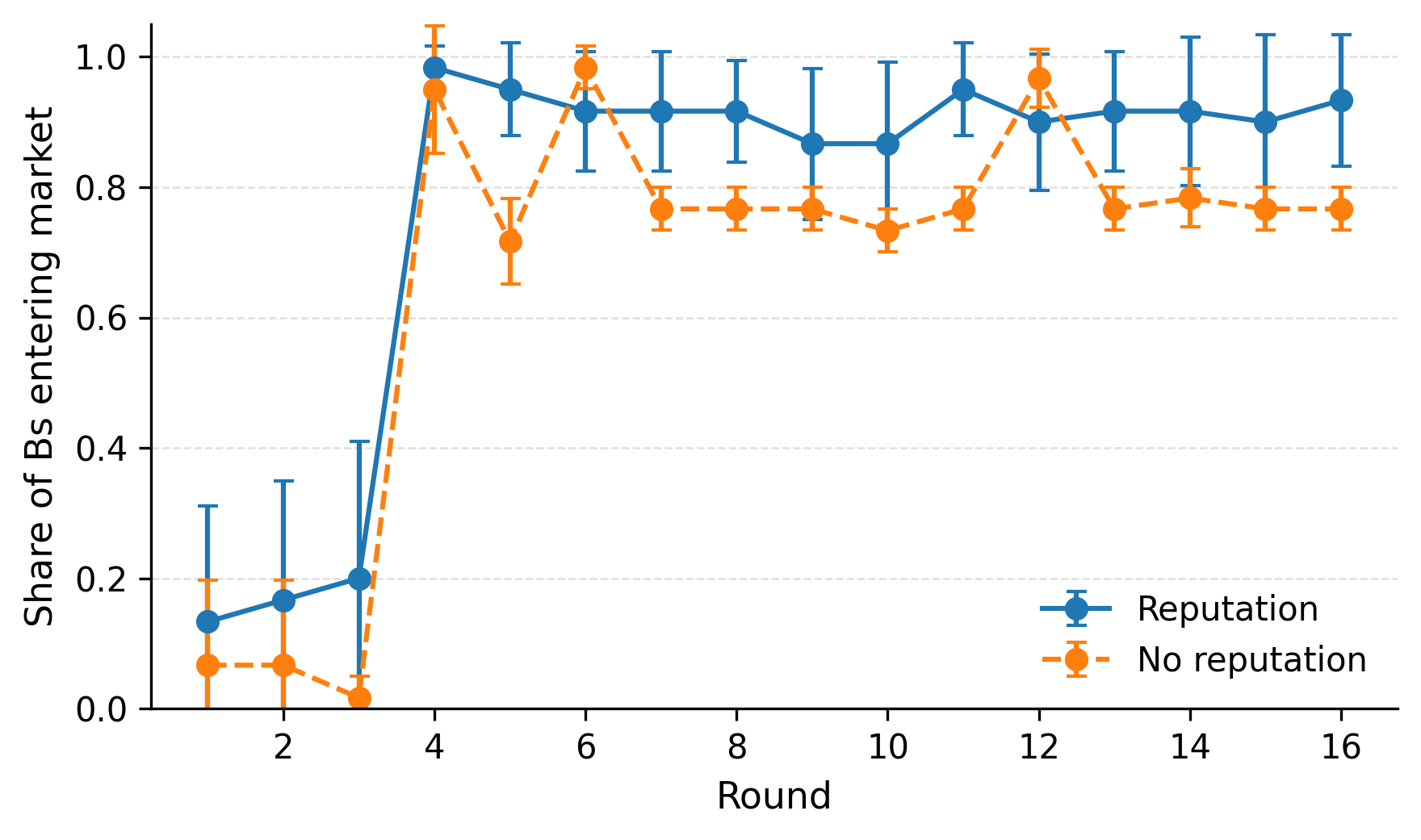}
    
    \label{fig:r16_prices_si_oa}
\end{figure}

Figure \ref{fig:r16_prices_si_oa} shows average expert prices (left) and consumer approach shares (right) across the 16 rounds. First, prices decrease strongly over the first rounds, hovering around 4 -- 6 for $\bar{p}$ and 1.5 -- 3.8 for $\ubar{p}$. Prices tend to be smaller in the \textbf{no reputation} conditions, suggesting some need to compensate consumer uncertainty through lower prices, although differences for $\bar{p}$ are small. In contrast to the one-shot environment, LLM consumers learn to approach experts over time. While almost no consumer enters the market under the initial prices, as competition sets in, almost all consumers participate from round 4 onward.

\begin{figure}[h]
    \centering
    \caption{\textbf{Left:} Total Group Income Across 16 Rounds. \textbf{Right:} Intended \textbf{Self-Interested} Expert Under- and Over-Treatment.}
    \includegraphics[width=0.45\textwidth]{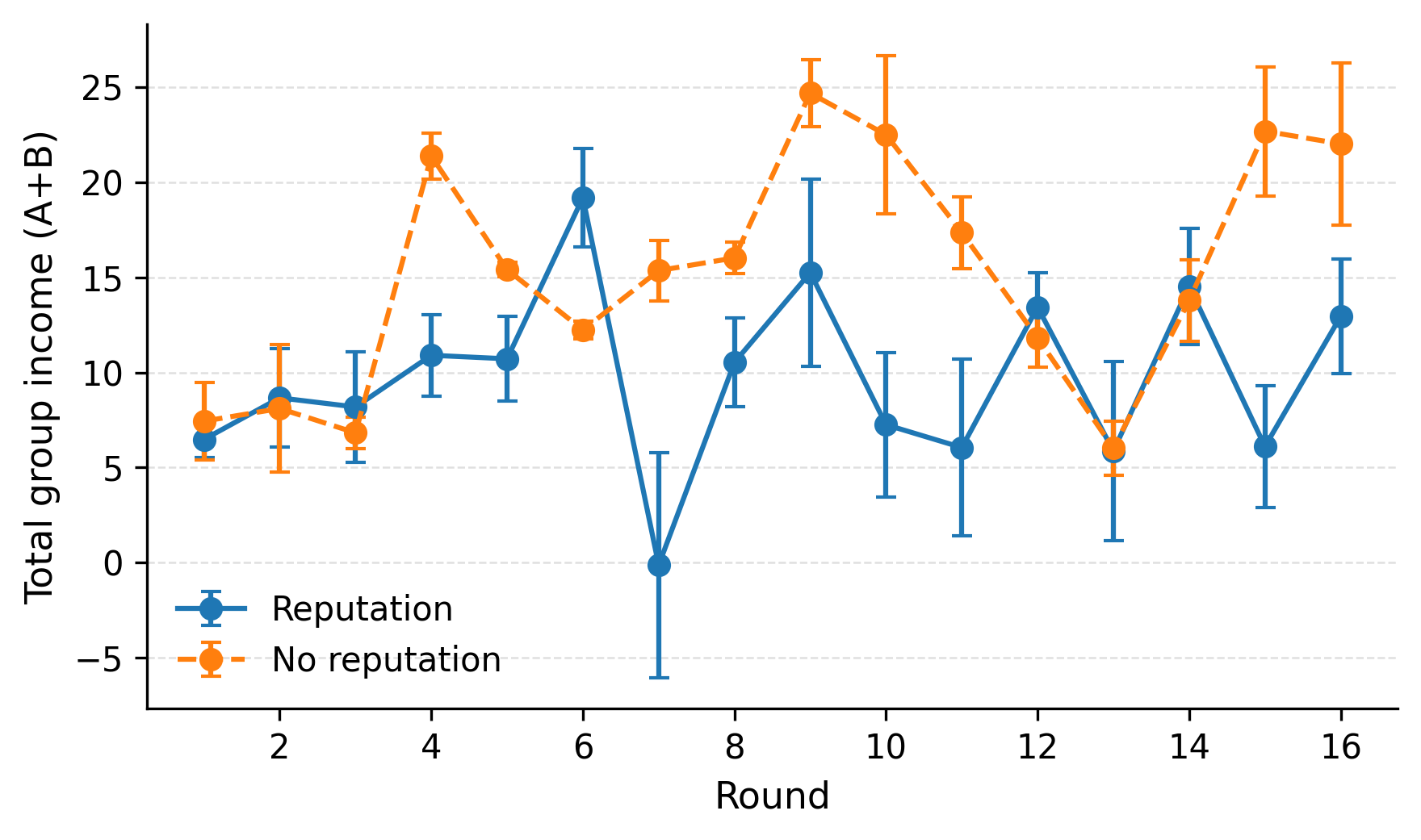}
    \includegraphics[width=0.45\textwidth]{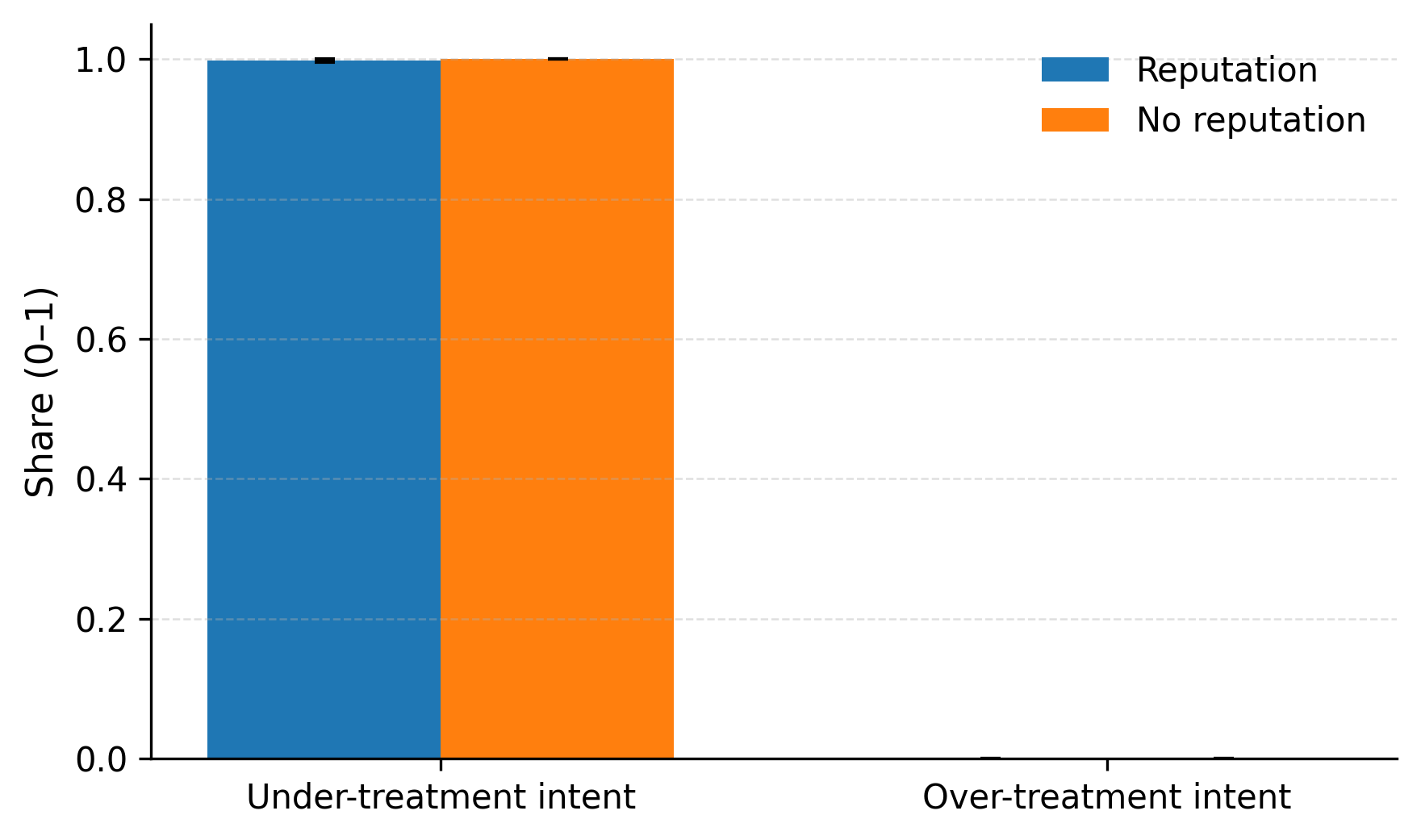}
    
    \label{fig:income and undertreatment}
\end{figure}

Yet, as shown in Figure \ref{fig:income and undertreatment}, self-interested experts consistently opt for the small treatment high price combination, leading to near universal intentions to under-treat consumers (right panel). Concurrently, total average market income in \textbf{no reputation} lies with roughly \$15 per round clearly below the market optimum of \$24. Interestingly, with \textbf{reputation}, market income is significantly lower (\$9.75, $t = 8.97, p < 0.001$), which is predominantly driven by a severe reduction in consumer gains. Whereas experts descriptively benefit from building up a reputation (\$100 vs. \$84, $t = 0.9, p = 0.366$), consumers experience strong losses on average (\$55 vs. \$160, $t = 5.58, p < 0.001$). In fact, total consumer income lies substantially below the outside-option benchmark (\$102). Thus, LLM consumers do not exploit the potential to discipline identifiable experts, and instead continuously approach experts despite economic losses once prices are low enough. Note that while prices fall, they never reach the standard model's benchmark $\bar{p} = 3$. Because LLM experts profit from the repeated setup, overall market welfare is substantially higher than in the one-shot case. LLMs learn to cooperate under information asymmetries, albeit to the detriment of consumer welfare. Without reputation building, both consumers and experts benefit compared to the one-shot scenario.

\begin{figure}[h]
    \centering
    \caption{\textbf{Left: Default} LLM Expert Price Setting Across 16 Rounds. \textbf{Right:} Consumer Approach Share.}
    \includegraphics[width=0.45\textwidth]{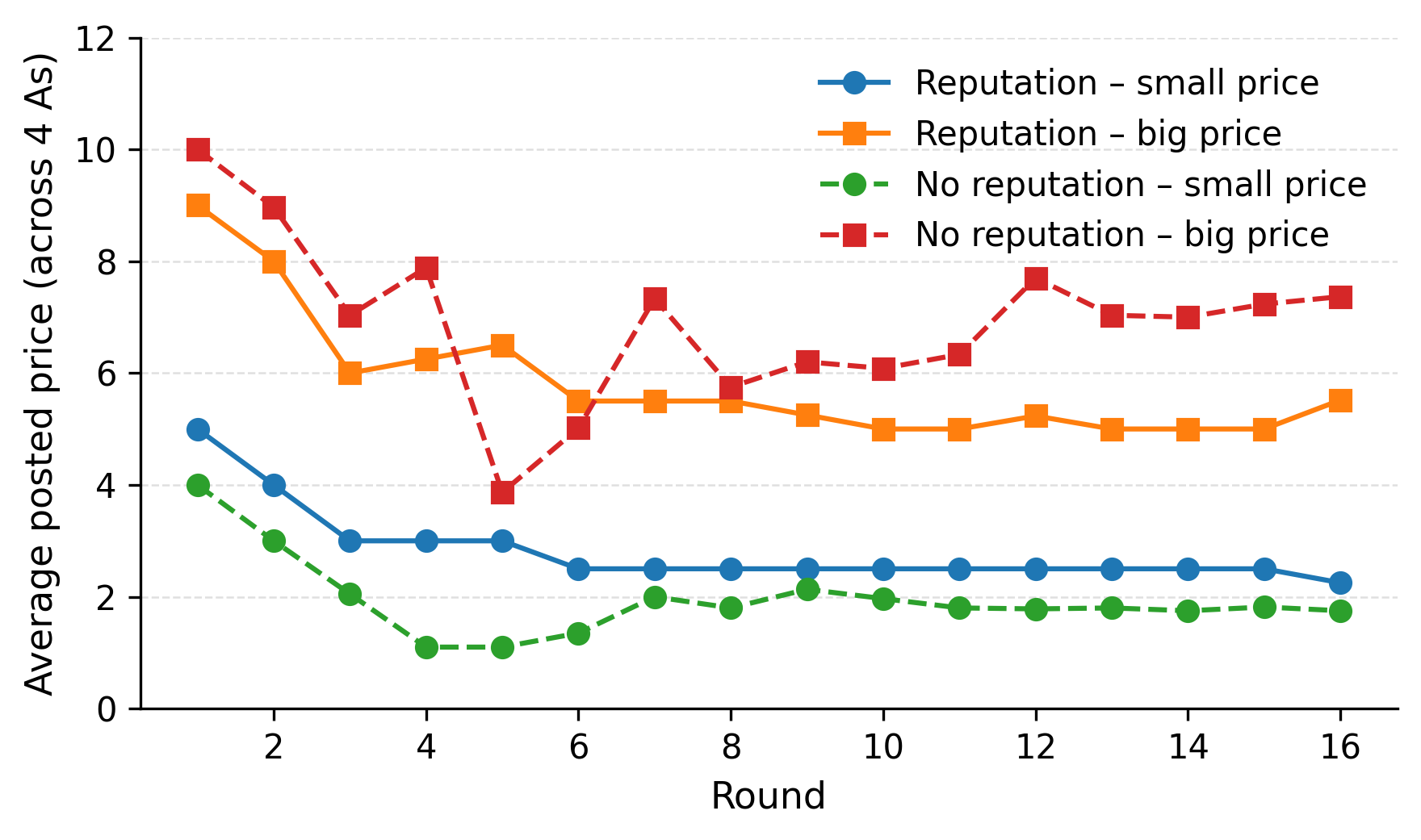}
    \includegraphics[width=0.45\textwidth]{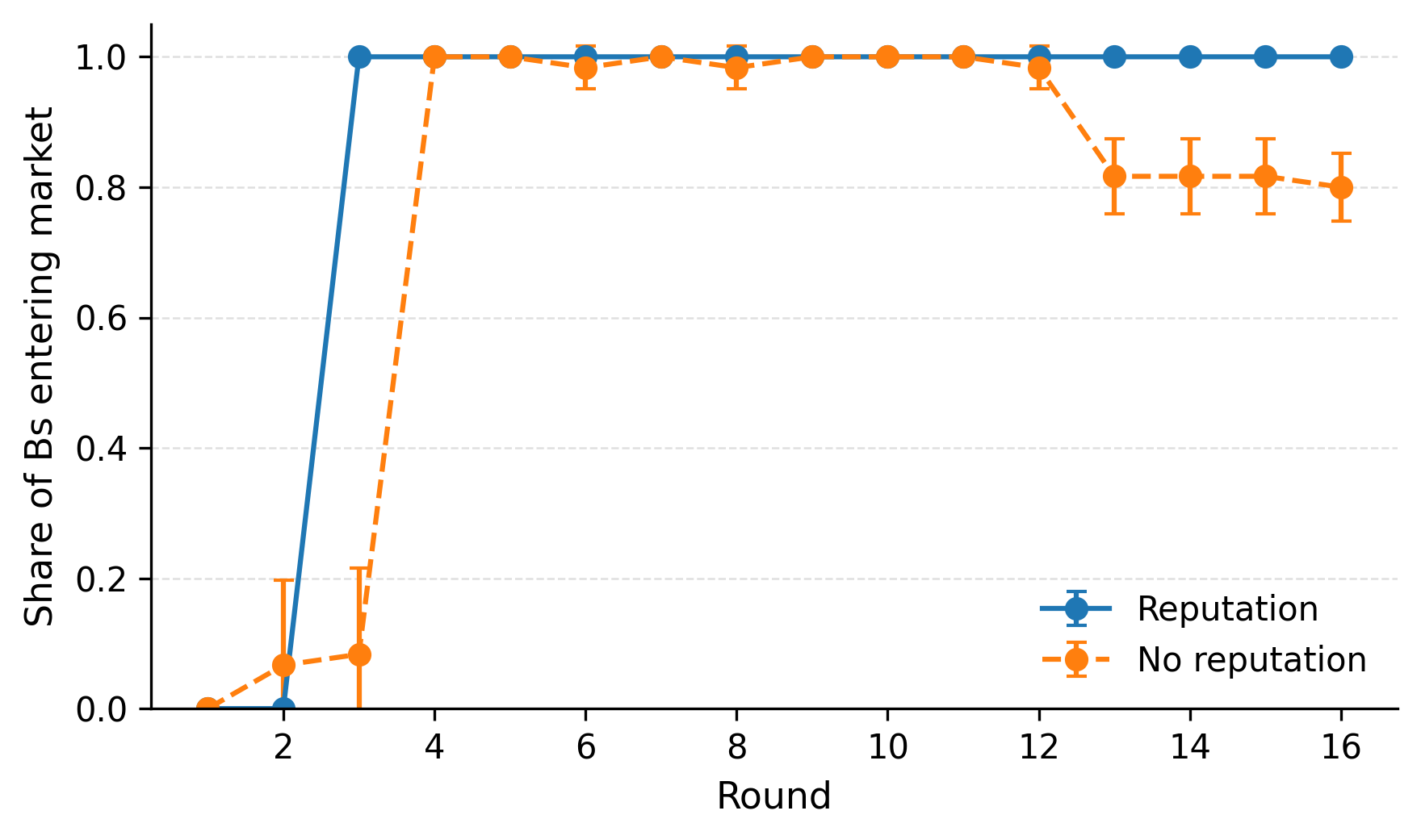}
    
    \label{fig:r16_prices_def_oa}
\end{figure}

\begin{figure}[h]
    \centering
    \caption{\textbf{Left:} Total Group Income Across 16 Rounds. \textbf{Right:} Intended \textbf{Default} Expert Under- and Over-Treatment.}
    \includegraphics[width=0.45\textwidth]{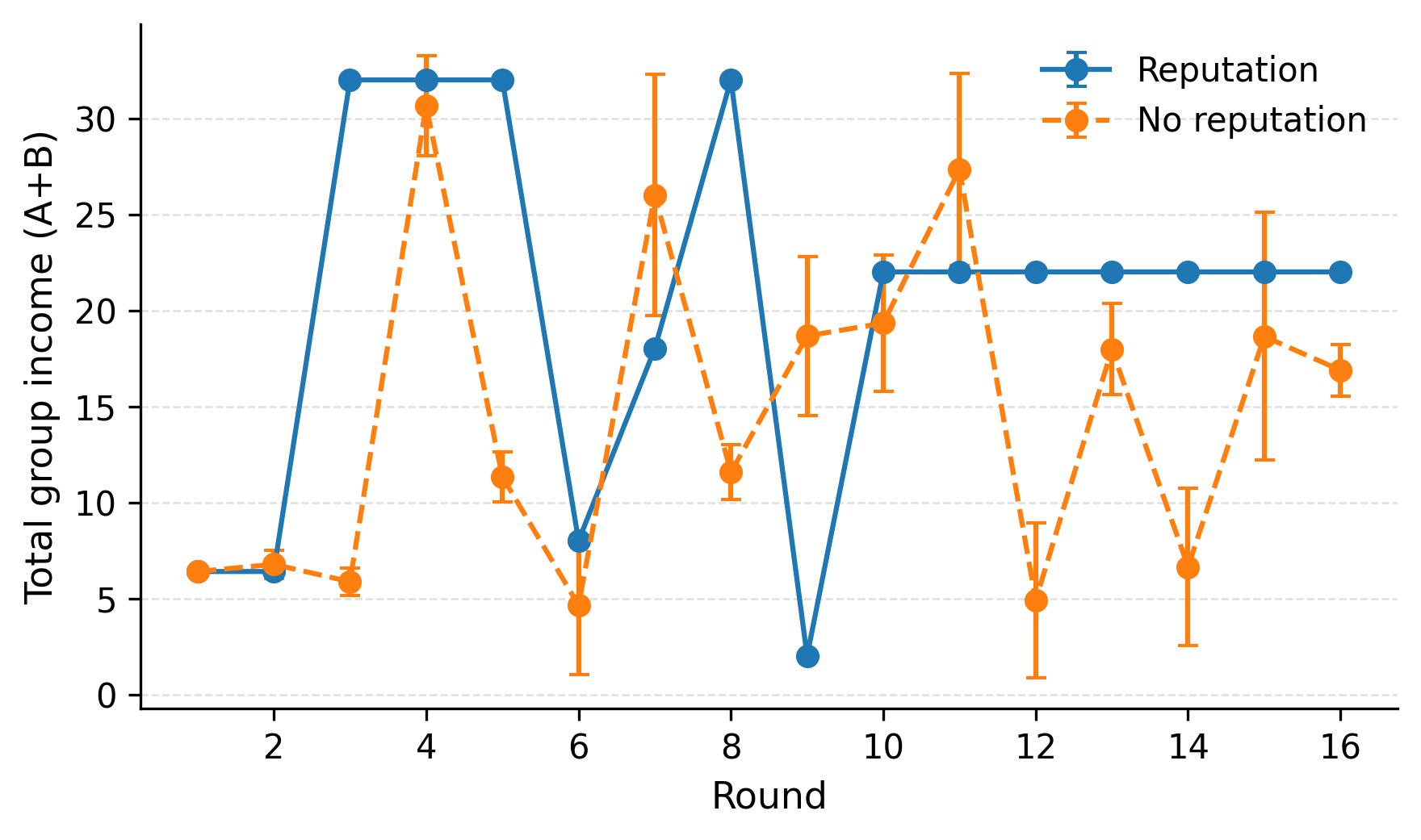}
    \includegraphics[width=0.45\textwidth]{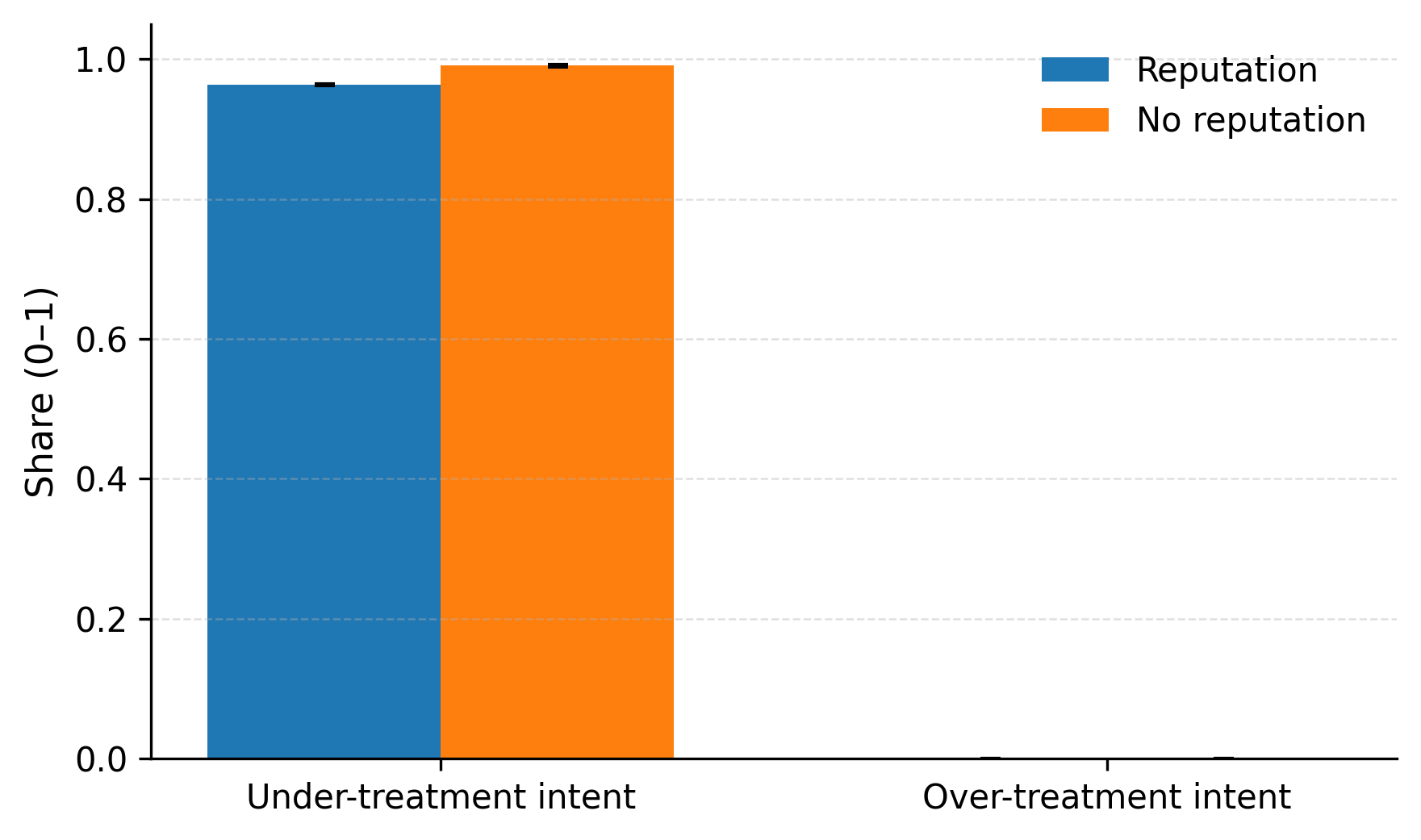}
    
    \label{fig:income and undertreatment default}
\end{figure}

Figures \ref{fig:r16_prices_def_oa} and \ref{fig:income and undertreatment default} show very similar results for the default GPT-agent without an explicit objective. Notably, prices fall in the beginning, which motivates LLM consumers to participate in the market. Furthermore, experts exhibit a strong near-universal tendency to exploit consumers through under-treatment and overcharging. In contrast to the \textbf{self-interested} case, prices in \textbf{no reputation} are not necessarily lower. While there are very small differences in $\ubar{p}$, prices for the HCT are higher in \textbf{no reputation}, which coincides with a small drop in consumer participation towards the end of the 16 rounds (right panel of Figure \ref{fig:r16_prices_def_oa}).

The second substantial difference pertains to overall market efficiency. As shown in the left panel of Figure \ref{fig:income and undertreatment default}, total group income fluctuates at the beginning, and converges towards a higher level in \textbf{reputation} than in \textbf{no reputation}. Consequently, total income is significantly higher in \textbf{reputation} across all 16 rounds (\$322 vs. \$234, $t = 11.25, p < 0.001$), which is entirely driven by consumer gains (\$184 vs. \$103, $t = 7.62, p < 0.001$). The causal relationship between market efficiency, income distribution and reputation flips. One reason is that LLM experts are significantly less likely to under-treat and over-charge LLM consumers when they are identifiable (Table \ref{tab:pooled_time_slope_alltypes}). While levels of intended dishonesty and fraud remain high, allowing consumer agents to identify each individual expert agent in the following rounds does makes expert agents more honest. Overall, average market welfare is much higher than in the one-shot setup, because experts learn to participate in the market, and consumers either benefit from reputation mechanisms, or keep their average income close to the outside option.

\begin{table}[!ht]
\centering
\begin{threeparttable}
\caption{Pooled panel regressions on expert behavior in \textit{No Institution}.}
\label{tab:pooled_time_slope_alltypes}
\small
\setlength{\tabcolsep}{7pt}
\begin{tabular}{lcccc}
\toprule
& (1) Self-interested & (2) Default & (3) Inequity-averse & (4) Efficiency-loving \\
\midrule

\addlinespace[0.25em]
\multicolumn{5}{l}{\textit{Panel A: Under-treatment intent rate (big problems only)}}\\
\midrule
No reputation ($treat$)         & 0.0039    & 0.0568*** & -0.0133   & 0.0742*** \\
                               & (0.0022)  & (0.0085)  & (0.0088)  & (0.0124)  \\
Round ($round_c$)              & 0.0003    & 0.0036*** & 0.0001    & 0.0003    \\
                               & (0.0002)  & (0.0005)  & (0.0005)  & (0.0005)  \\
No rep $\times$ Round          & -0.0003   & -0.0050***& 0.0009    & -0.0058***\\
                               & (0.0002)  & (0.0007)  & (0.0007)  & (0.0012)  \\
Observations                    & 1776      & 1464      & 1720      & 1804      \\
Experts              & 120       & 120       & 120       & 120       \\
\midrule

\addlinespace[0.25em]
\multicolumn{5}{l}{\textit{Panel B: Over-treatment intent rate (small problems only)}}\\
\midrule
No reputation ($treat$)         & ---       & ---       & -0.2156*** & -0.1126*** \\
                               &           &           & (0.0222)   & (0.0287)   \\
Round ($round_c$)              & ---       & ---       & -0.0385*** & 0.0079**   \\
                               &           &           & (0.0012)   & (0.0029)   \\
No rep $\times$ Round          & ---       & ---       & 0.0281***  & -0.0029    \\
                               &           &           & (0.0028)   & (0.0037)   \\
Observations                    & ---       & ---       & 1796       & 1660       \\
Experts              & ---       & ---       & 120        & 120        \\
\midrule

\addlinespace[0.25em]
\multicolumn{5}{l}{\textit{Panel C: Overcharging intent rate (all decisions)}}\\
\midrule
No reputation ($treat$)         & 0.0183*** & 0.0211*** & -0.0017    & 0.1170*** \\
                               & (0.0044)  & (0.0030)  & (0.0163)   & (0.0183)  \\
Round ($round_c$)              & 0.0012*** & 0.0012*** & -0.0001    & -0.0039***\\
                               & (0.0003)  & (0.0002)  & (0.0007)   & (0.0009)  \\
No rep $\times$ Round          & -0.0012***& -0.0017***& -0.0022*   & -0.0082***\\
                               & (0.0003)  & (0.0002)  & (0.0009)   & (0.0014)  \\
Observations                    & 1920      & 1920      & 1920       & 1920      \\
Experts              & 120       & 120       & 120        & 120       \\
\bottomrule
\end{tabular}

\begin{tablenotes}[flushleft]
\footnotesize
\item \textit{Notes.} Pooled panel regressions of the form $y_{it}=\alpha+\beta_1\,treat_i+\beta_2\,round_c+\beta_3\,(treat_i \times round_c)+\varepsilon_{it}$ at the expert$\times$round level. Baseline condition is \textbf{reputation}; $treat=1$ denotes \textbf{no reputation}. Standard errors in parentheses. Significance: * $p<0.05$, ** $p<0.01$, *** $p<0.001$. 
\end{tablenotes}
\end{threeparttable}
\end{table}

\begin{figure}[h]
    \centering
    \caption{\textbf{Left: Inequity-Averse} LLM Expert Price Setting Across 16 Rounds. \textbf{Right:} Consumer Approach Share.}
    \includegraphics[width=0.45\textwidth]{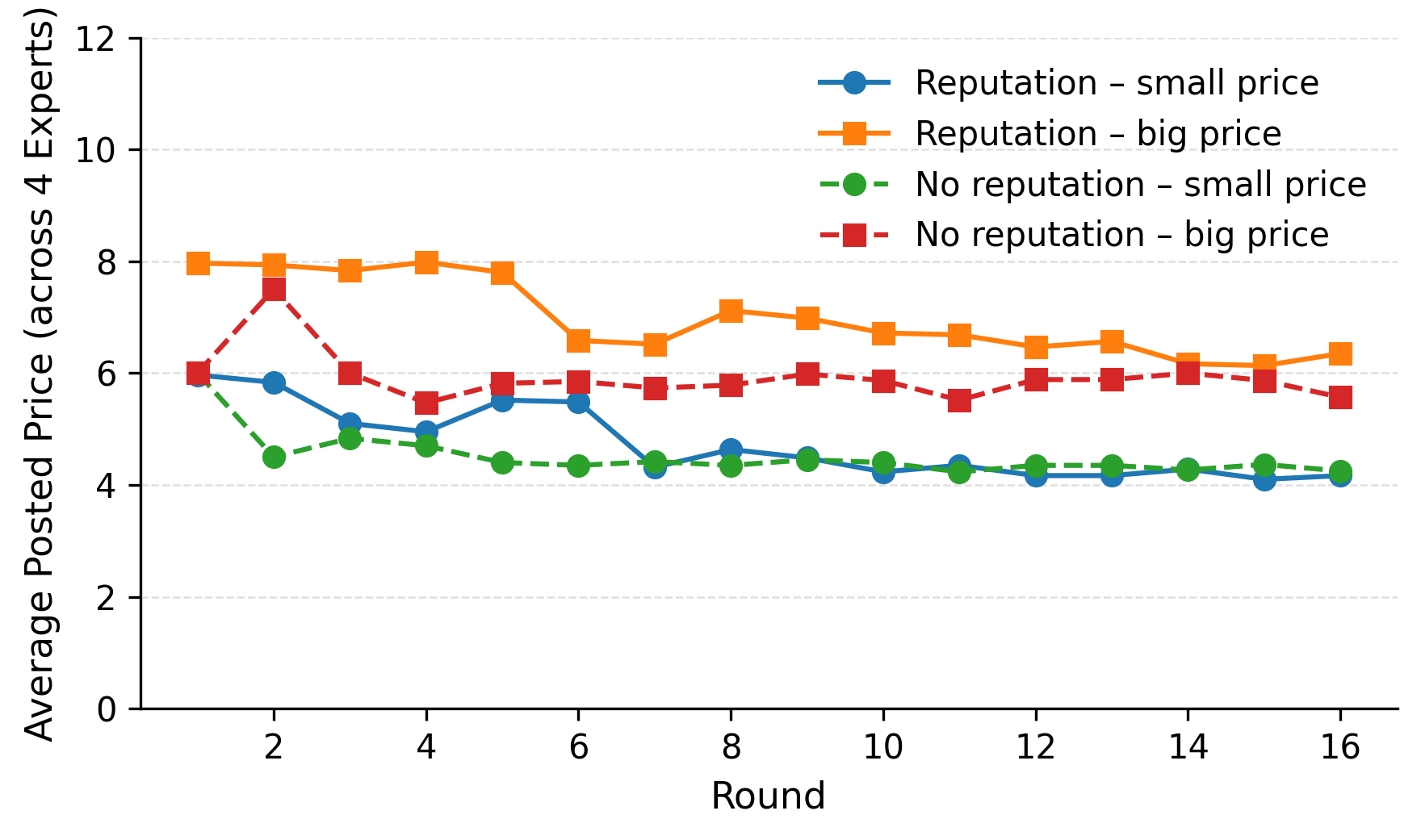}
    \includegraphics[width=0.45\textwidth]{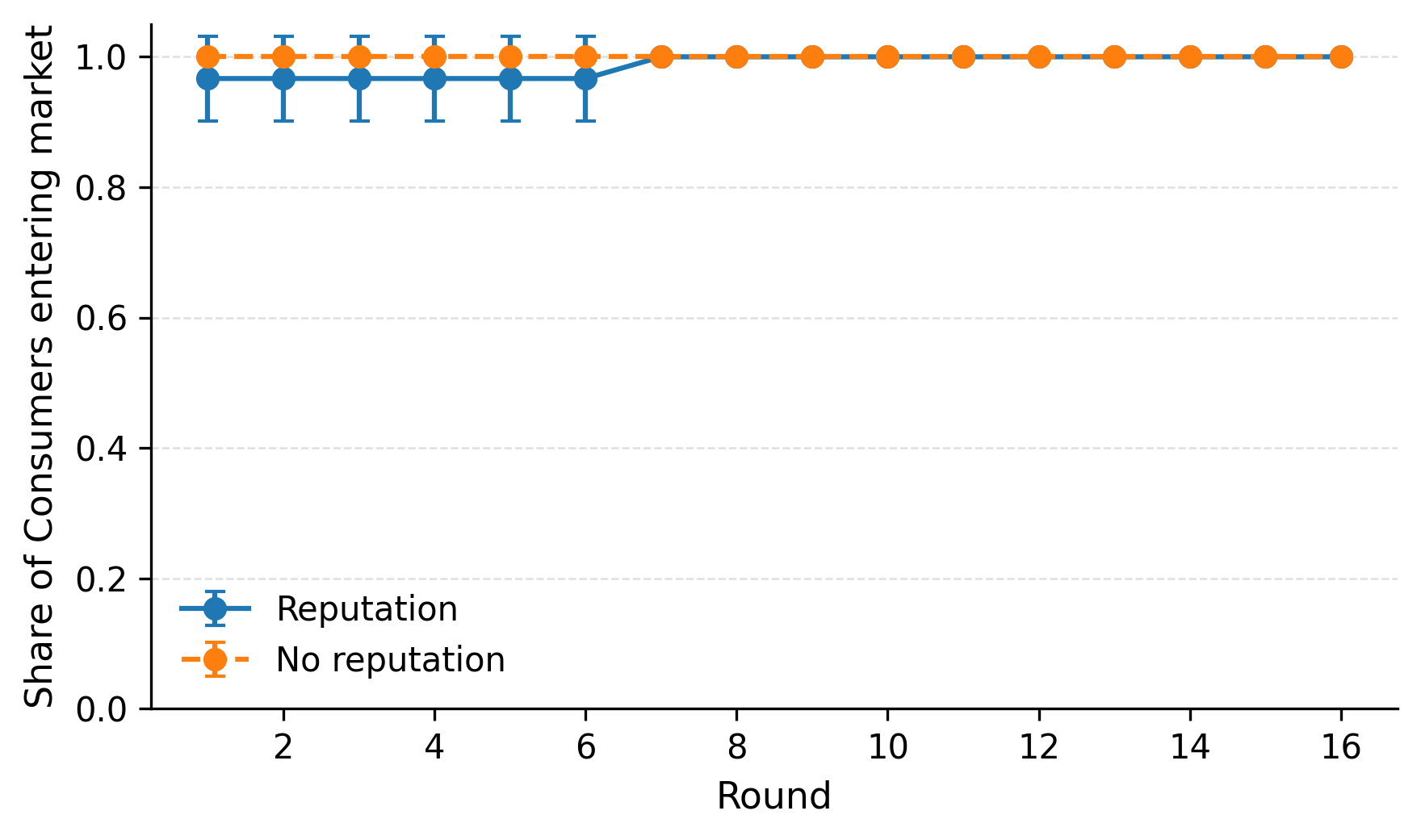}
    
    \label{fig:r16_prices_ia_oa}
\end{figure}

\begin{figure}[h]
    \centering
    \caption{\textbf{Left:} Total Group Income Across 16 Rounds. \textbf{Right:} Intended \textbf{Inequity-Averse} Expert Under- and Over-Treatment.}
    \includegraphics[width=0.45\textwidth]{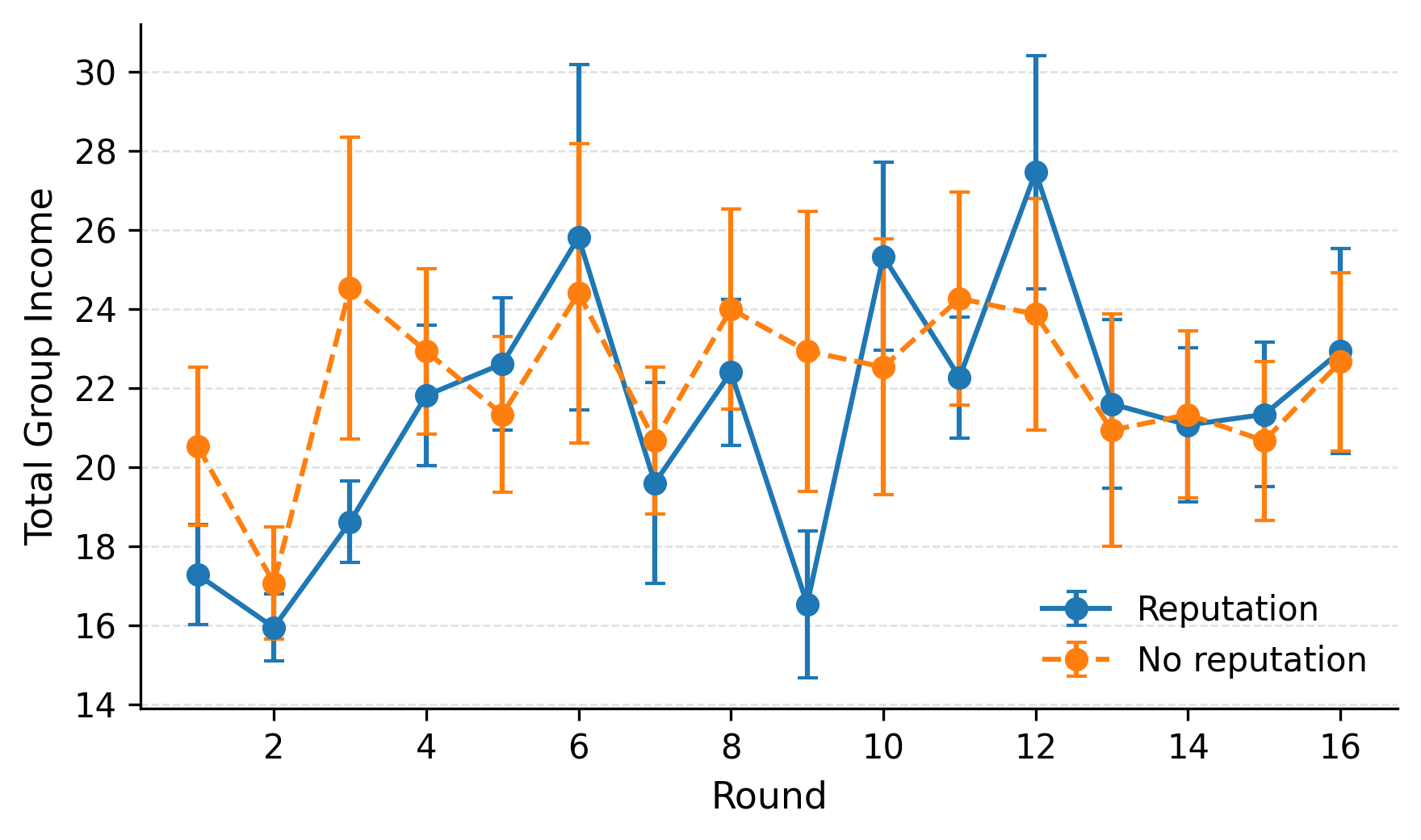}
    \includegraphics[width=0.45\textwidth]{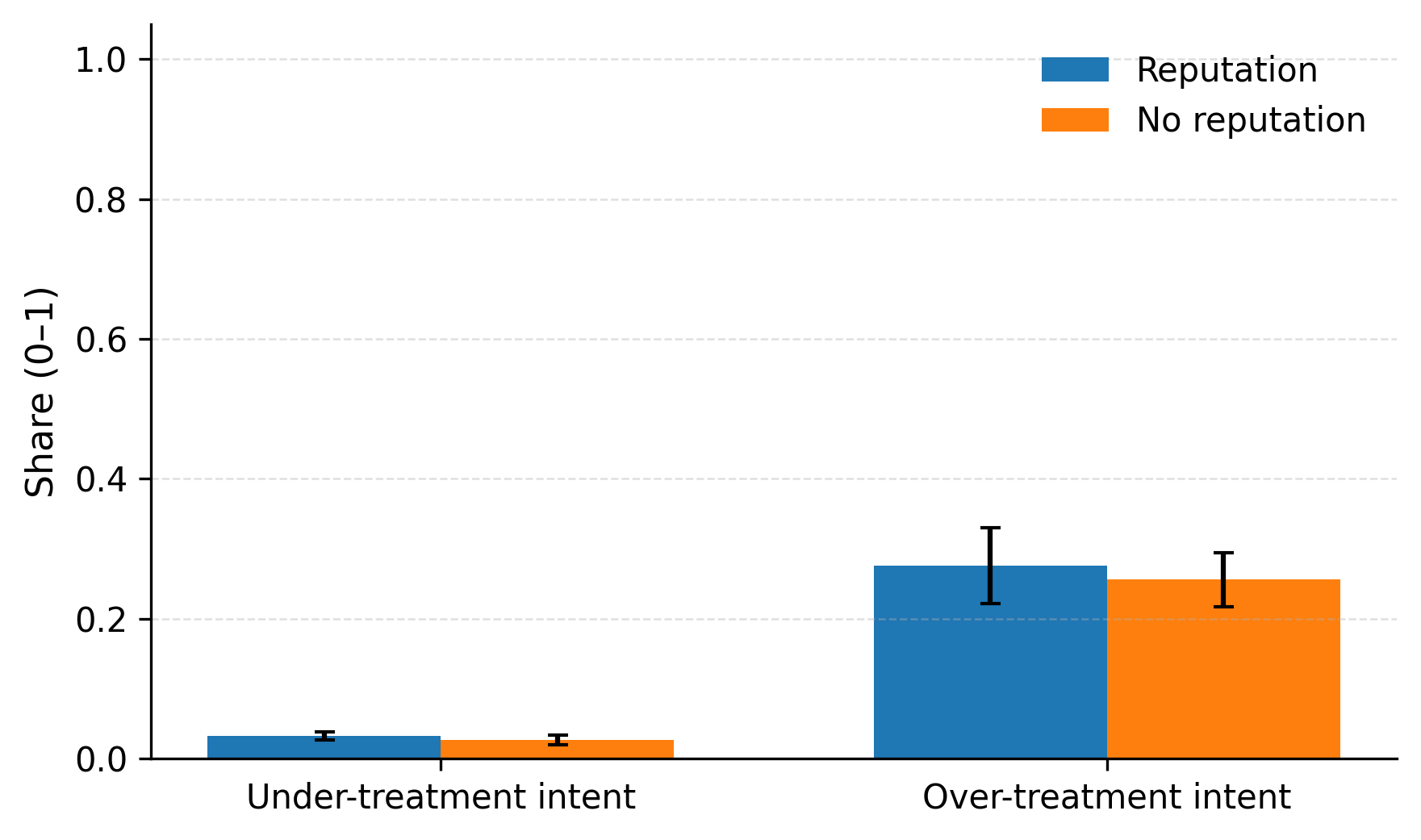}
    
    \label{fig:income and undertreatment ia}
\end{figure}

Figures \ref{fig:r16_prices_ia_oa} and \ref{fig:income and undertreatment ia} show the respective figure for \textbf{inequity-averse}. In contrast to the one-shot environment, consumers enter the market from round 1 onward. That is despite the familiar \{6,8\} price pair in \textbf{reputation}. This points towards a tendency for exploration. LLM experts under inequity aversion exhibit low levels of under-treatment, ensuring consistent consumer participation. In line with LLMs trying to reduce payoff differences between experts and consumers, over-treatment rates are much higher than in any prior setup, which makes sense as consumer-surplus for a successful LCT is higher than for a successful HCT. Furthermore, over-treatment (but not under-treatment) is significantly higher in \textbf{reputation}, once again pointing towards an ambiguous effect of reputation mechanisms on AI agent behavior.

Finally, consumer participation in \textbf{efficiency-loving} is always at 100\% and prices converge roughly towards \{3,5\} which is higher than predicted for \textit{No Institution} but also less damaging for experts in case of honest treatment behavior. Like in the one-shot setup, average expert income across 16 rounds is negative (-\$60), while average consumer income eclipses all other social preference conditions (\$405), with no significant difference between reputation conditions. Reputation significantly decreases under-treatment and overcharging, but increases over-treatment by efficiency-loving LLM experts (Table \ref{tab:pooled_time_slope_alltypes}). 

\begin{figure}[t]
    \centering
    \caption{Average total, consumer and expert payoffs across all 16 rounds.}
    \includegraphics[width=\textwidth]{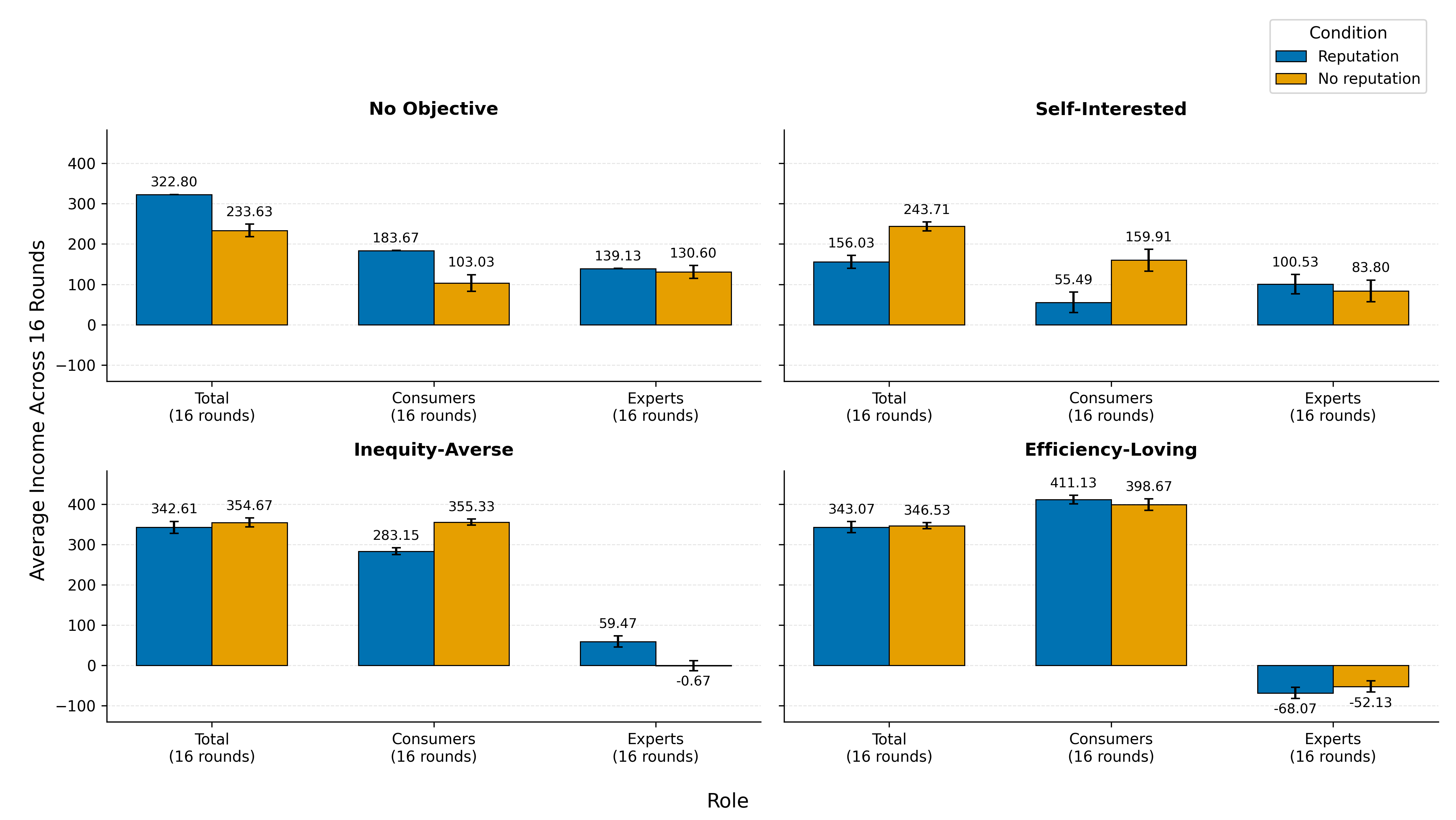}
    \label{fig:llm_16rounds_welfare}
\end{figure}

Looking at aggregate and role-specific incomes across the 16 rounds (Figure \ref{fig:llm_16rounds_welfare}) highlights how other-regarding LLM expert preferences positively affect total and consumer-specific welfare while hurting experts. In their absence (\textbf{self-interested} and \textbf{no objective}), total income hovers between 150 and 320, depending on condition, with relatively small expert--consumer differences that depend on the reputation regime. Generally, self-interested LLM experts minimize welfare. On the other hand, market income in \textbf{inequity-averse} and \textbf{efficiency-loving} consistently reaches 340--350, and heavily favors consumers. Welfare gains compared to the one-shot setup predominantly accrue through increased expert incomes following higher consumer participation, while in \text{efficiency-loving}, effects are mostly distributive due to on average higher expert prices $\ubar{p}$. Repeated interactions do not meaningfully restrain expert dishonesty, because LLM consumers do not enforce it.

\begin{figure}[h]
    \centering
    \caption{\textbf{Left: Efficiency-Loving} LLM Expert Price Setting Across 16 Rounds. \textbf{Right:} Consumer Approach Share.}
    \includegraphics[width=0.45\textwidth]{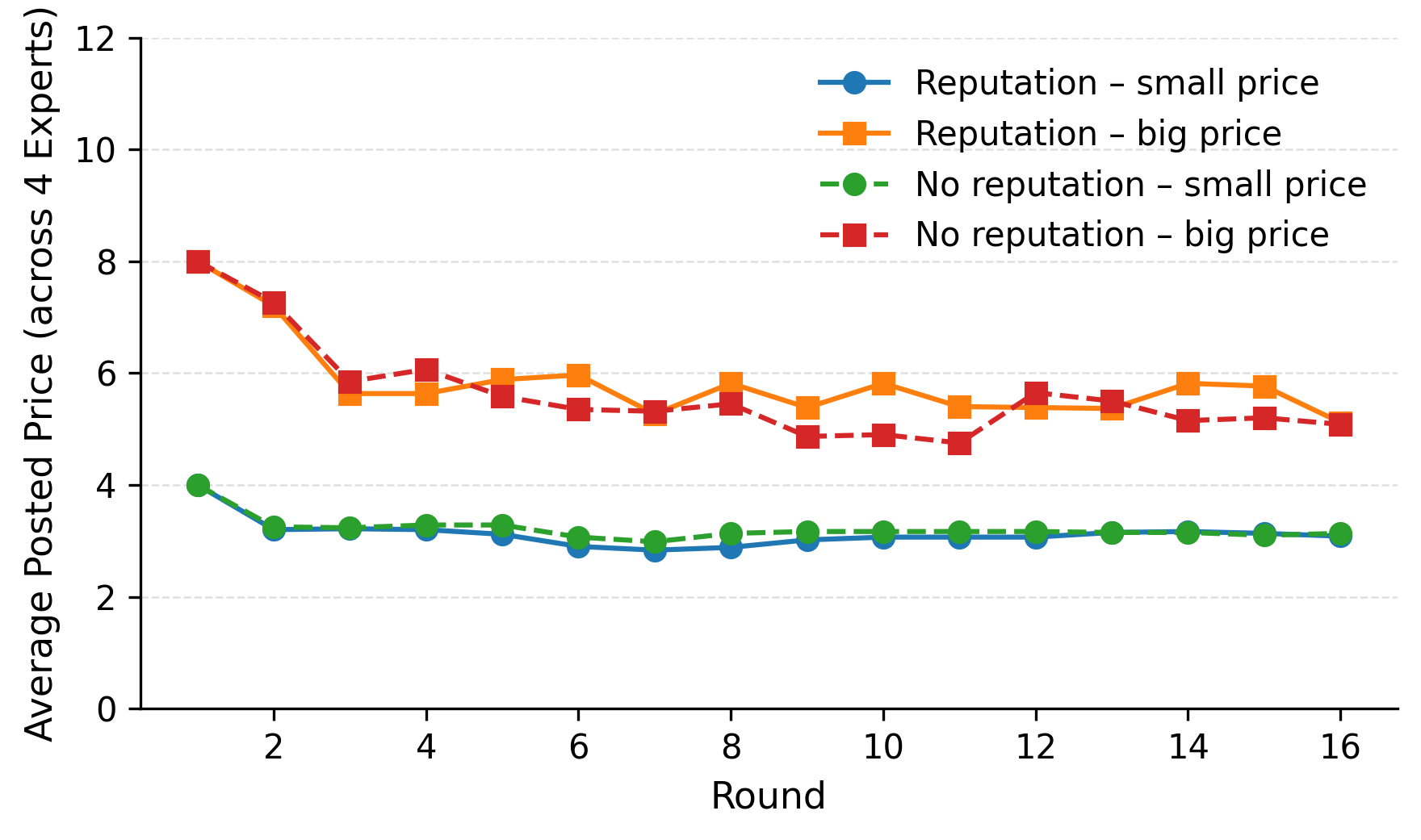}
    \includegraphics[width=0.45\textwidth]{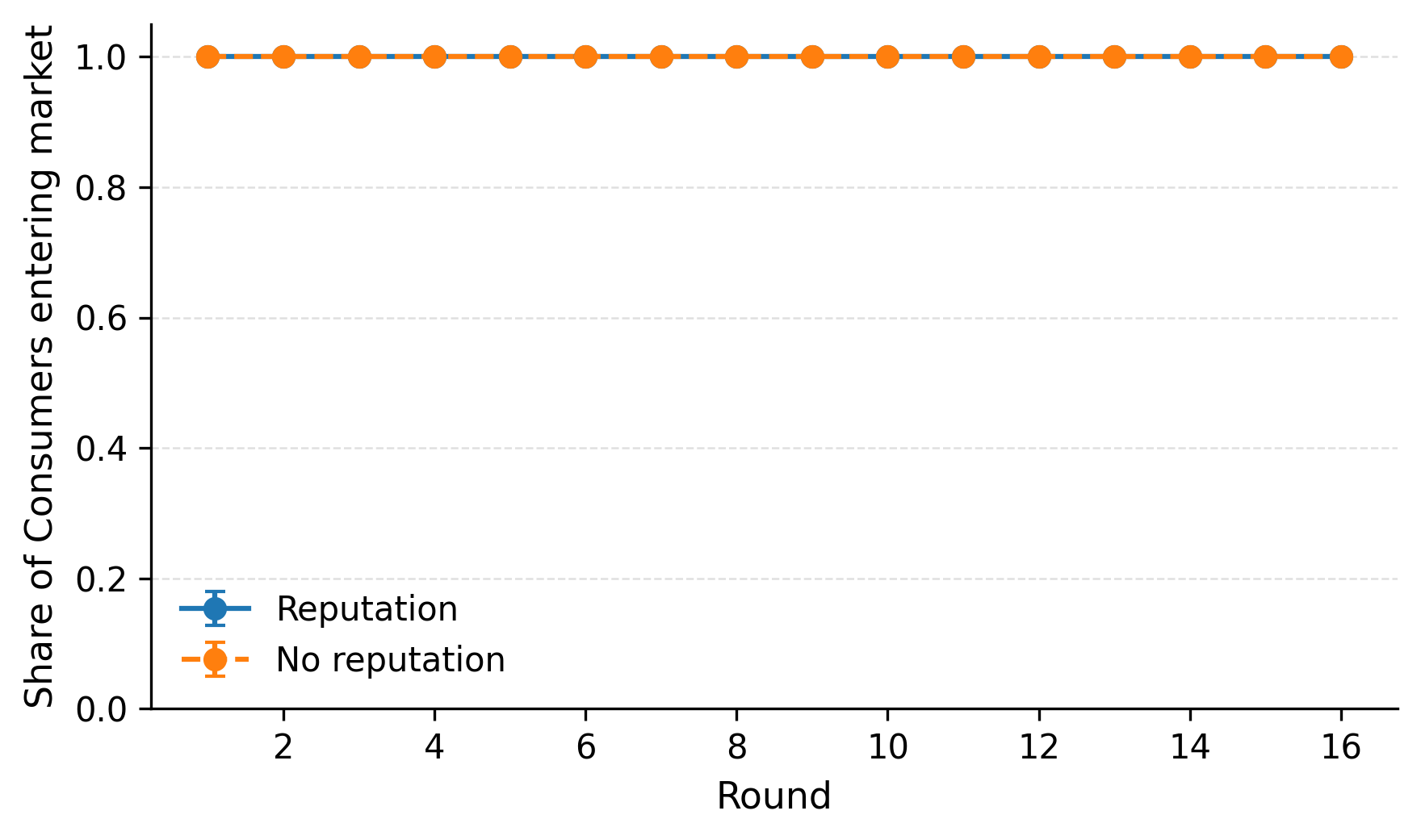}
    
    \label{fig:r16_prices_el_oa}
\end{figure}

\begin{figure}[h]
    \centering
    \caption{\textbf{Left:} Total Group Income Across 16 Rounds. \textbf{Right:} Intended \textbf{Efficiency-Loving} Expert Under- and Over-Treatment.}
    \includegraphics[width=0.45\textwidth]{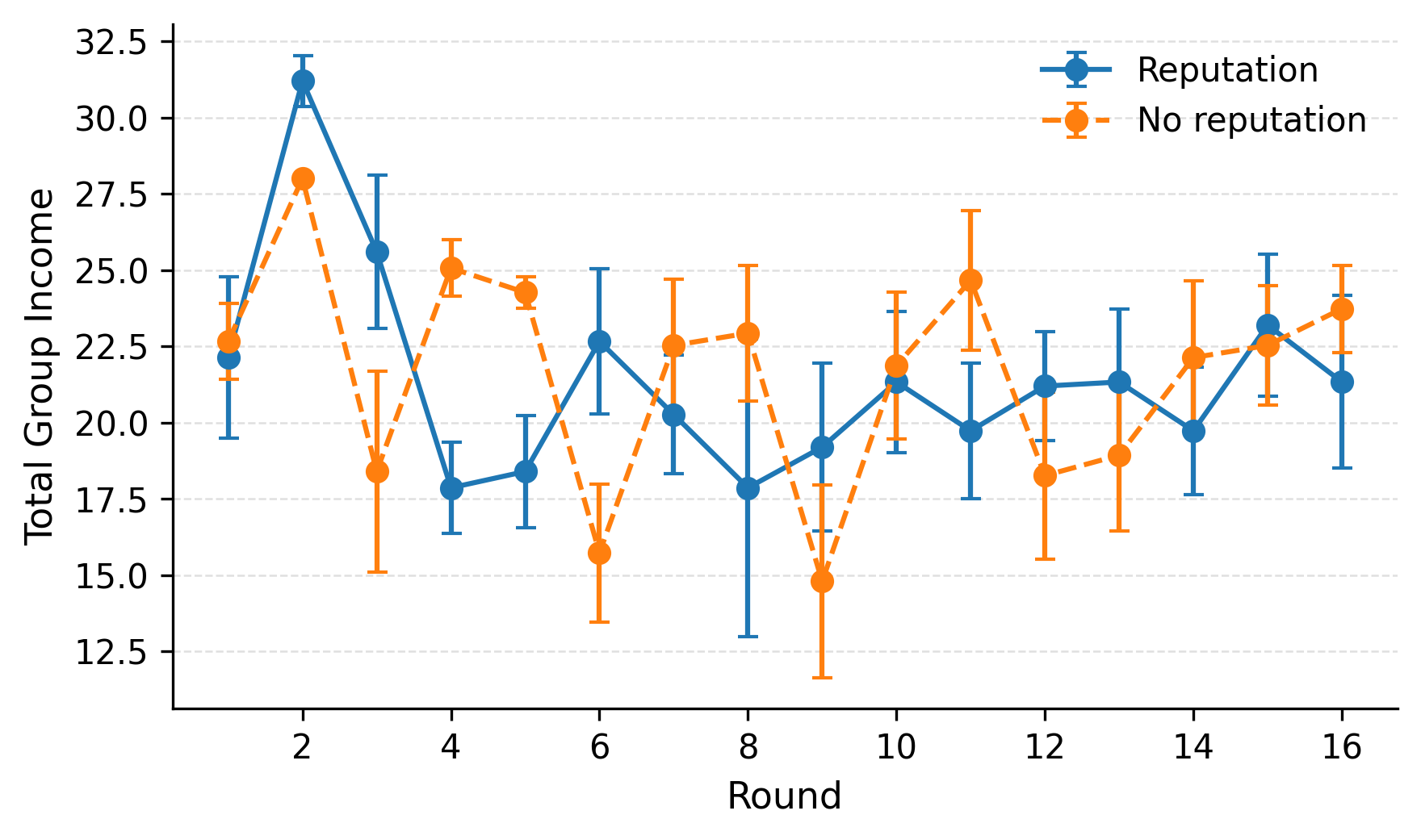}
    \includegraphics[width=0.45\textwidth]{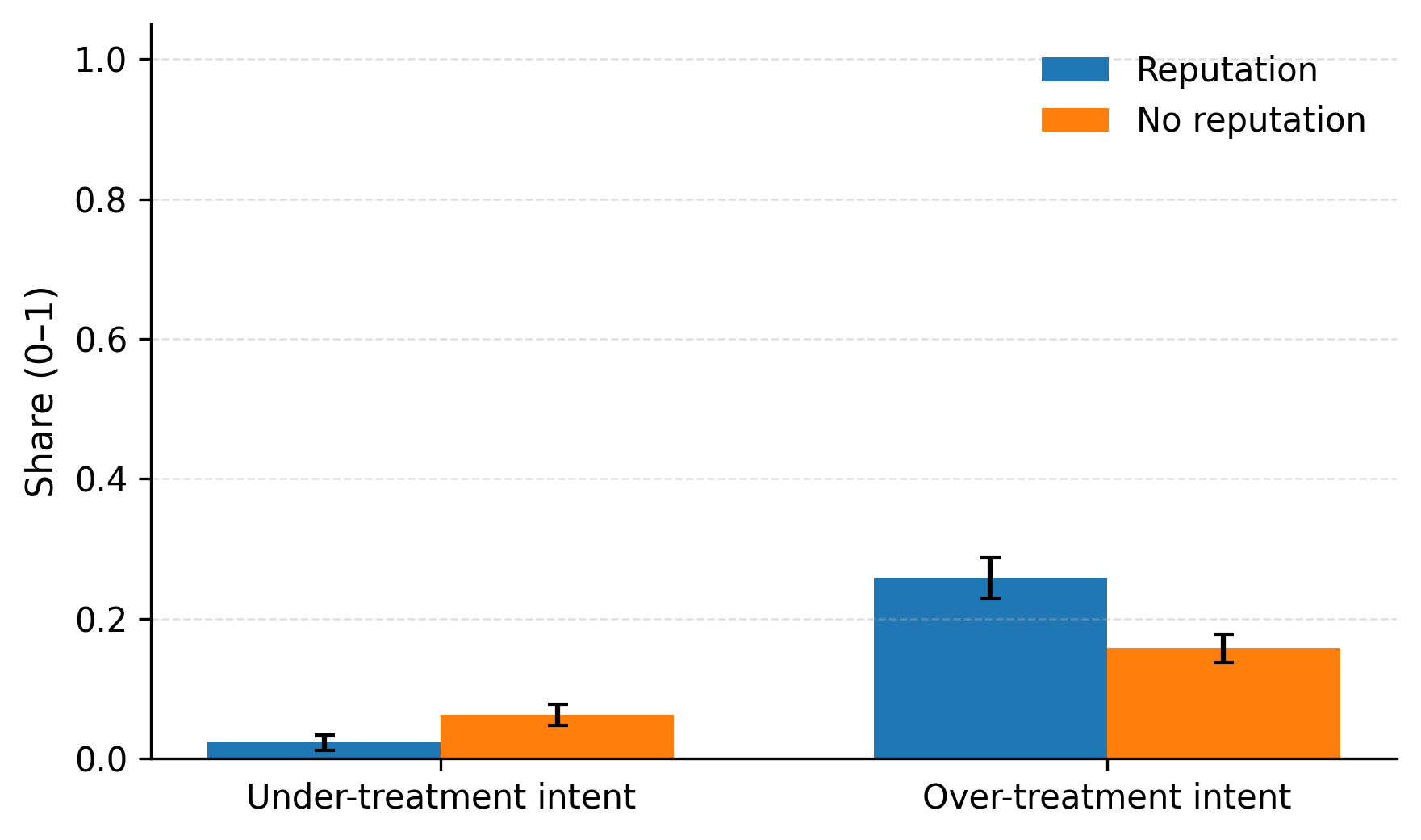}
    
    \label{fig:income and undertreatment el}
\end{figure}

\clearpage
\subsubsection{Verifiability}
\begin{figure}[h]
    \centering
    \caption{\textbf{Left: Self-Interested} LLM Expert Price Setting Across 16 Rounds. \textbf{Right:} Consumer Approach Share. Verifiability.}
    \includegraphics[width=0.45\textwidth]{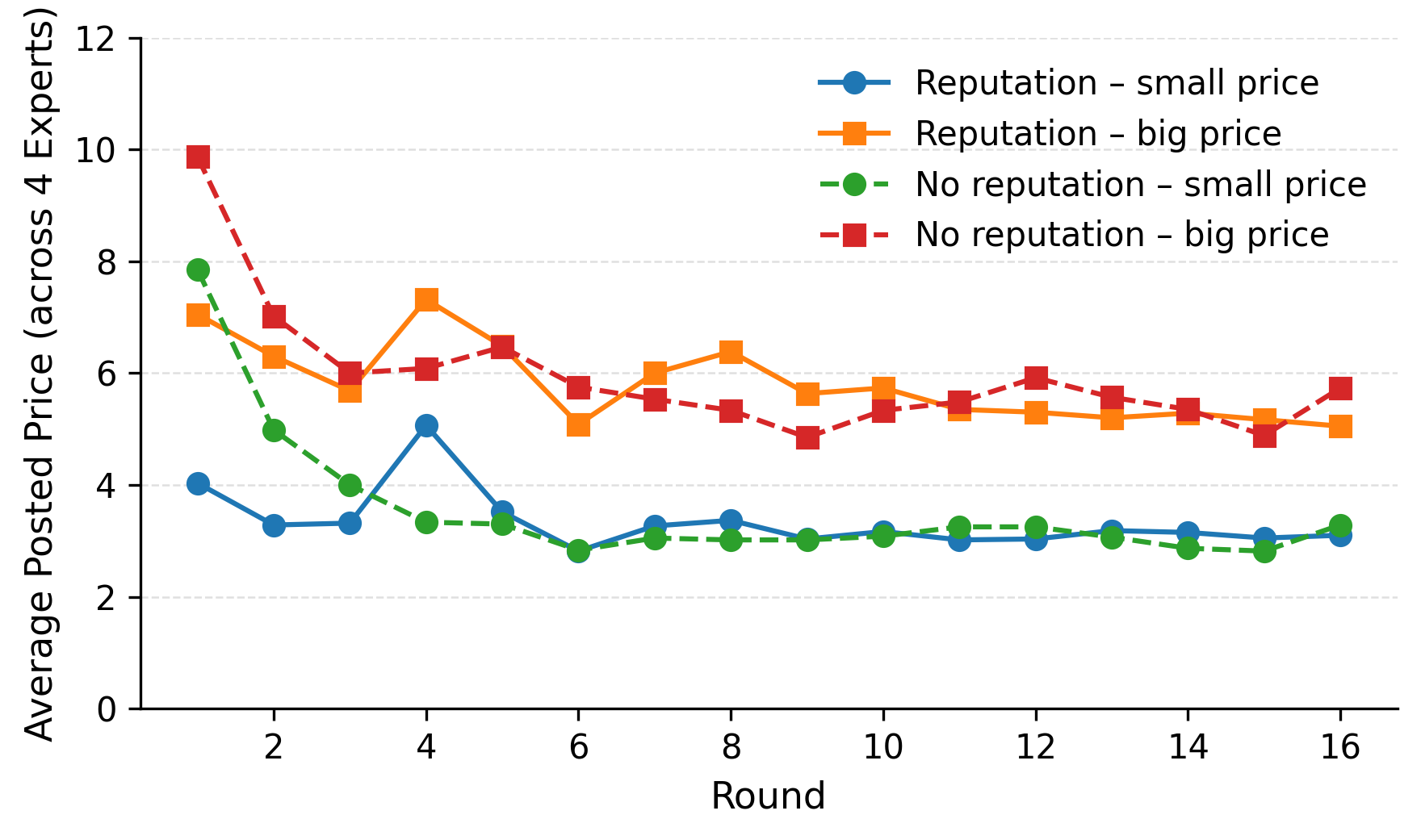}
    \includegraphics[width=0.45\textwidth]{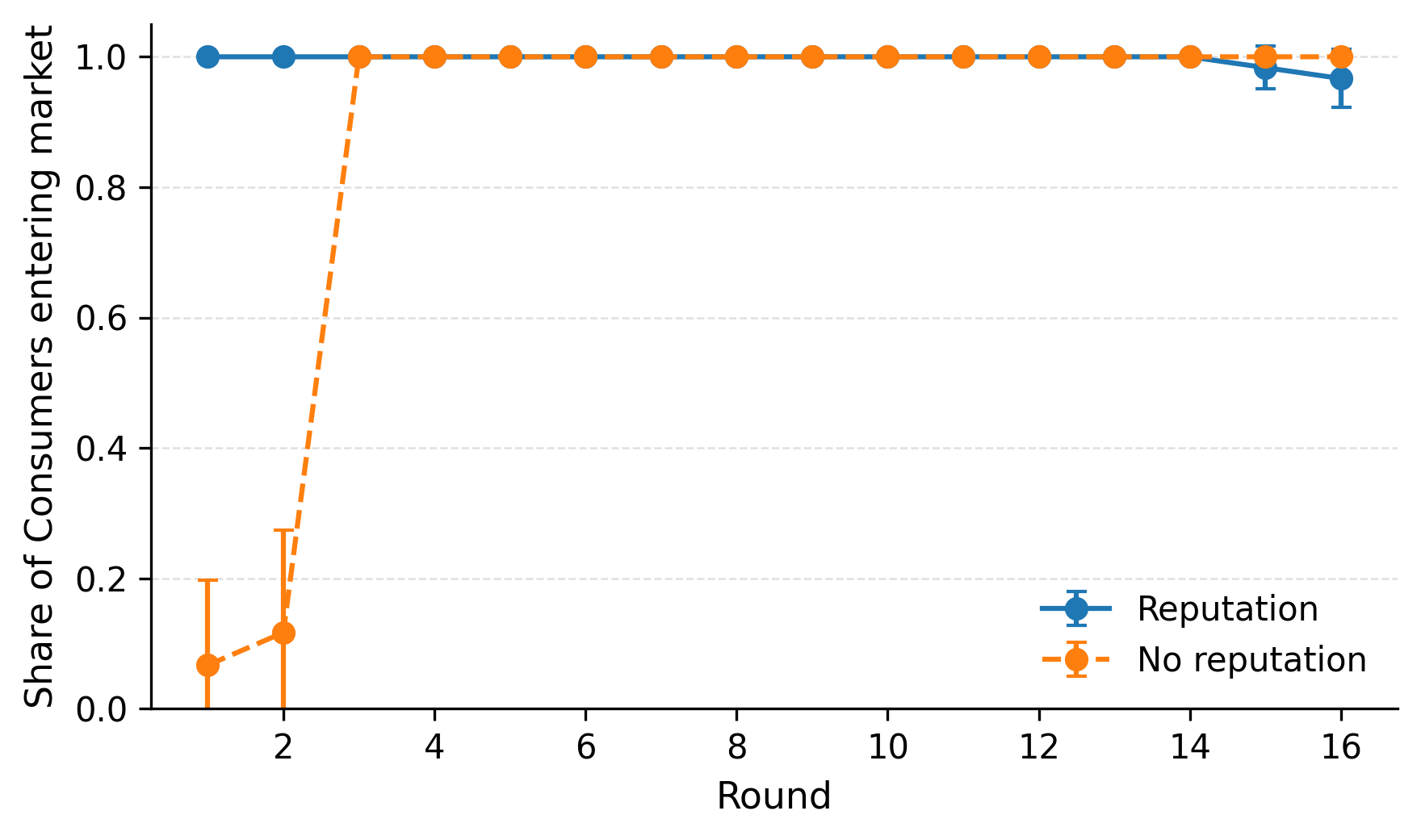}
    
    \label{fig:r16_prices_si_veri}
\end{figure}

\begin{figure}[h]
    \centering
    \caption{\textbf{Left:} Total Group Income Across 16 Rounds. \textbf{Right:} Intended \textbf{Self-Interested} Expert Under- and Over-Treatment. Verifiability.}
    \includegraphics[width=0.45\textwidth]{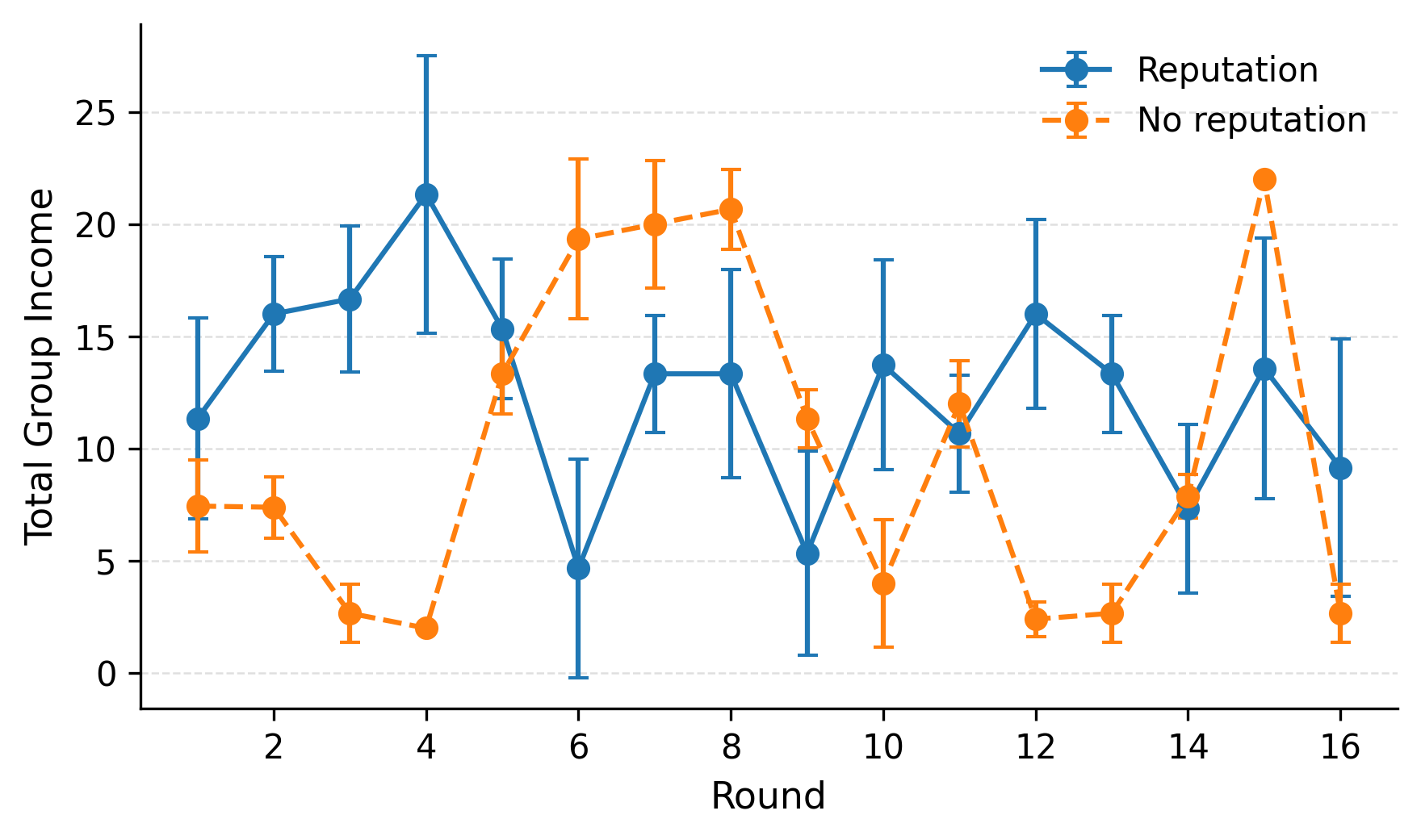}
    \includegraphics[width=0.45\textwidth]{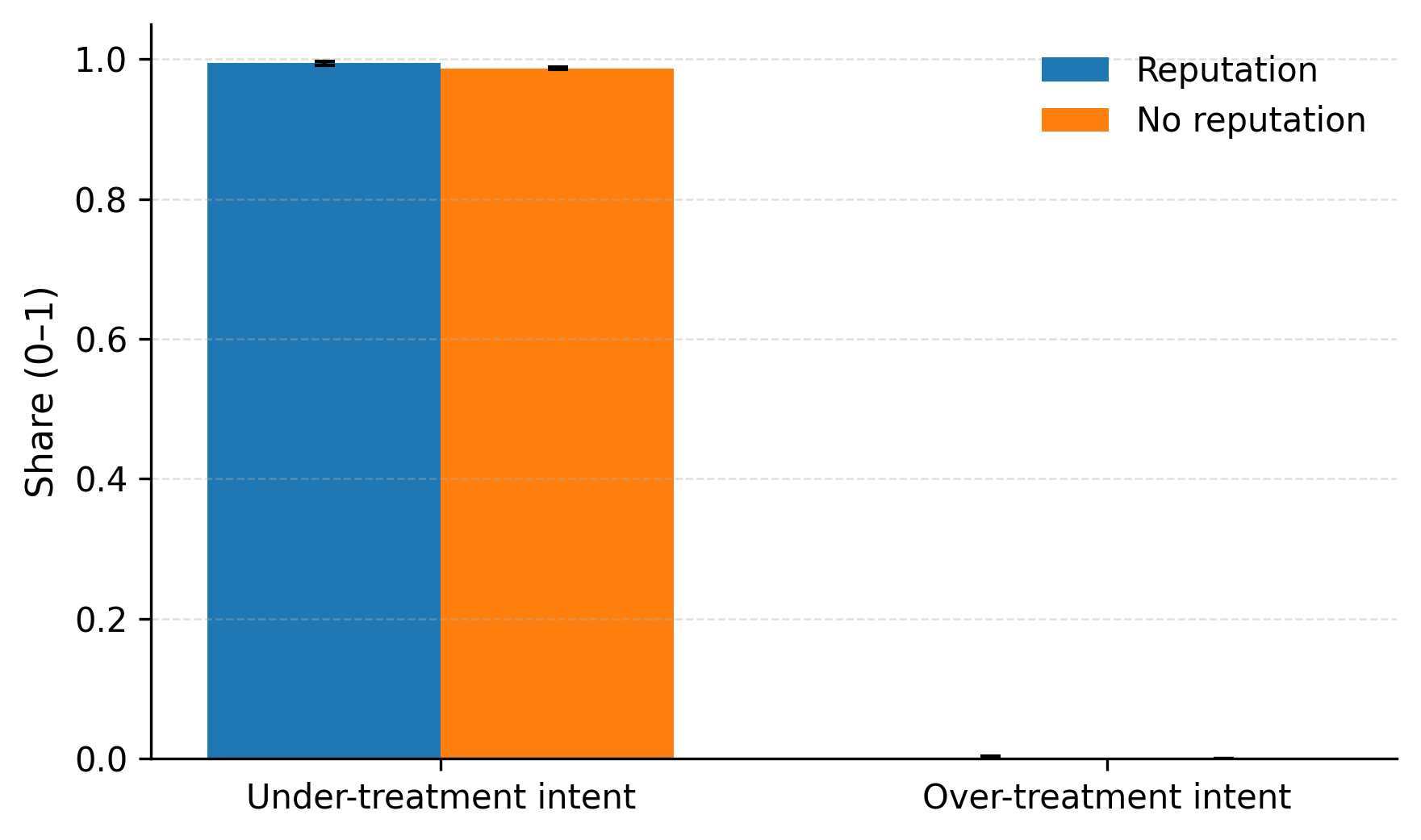}
    
    \label{fig:income_under_si_veri}
\end{figure}

Preventing LLM experts from overcharging consumers by introducing a verifiability institution does not fundamentally alter market dynamics under information asymmetry. All figures are in the appendix. Pricing and consumer approach behavior for \textbf{self-interested} LLM experts confirms that (1) our LLM consumer agents do not use the markup-logic to infer incentives and thus expert strategies but instead (2) enforce (relatively) low prices to alleviate the negative consequences of potential expert dishonesty. Consumers discipline experts towards prices around $\ubar{p} = 3$ and $\bar{p} < 6$, such that consumers' expected payoff $\pi^c_v = 2 > 1.6$ (see Figures \ref{fig:r16_prices_si_veri} and \ref{fig:income_under_si_veri}). Here, expected (and realized) total group income per round is below the standard model's benchmark ($12 < 24$). In \textbf{reputation}, consumers immediately enter the market, otherwise, LLMs need a short learning period to sufficiently decrease prices. There are little difference regarding expert honesty, but the regression points to a marginal positive effect of \textbf{reputation} (Table \ref{tab:pooled_time_slope_alltypes_veri}).

\begin{figure}[h]
    \centering
    \caption{Average total, consumer and expert payoffs across all 16 rounds in \textit{Verifiability}.}
    \includegraphics[width=\textwidth]{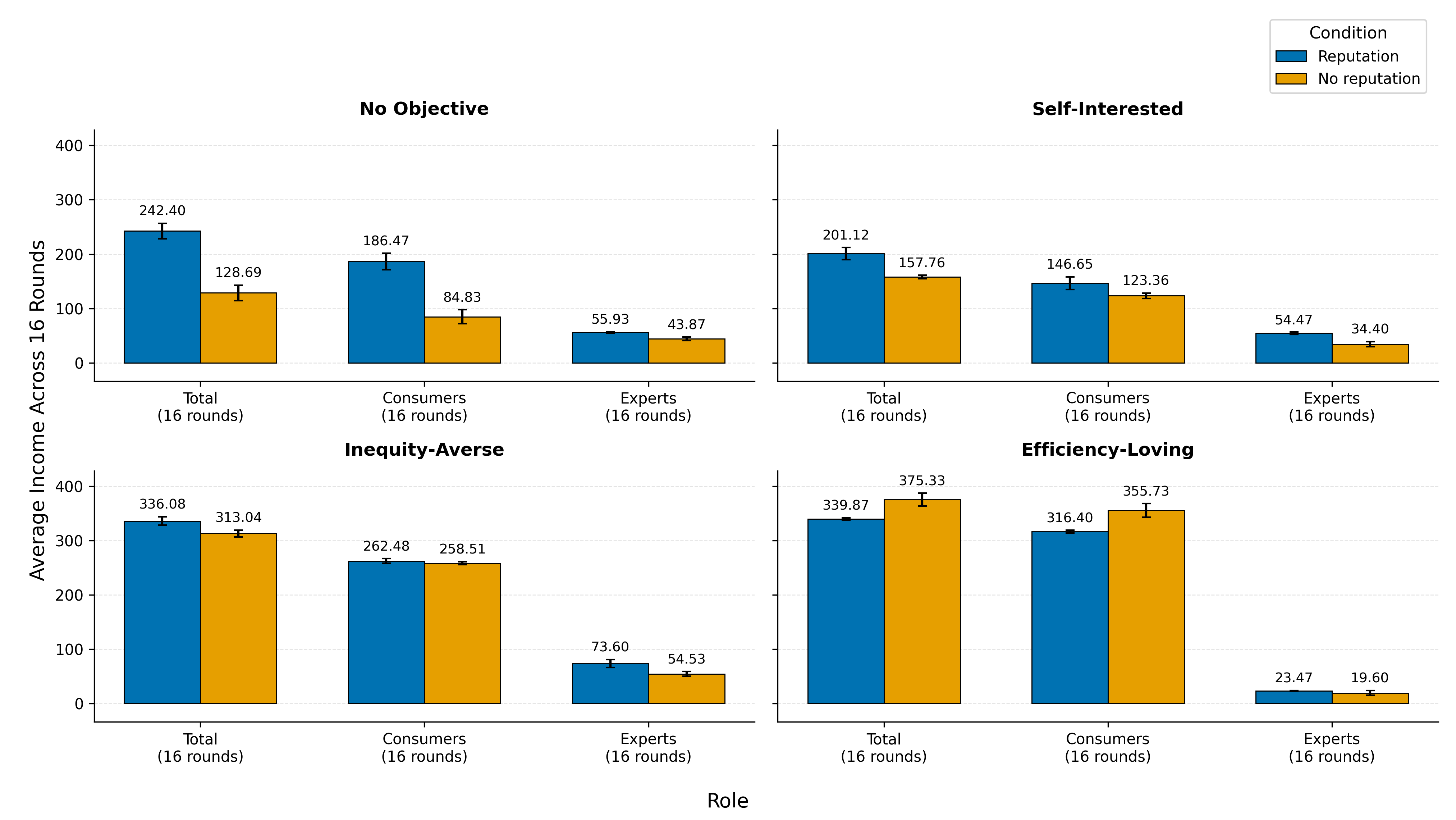}
    \label{fig:llm_16rounds_welfare_veri}
\end{figure}

Total income is higher in \textbf{reputation} (\$201 vs. \$158, $t = 7.3, p < 0.001$), which is mostly driven by the initially higher market participation of consumers due to lower expert prices. Comparing total market income in \textit{Verifiability} (Figure \ref{fig:llm_16rounds_welfare_veri}) with total market income in \textit{No Institution} reveals a significantly higher aggregate income in \textit{No Institution} for \textbf{no reputation} ($t = 14.89, p < 0.001$), which reverses in \textbf{reputation} ($t = 4.57, p < 0.001$). Hence, verifiability does not automatically increase welfare in agent-based credence markets. This is only further highlighted by looking at results for the default \textbf{no objective} model. Here, aggregate income is higher for \textit{No Institution} across both reputation conditions ($t = 6.23, p < 0.001)$, driven primarily by substantial expert gains. Consequently, market efficiency for the \textbf{no objective} agents in \textit{Verifiability} is very low (9\%). Otherwise, results mirror the ones from above. LLM experts tend towards high levels of dishonesty, reputation slightly alleviates intention-to-under-treat, and income is significantly higher in \textbf{reputation} (\$242 vs. \$129, $t = 10.97, p < 0.001$). Regarding \textbf{inequity-averse} and \textbf{efficiency-loving}, the results are very similar to \textit{No Institution}, without meaningful systematic changes. In these conditions, LLM expert agents tend towards honesty, although \textbf{inequity-averse} preferences motivate relatively high rates of over-treatment, especially without any reputation mechanism. LLM experts do not make use of equal markup prices to signal honest intentions, but compete via relatively low prices. Total market income is higher in \textit{No Institution} for \textbf{inequity-averse} ($t = 4.18, p < 0.001$), and slightly higher in \textit{Verifiability} for \textbf{efficiency-loving} ($t = 2.15, p = 0.036$). Overall, \textit{Verifiability} provides little benefit for agent-based credence goods markets. Because the markup-logic of the standard model does not play a role, and the elimination of overcharging leads to on average higher prices, while experts continue to make dishonest choices in the absence of explicit other-regarding preferences, aggregate welfare fluctuates and often even decreases. In the other-regarding conditions, experts tend to benefit from verifiability. Otherwise, expert income is higher for the unregulated market. Notably, in \textit{Verifiability}, expert income is maximized under inequity-averse preferences, potentially inducing incentives for experts to consider consumer outcomes. Overall, because repeated interactions induce almost full consumer participation across the board, while social preferences mostly determine expert honesty, verifiability has little influence on aggregate welfare.

\subsection{Comparison with Human Experiments}

\begin{table}[!t]
\centering
\caption{Key outcome comparison: Human-Markets vs.\ AI-Agent-Markets}
\label{tab:key_outcomes_2x2}
\scriptsize
\setlength{\tabcolsep}{3pt}
\renewcommand{\arraystretch}{1.15}

\newcommand{\KeyHdr}{%
 & \makecell{Trade\\(C/S)}
 & \makecell{Avg.\ cons\\per active\\seller}
 & \makecell{Paid\\price}
 & \makecell{Fraud\\(UT/OC)}
 & \makecell{Profits\\(S/C)}
 & \makecell{Eff.}\\
}

\begin{subtable}[t]{0.49\textwidth}
\centering
\caption{C/N (competition, no reputation, no verifiability)}
\begin{adjustbox}{max width=\linewidth}
\begin{tabular}{lcccccc}
\toprule
\KeyHdr
\midrule
Humans      & 0.730/0.470 & 1.540 & 5.350 & 0.730/0.790 & 2.360/1.210 & 0.130 \\
No obj.     & 0.772/0.256 & 3.030 & 4.656 & 0.998/0.615 & 2.041/1.610 & 0.466 \\
Self-int.   & 0.666/0.211 & 3.187 & 3.948 & 1.000/0.994 & 1.309/2.499 & 0.502 \\
Ineq.-averse& 1.000/0.257 & 3.895 & 4.365 & 0.017/0.038 & -0.010/5.552 & 0.891 \\
Eff.-loving & 1.000/0.292 & 3.475 & 3.427 & 0.064/0.071 & -0.815/6.229 & 0.866 \\
\bottomrule
\end{tabular}
\end{adjustbox}
\end{subtable}\hfill
\begin{subtable}[t]{0.49\textwidth}
\centering
\caption{CR/N (competition, reputation, no verifiability)}
\begin{adjustbox}{max width=\linewidth}
\begin{tabular}{lcccccc}
\toprule
\KeyHdr
\midrule
Humans      & 0.850/0.550 & 1.540 & 5.800 & 0.640/0.620 & 2.650/0.940 & 0.140 \\
No obj.     & 0.875/0.282 & 3.100 & 4.627 & 0.867/1.000 & 2.174/2.870 & 0.783 \\
Self-int.   & 0.777/0.298 & 2.697 & 4.106 & 1.000/0.954 & 1.571/0.867 & 0.190 \\
Ineq.-averse& 0.988/0.347 & 2.901 & 5.323 & 0.046/0.206 & 0.929/4.424 & 0.852 \\
Eff.-loving & 1.000/0.324 & 3.135 & 3.420 & 0.034/0.070 & -1.064/6.424 & 0.849 \\
\bottomrule
\end{tabular}
\end{adjustbox}
\end{subtable}

\vspace{0.7em}

\begin{subtable}[t]{0.49\textwidth}
\centering
\caption{C/V (competition, no reputation, verifiability)}
\begin{adjustbox}{max width=\linewidth}
\begin{tabular}{lcccccc}
\toprule
\KeyHdr
\midrule
Humans      & 0.880/0.540 & 1.630 & 5.190 & 0.530/\textemdash & 1.960/2.180 & 0.340 \\
No obj.     & 0.985/0.294 & 3.371 & 2.695 & 1.000/\textemdash & 0.685/1.325 & 0.094 \\
Self-int.   & 0.886/0.279 & 3.179 & 2.668 & 0.970/\textemdash & 0.537/1.927 & 0.197 \\
Ineq.-averse& 0.942/0.235 & 4.000 & 5.800 & 0.002/\textemdash & 0.852/4.039 & 0.748 \\
Eff.-loving & 1.000/0.265 & 3.789 & 4.265 & 0.033/\textemdash & 0.306/5.558 & 0.950 \\
\bottomrule
\end{tabular}
\end{adjustbox}
\end{subtable}\hfill
\begin{subtable}[t]{0.49\textwidth}
\centering
\caption{CR/V (competition, reputation, verifiability)}
\begin{adjustbox}{max width=\linewidth}
\begin{tabular}{lcccccc}
\toprule
\KeyHdr
\midrule
Humans      & 0.930/0.580 & 1.600 & 5.580 & 0.360/\textemdash & 1.900/2.590 & 0.460 \\
No obj.     & 0.995/0.381 & 2.637 & 2.892 & 0.994/\textemdash & 0.874/2.914 & 0.497 \\
Self-int.   & 0.997/0.397 & 2.523 & 2.858 & 1.000/\textemdash & 0.851/2.291 & 0.351 \\
Ineq.-averse& 0.998/0.315 & 3.200 & 5.779 & 0.023/\textemdash & 1.150/4.101 & 0.830 \\
Eff.-loving & 0.943/0.327 & 2.897 & 4.244 & 0.125/\textemdash & 0.367/4.944 & 0.843 \\
\bottomrule
\end{tabular}
\end{adjustbox}
\end{subtable}

\vspace{0.4em}
\par\footnotesize
\emph{Notes:} Human data from \citet{dulleck2011economics}. Trade (C/S) reports consumer-side trade frequency and seller-side trade frequency (share of experts approached at least once per period).
“Avg.\ consumers per active seller” is conditional on at least one consumer entering and captures market concentration.
Fraud reports undertreatment (UT) and overcharging (OC); OC is mechanically prevented by verifiability.
Profits (S/C) are per seller-period and per consumer-period.
Eff. is normalized by the outside option $\sigma = 1.6$. Note that in \citet{dulleck2011economics}, both experts and consumers have the respective outside options.
\end{table}

We contextualize the simulated results with human experimental data from \citep{dulleck2011economics} who use the same setup.\footnote{The only difference is that in their experiments, experts also had a positive outside option $\sigma = \$1.6$. This may affect expert behavior. On the other hand, LLM expert agents in our simulations frequently disregard the outside option 0 in \textbf{efficiency-loving} as exemplified by negative expert returns, and LLM consumer agents continue to approach experts if they average earnings lie below their outside option (e.g., \textbf{self-interested} $\times$ \textbf{reputation} in \textit{No Institution}), suggesting a limited effect of outside options on LLM behavior.} A full table comparing all results is shown in the Appendix (Figure \ref{tab:dulleck_vs_sims_full}). Here, we concentrate on a few selected key comparisons (Table \ref{tab:key_outcomes_2x2}). First, in \citet{dulleck2011economics}, consumer participation is fairly high while seller participation hovers around 50\%-60\%, suggesting frequent trades with relatively low market concentration. In agent markets, participation is generally higher -- often near 1 --, but simultaneously much more concentrated. For example, in \textit{No Institution} with agents that have no pre-defined objective function, approximately 3.1 consumers trade on average, while the average approached expert has 3.03 consumers, almost monopolizing the market. In every single LLM-agent condition, a much higher share of trades goes to only one seller. This kind of Bertrand-like choice behavior ties well into the second noticeable difference: expert prices in agentic credence goods market are often much lower than in human markets. In \citet{dulleck2011economics}, transaction prices evolve around 5.2--5.80. For AI-markets, these prices are only matched by \textbf{inequity-averse} LLM experts. Otherwise, competition drives prices towards 2.7--4.7. Note that in \textbf{self-interested} and \textbf{no objective}, expert fraud remains high, often approaching 1. Compared to human markets, expert fraud is much more polarized by the objective function. While under-treatment and over-charging is substantial and present in humans, it approaches either 1 or 0 for LLM-agents. Furthermore, whereas reputation consistently reduces overcharging for humans, effects are context-dependent for agentic interactions. These behavioral patterns lead to sharp shifts in the distribution of surplus. In \citet{dulleck2011economics}, sellers often out-earn consumers on average. In our LLM interactions, consumers almost always secure substantially higher surplus, a pattern that is broken in only two conditions (CR/N self-interested and C/N with default agents). Finally, looking at welfare and institutions, verifiability consistently and strongly improves market efficiency for human subjects, but not for AI agents. Overall, LLM-interactions with experts who are endowed with other-regarding preferences exhibit the highest efficiency levels by far. The effects for \textbf{self-interested} and \textbf{no objective} depend on the institutional context -- efficiency sometimes collapses (e.g., C/V), but often out-competes human markets (e.g., C/N), because (1) the human baseline can be quite inefficient and (2) prices are much lower.

\section{Conclusion}
This article provides initial results about economic outcomes and behavioral trends in agentic credence goods markets. We find that LLM agents do not follow the strategic behavior dictated by the standard model. For one-shot settings, this causes market breakdown in the absence of liability. In the repeated setting, LLM experts get away with widespread consumer fraud because LLM consumers appear to condition their behavior mostly on prices. In comparison to human data, market concentration increases, and the welfare effects of institutions becomes much more ambiguous.

\noindent
\textbf{One-Shot.} LLM Agents (here GPT-5.1) do not consider incentives through markups in their pricing choices and therefore regularly offer negative expected value deals to consumers. This leads to widespread market breakdown. However, even when prices follow theory (e.g., equal markups in self-interested $\times$ verifiability), LLM consumers do not approach LLM experts because they do not trust indifferent experts to not defraud them. AI-agents behave dishonestly by default and when prompted to follow the objective of maximizing their own economic position. Inducing other-regarding preferences alleviates LLM dishonesty, but does not improve price-setting. Notably, efficiency-loving LLM experts consistently set prices that are so low that they always attract consumers but hurt the expert's income through negative returns. Apparently, the LLM prioritizes consumer participation, which is indeed the most important factor for aggregate welfare. However, because experts lose on average, this is not a stable strategy without some re-distributive institution. Compared to recent behavior evidence from one-shot experiments with humans \citep{erlei2025digital}, LLM agents are generally substantially less successful than humans in establishing cooperation on credence goods markets. 

\noindent
\textbf{Repeated.} Allowing LLM agents to interact for multiple rounds within a finite time horizon solves the consumer participation problem because (1) LLM experts quickly learn to lower their prices and (2) LLM consumers are quick to enter once prices reach a certain threshold -- despite adverse expert incentives hidden via different treatment-markups. This makes LLM consumers vulnerable to dishonesty, which happens frequently, and is not sufficiently punished. Hence, expert dishonesty becomes entrenched in the absence of explicit other-regarding preferences. Experts' treatment and price-charging behavior is thoroughly dependent on their social preferences. Because consumers play no role in disciplining experts, the patterns mostly mirror the ones from the one-shot setting, with experts consistently opting for the high markup treatment (LCT) while overcharging consumers ($\bar{p}$). Private reputation institutions have small but positive effects on expert fraud, while verifiability is, at best, neutral. In the absence of other-regarding preferences, it hurts aggregate income. Neither expert agents nor consumer agents exploit the potential of verifiability, which is consistent with them not acting strategically by signaling (or inferring) intentions through markups. In comparison to the one-shot setting, market efficiency is much higher. In particular, inequity-averse and efficiency-loving experts cooperate with consumers to achieve almost full market efficiency. Yet, even with self-interested experts, the market produces much higher levels of income that in one-shot, which is driven by very high consumer participation. Compared to human benchmark data, prices are lower, fraud is much more polarized, and efficiency is generally higher, while the effect of institutions is less predictable and more context-dependent. Finally, market concentration increases sharply with the introduction of AI agents, with singular experts attracting the vast majority of consumers.

\noindent
\textbf{Implications.} Our results highlight how information asymmetries that induce incentives for fraud remain widespread in agent-based interactions. Even more so, they may become entrenched parts of repeated interactions. Markets converge towards states with rampant fraud that is compensated by somewhat low prices. In reality, this may be difficult to accept for human stakeholders. However, it does lead to sustained expert-consumer interactions, which -- even when self-interested experts consistently intend to defraud consumers -- strongly outperforms one-shot settings and reaches efficiency rates of roughly 31\% on average. Notably, market efficiency climbs to 88\% with efficiency-loving objective prompts and 83\% for inequity-averse prompts. This coincides with strong expert losses, leading to negative aggregate income in efficiency-loving $\times$ No Institution, except for inequity-averse experts under verifiable treatment outcomes. Therefore, \textit{Verifiability} provides an institutional structure that rewards experts for considering consumer outcomes. This may be one promising avenue when designing agentic credence goods markets. Similarly, private reputation institutions like verified identities appear more consistently useful under verifiability, whereas in \textit{No Institution}, the effects are very ambiguous, often harming overall welfare. In conclusion, under complicated market structures that involve information asymmetries and incentives for fraud, LLM behavioral structurally diverges from standard theory, leading to novel empirical patterns like entrenched fraud. Social preferences can alleviate dishonesty, but often at the cost of expert welfare. Verifiability provides some promise, as it is the only environment that offers experts incentives to deviate from self-interest, and may positively affect the value of private reputation institutions that reduce under-treatment and over-charging. However, in many conditions, verifiability consistently worsens market outcomes, suggesting that it is far from a panacea. In general, the impact of institutions like reputation and verifiability is very context-dependent and difficult to predict. This stands in contrast to human subject experiments, in which verifiability and reputation consistently produce either positive or null outcomes. Future research may consider how these results change with the introduction of human oversight, as both pricing and (dis-) honesty patterns are markedly different from human behavior as observed in standard credence goods experiments.

\clearpage
\bibliography{\bib}

\clearpage
\section*{Appendix}

\begin{figure}[h]
    \centering
    \caption{\textbf{Left: Default} LLM Expert Price Setting Across 16 Rounds. \textbf{Right:} Consumer Approach Share. Verifiability.}
    \includegraphics[width=0.45\textwidth]{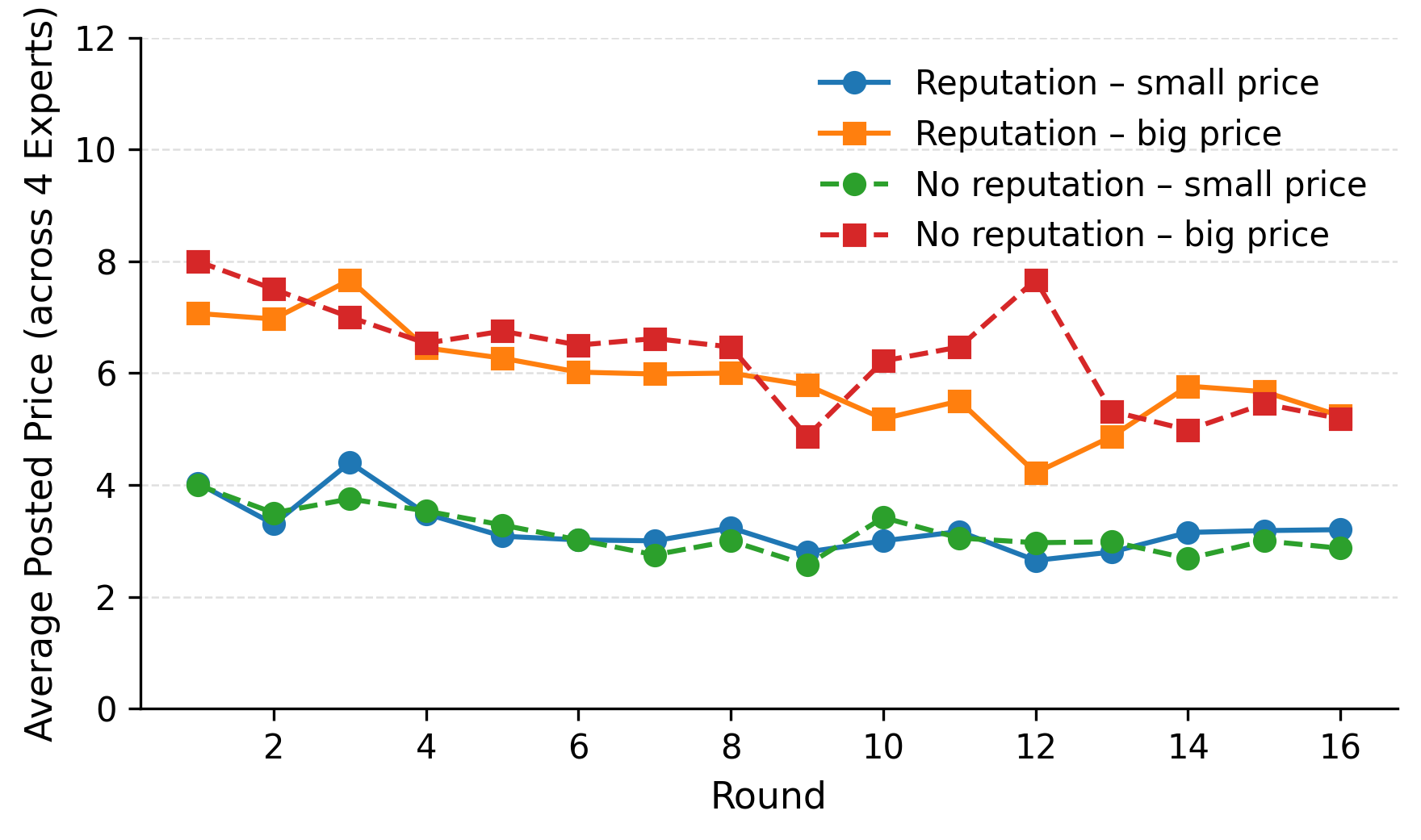}
    \includegraphics[width=0.45\textwidth]{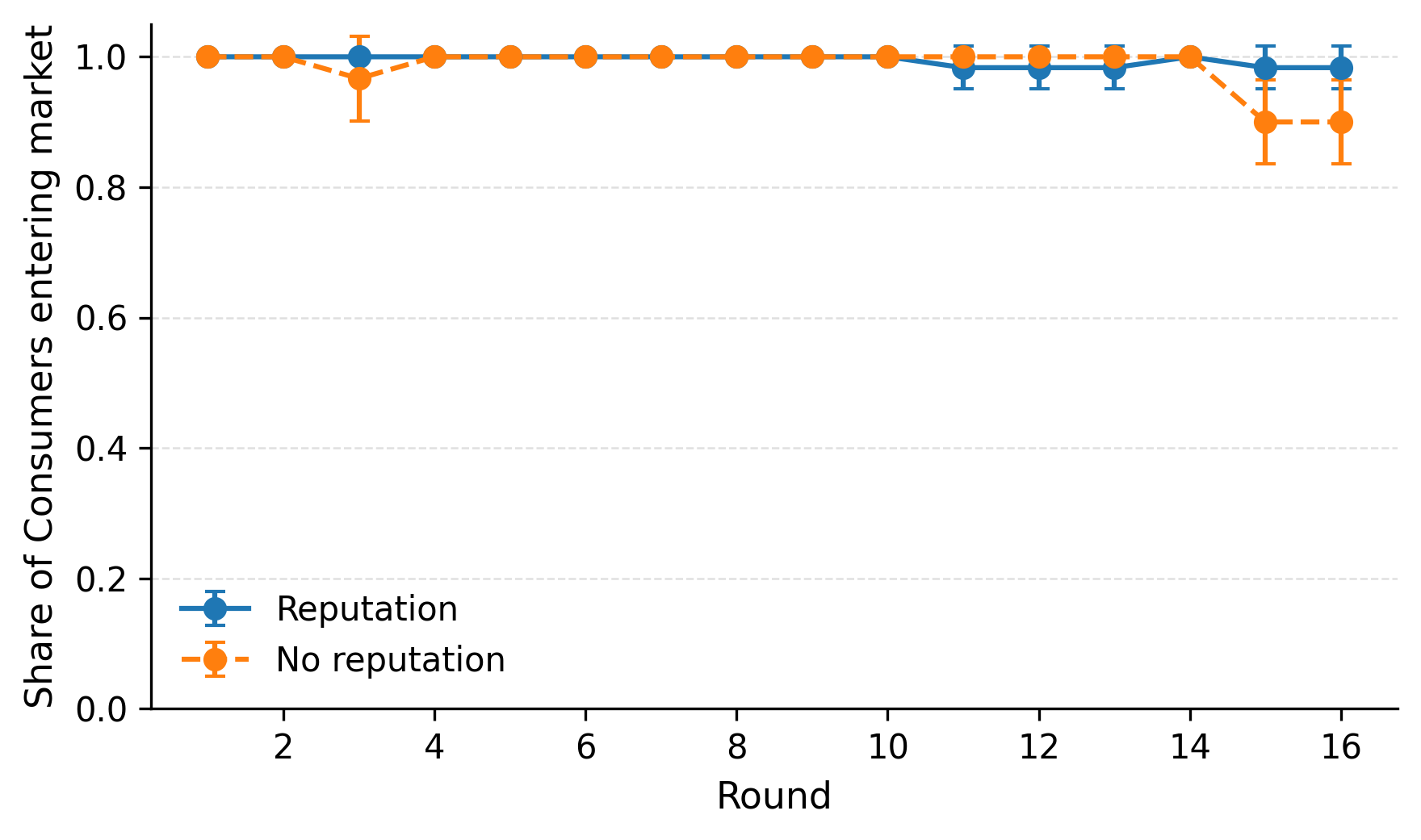}
    
    \label{fig:r16_prices_d_veri}
\end{figure}
\begin{figure}[h]
    \centering
    \caption{\textbf{Left:} Total Group Income Across 16 Rounds. \textbf{Right:} Intended \textbf{Default} Expert Under- and Over-Treatment. Verifiability.}
    \includegraphics[width=0.45\textwidth]{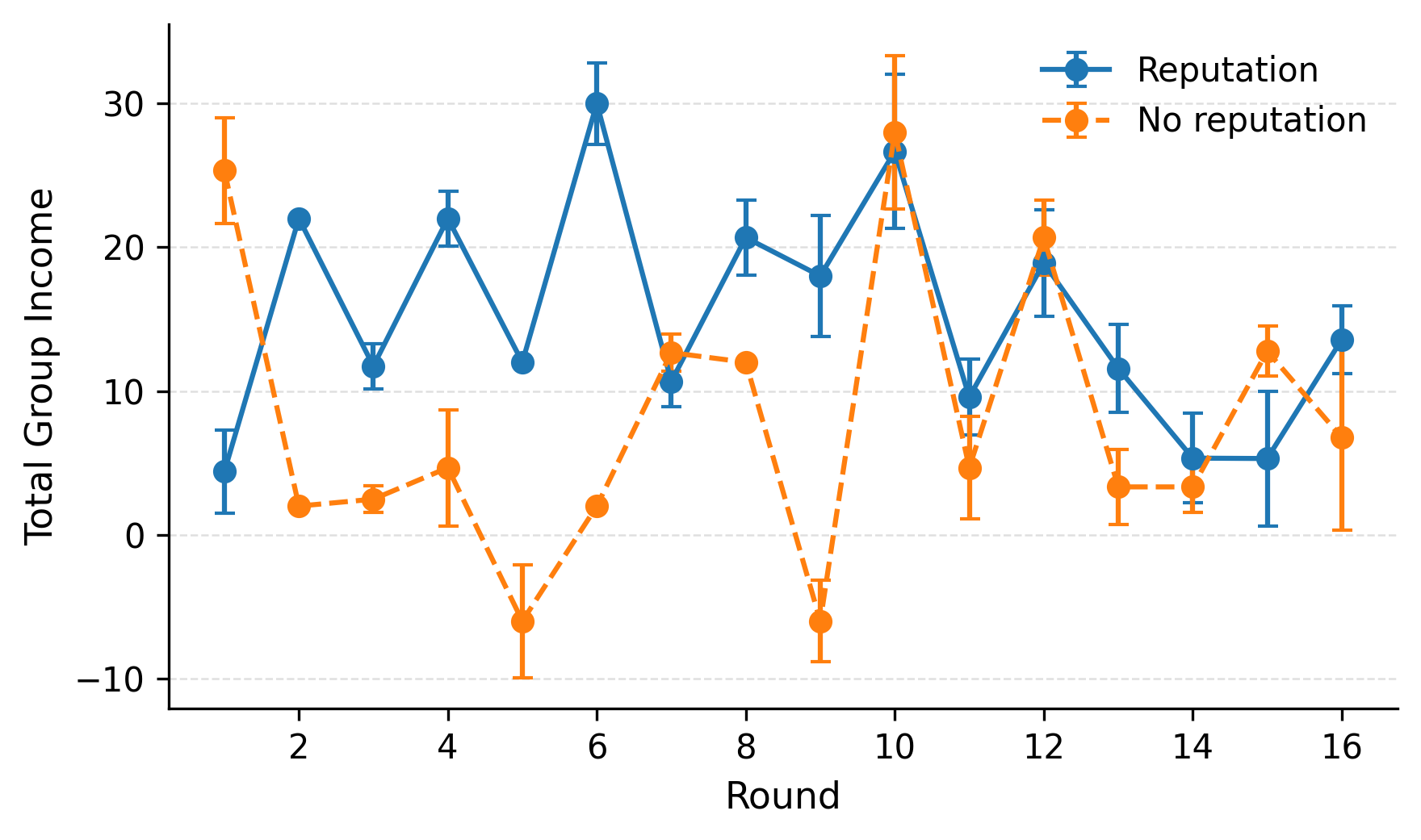}
    \includegraphics[width=0.45\textwidth]{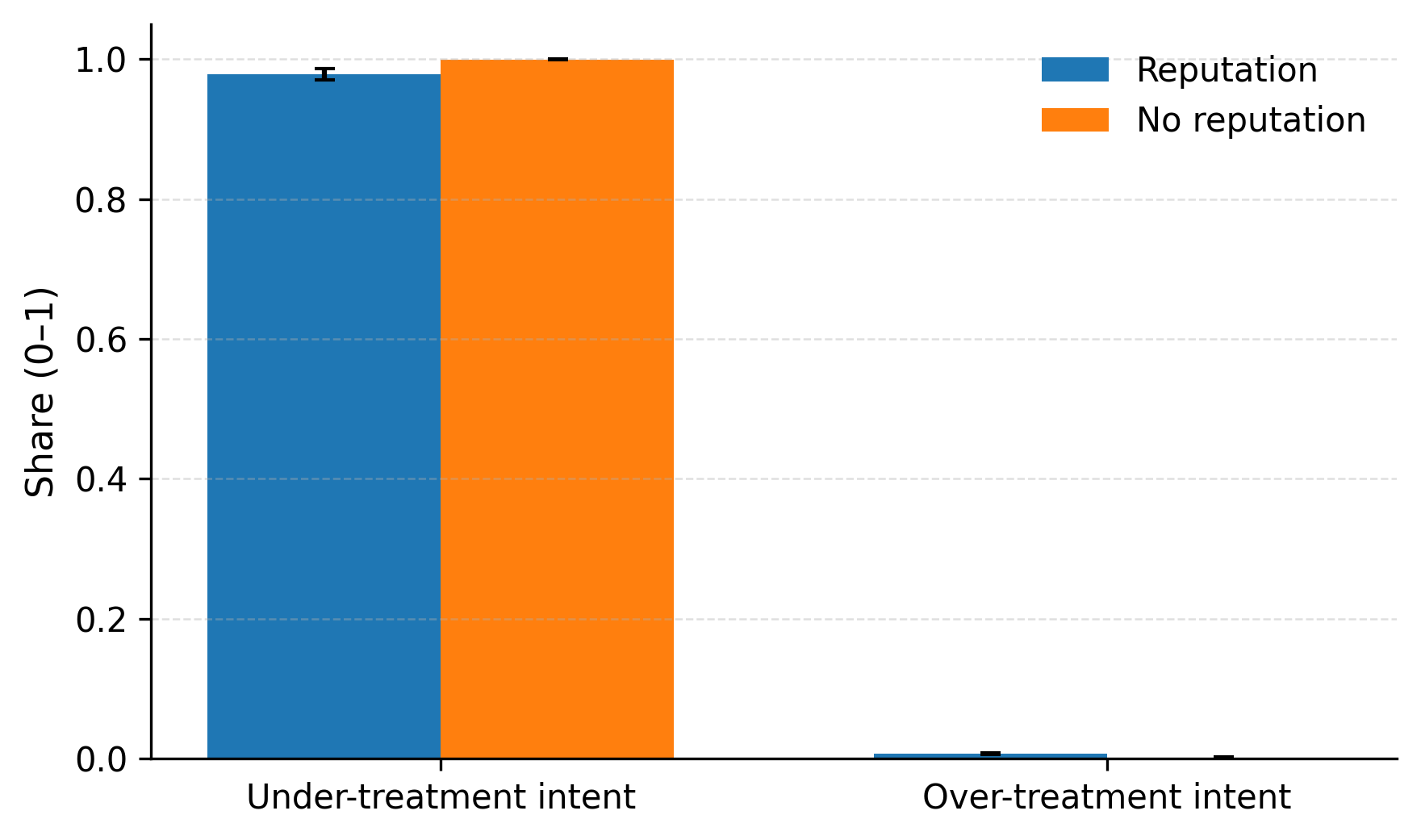}
    
    \label{fig:income_under_d_veri}
\end{figure}

\begin{figure}[h]
    \centering
    \caption{\textbf{Left: Inequity-averse} LLM Expert Price Setting Across 16 Rounds. \textbf{Right:} Consumer Approach Share. Verifiability.}
    \includegraphics[width=0.45\textwidth]{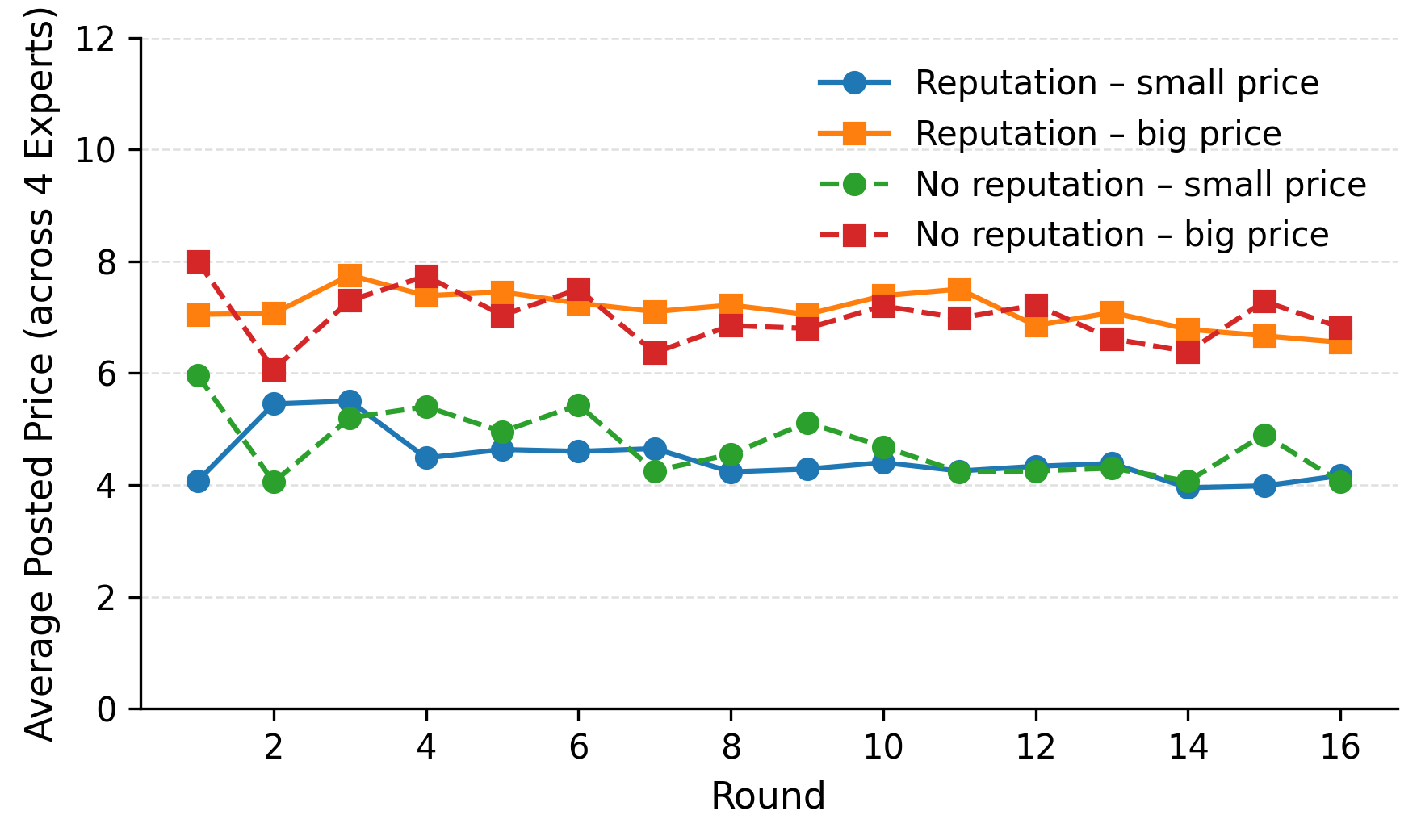}
    \includegraphics[width=0.45\textwidth]{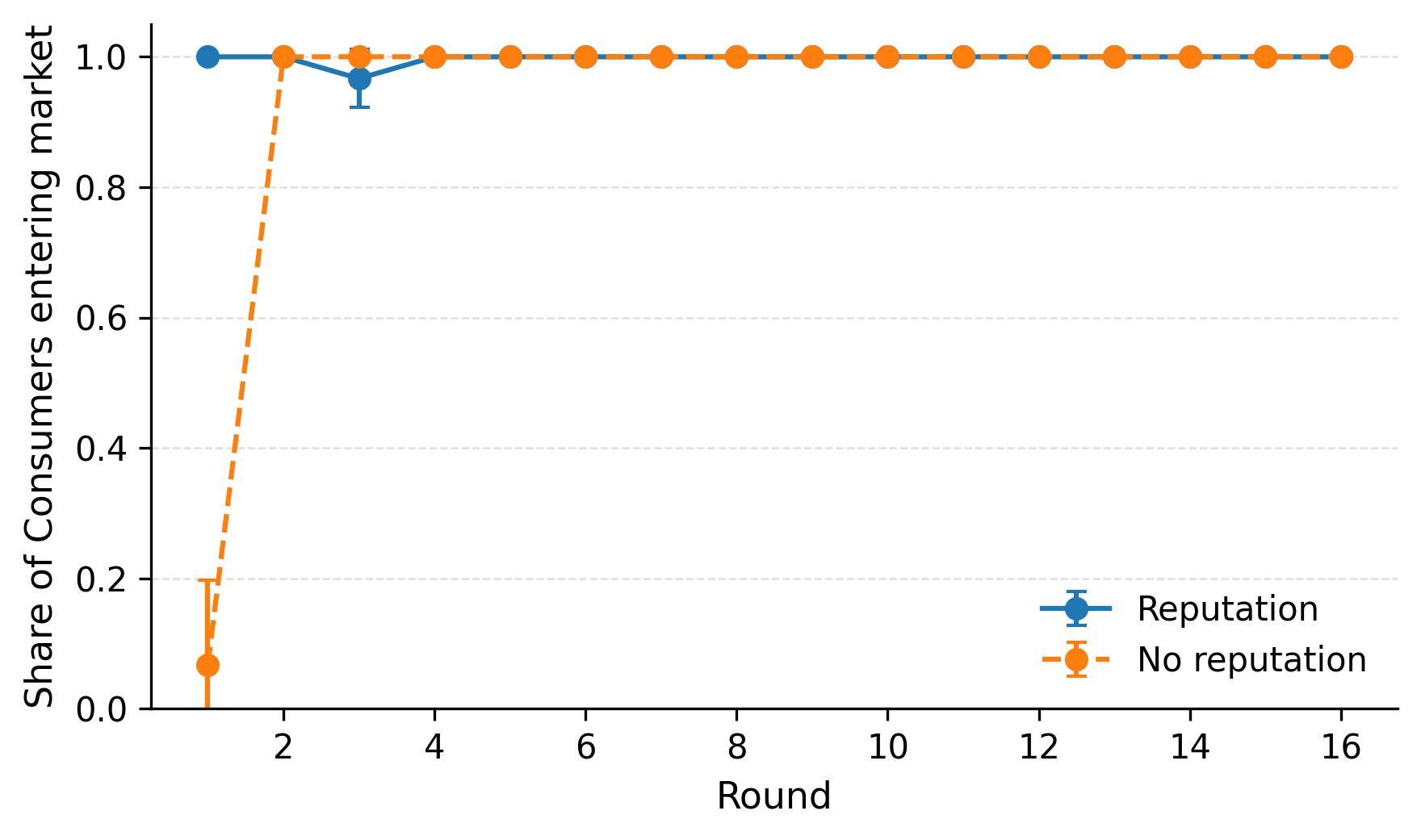}
    
    \label{fig:r16_prices_ia_veri}
\end{figure}
\begin{figure}[h]
    \centering
    \caption{\textbf{Left:} Total Group Income Across 16 Rounds. \textbf{Right:} Intended \textbf{Inequity-averse} Expert Under- and Over-Treatment. Verifiability.}
    \includegraphics[width=0.45\textwidth]{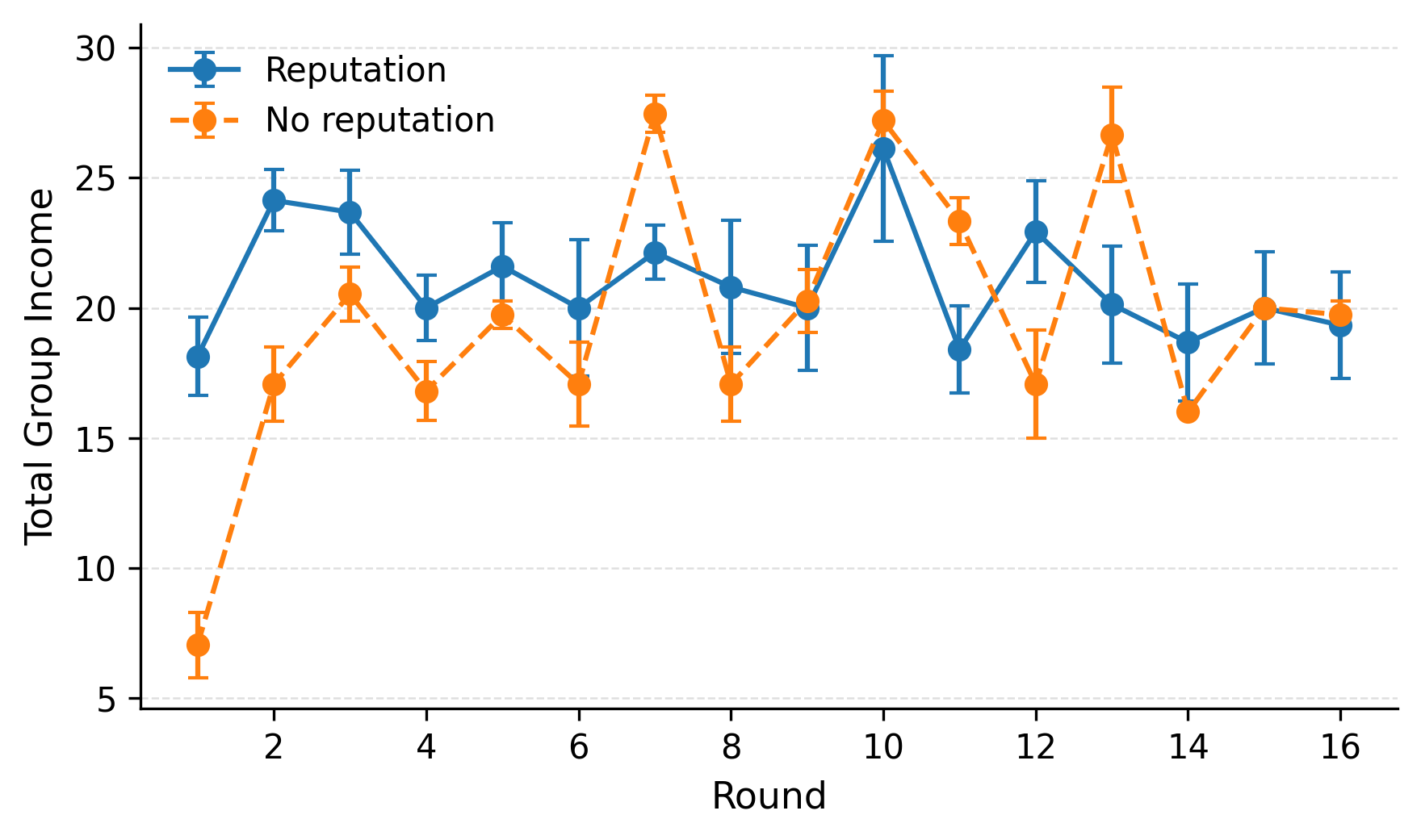}
    \includegraphics[width=0.45\textwidth]{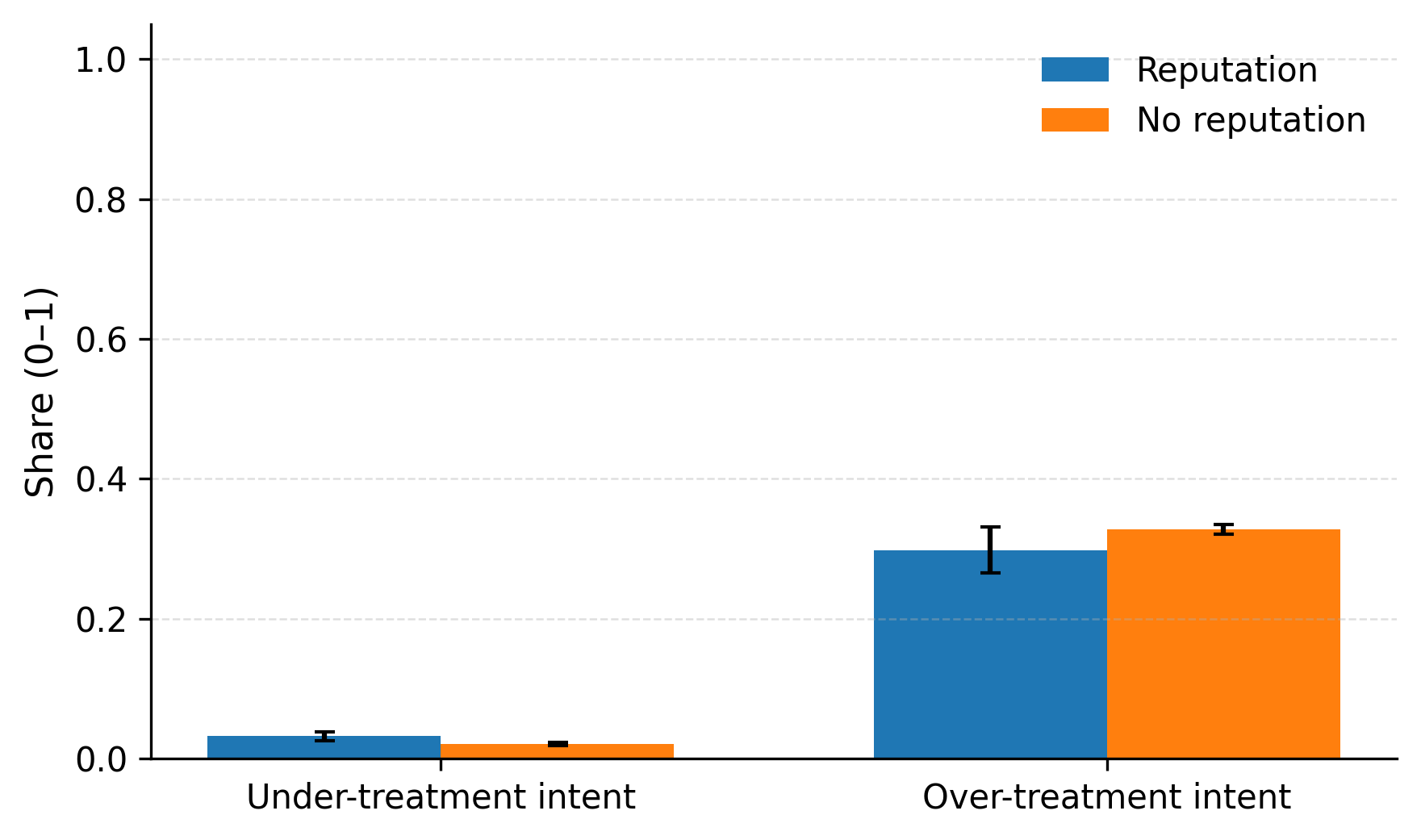}
    
    \label{fig:income_under_ia_veri}
\end{figure}

\begin{figure}[h]
    \centering
    \caption{\textbf{Left: Efficiency-loving} LLM Expert Price Setting Across 16 Rounds. \textbf{Right:} Consumer Approach Share. Verifiability.}
    \includegraphics[width=0.45\textwidth]{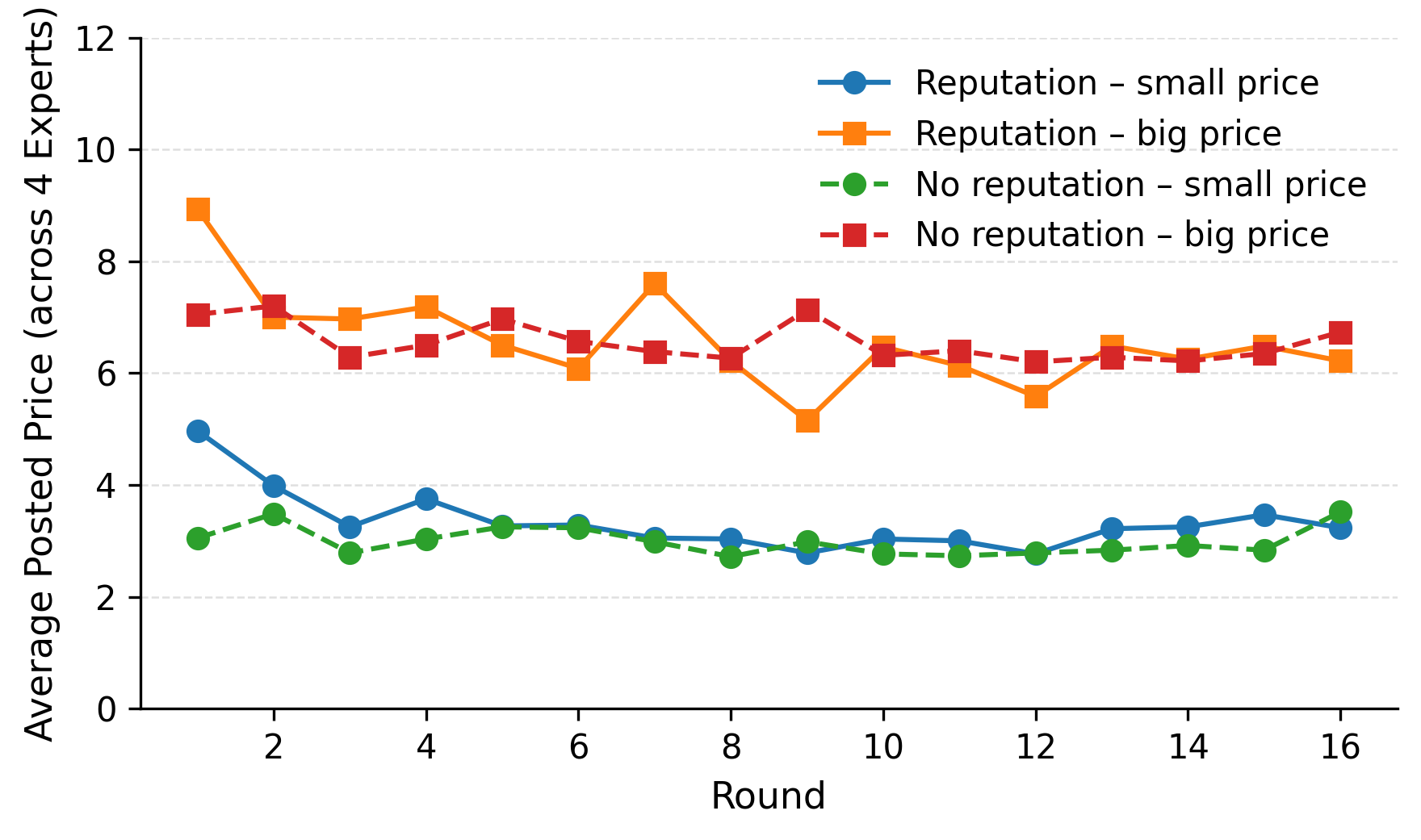}
    \includegraphics[width=0.45\textwidth]{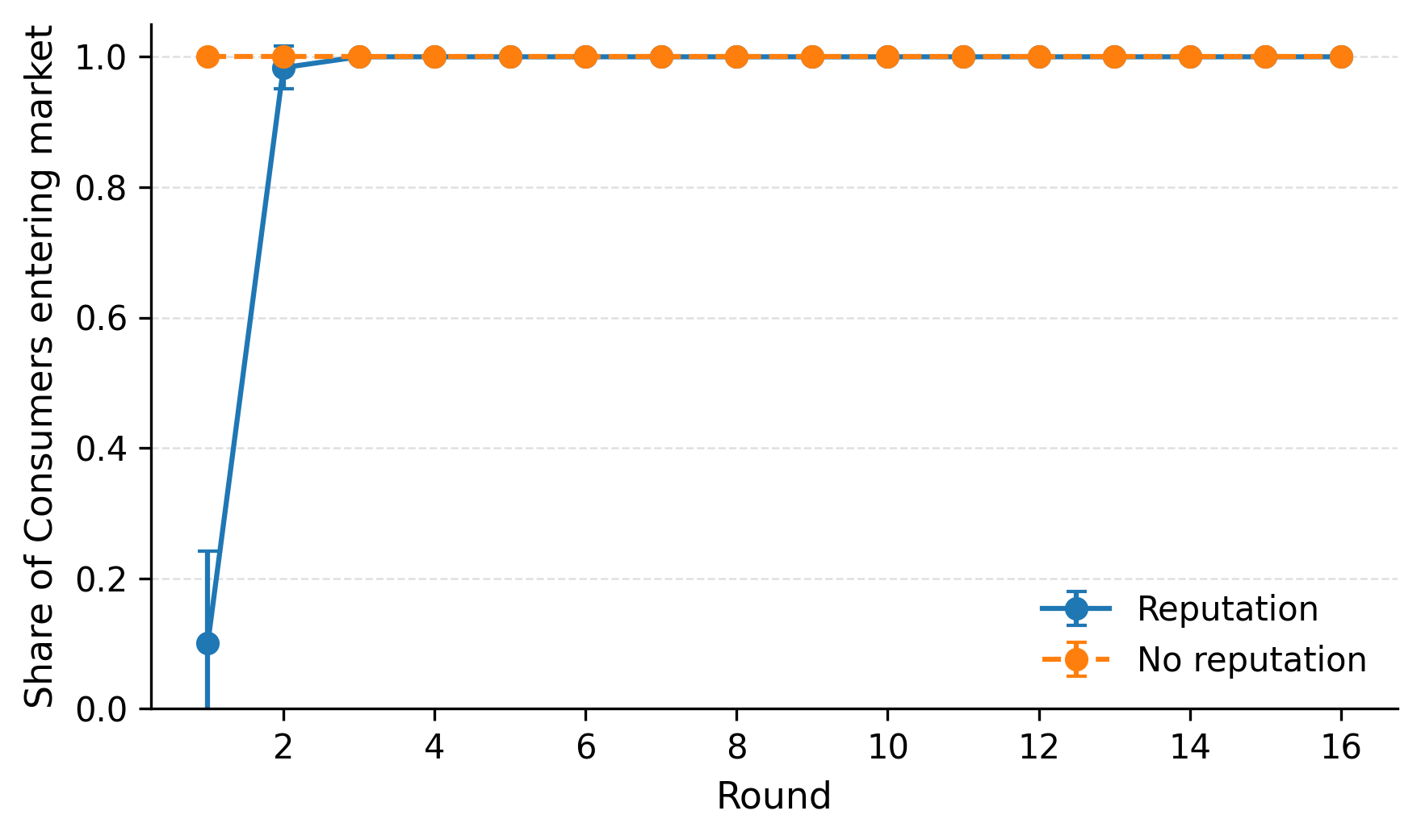}
    
    \label{fig:r16_prices_d_veri}
\end{figure}
\begin{figure}[h]
    \centering
    \caption{\textbf{Left:} Total Group Income Across 16 Rounds. \textbf{Right:} Intended \textbf{Efficiency-loving} Expert Under- and Over-Treatment. Verifiability.}
    \includegraphics[width=0.45\textwidth]{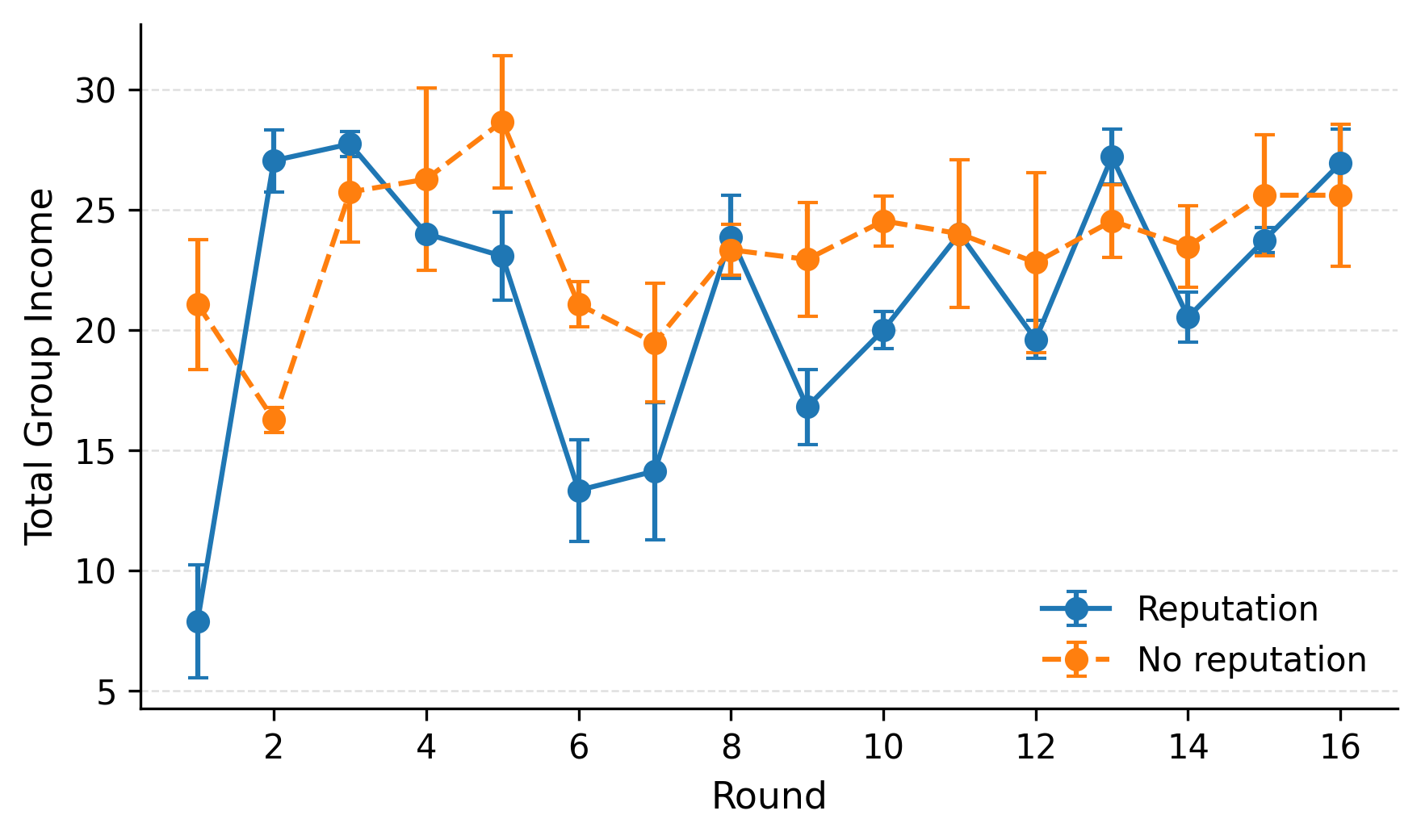}
    \includegraphics[width=0.45\textwidth]{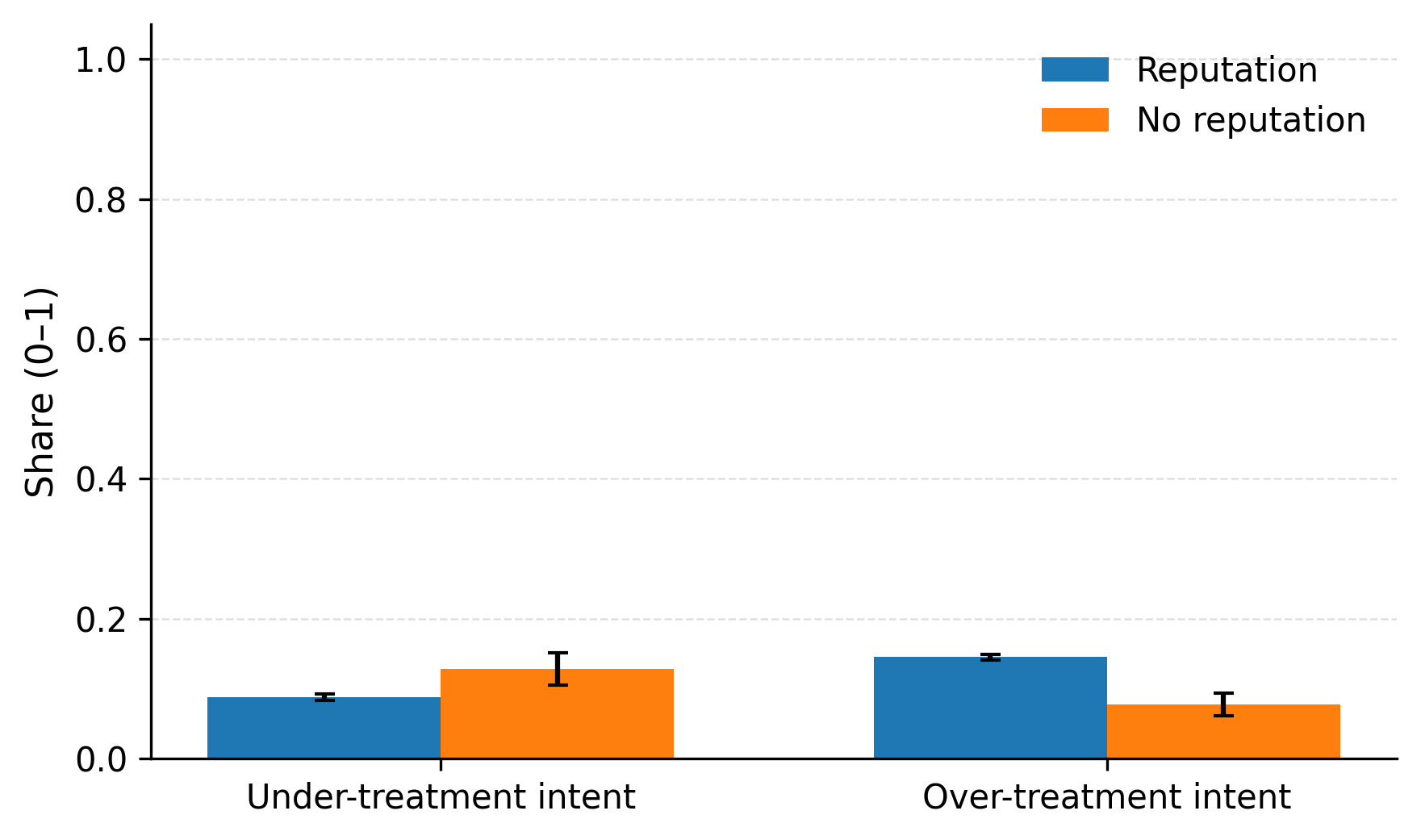}
    
    \label{fig:income_under_el_veri}
\end{figure}

\begin{table}[!ht]
\centering
\begin{threeparttable}
\caption{Pooled panel regressions on expert behavior in \textit{Verifiability}.}
\label{tab:pooled_time_slope_alltypes_veri}
\small
\setlength{\tabcolsep}{7pt}
\begin{tabular}{lcccc}
\toprule
& (1) Self-interested & (2) Default & (3) Inequity-averse & (4) Efficiency-loving \\
\midrule

\addlinespace[0.25em]
\multicolumn{5}{l}{\textit{Panel A: Under-treatment intent rate (big problems only)}}\\
\midrule
No reputation ($treat$)         & 0.0109***  & 0.0413***  & -0.1286*** & 0.0765* \\
                               & (0.0028)   & (0.0099)   & (0.0129)   & (0.0328) \\
Round ($round_c$)              & -0.0008*   & 0.0016     & -0.0135*** & -0.0066*** \\
                               & (0.0003)   & (0.0010)   & (0.0011)   & (0.0018) \\
No rep $\times$ Round          & -0.0021*** & -0.0020    & 0.0120***  & -0.0045 \\
                               & (0.0005)   & (0.0011)   & (0.0011)   & (0.0029) \\
Observations                    & 1856       & 1736       & 1868       & 1804 \\
Experts              & 120        & 120        & 120        & 120 \\
\midrule

\addlinespace[0.25em]
\multicolumn{5}{l}{\textit{Panel B: Over-treatment intent rate (small problems only)}}\\
\midrule
No reputation ($treat$)         & -0.0000    & -0.0088*** & 0.2908***  & -0.0479** \\
                               & (0.0004)   & (0.0026)   & (0.0310)   & (0.0180) \\
Round ($round_c$)              & 0.0002     & -0.0001    & 0.0273***  & 0.0093*** \\
                               & (0.0001)   & (0.0002)   & (0.0029)   & (0.0020) \\
No rep $\times$ Round          & -0.0002    & 0.0002     & -0.0375*** & -0.0018 \\
                               & (0.0001)   & (0.0003)   & (0.0040)   & (0.0029) \\
Observations                    & 1836       & 1768       & 1788       & 1736 \\
Experts               & 120        & 120        & 120        & 120 \\
\midrule

\addlinespace[0.25em]
\multicolumn{5}{l}{\textit{Panel C: Overcharging intent rate (all decisions)}}\\
\midrule
No reputation ($treat$)         & ---        & ---        & ---        & --- \\
                               &            &            &            &     \\
Round ($round_c$)              & ---        & ---        & ---        & --- \\
                               &            &            &            &     \\
No rep $\times$ Round          & ---        & ---        & ---        & --- \\
                               &            &            &            &     \\
Observations                    & ---        & ---        & ---        & --- \\
Experts              & ---        & ---        & ---        & --- \\
\bottomrule
\end{tabular}

\begin{tablenotes}[flushleft]
\footnotesize
\item \textit{Notes.} Pooled panel regressions of the form $y_{it}=\alpha+\beta_1\,treat_i+\beta_2\,round_c+\beta_3\,(treat_i \times round_c)+\varepsilon_{it}$ at the expert$\times$round level. Baseline condition is \textbf{reputation}; $treat=1$ denotes \textbf{no reputation}. Standard errors in parentheses. Significance: * $p<0.05$, ** $p<0.01$, *** $p<0.001$. 
\end{tablenotes}
\end{threeparttable}
\end{table}

\clearpage

\subsection*{Comparison with Dulleck et al. (2011)}

\begingroup
\footnotesize
\setlength{\tabcolsep}{5pt}

\begin{longtable}{lcccc}
\caption{Comparison of simulated LLM-agent market results and human subject experiment from \citet{dulleck2011economics} across treatments.}
\label{tab:dulleck_vs_sims_full}\\
\toprule
& \multicolumn{4}{c}{Treatment} \\
\cmidrule(lr){2-5}
 & C/N & CR/N & C/V & CR/V \\
\midrule
\endfirsthead

\toprule
& \multicolumn{4}{c}{Treatment} \\
\cmidrule(lr){2-5}
 & C/N & CR/N & C/V & CR/V \\
\midrule
\endhead

\midrule
\multicolumn{5}{r}{\footnotesize Continued on next page} \\
\endfoot

\bottomrule
\endlastfoot

\multicolumn{5}{l}{\textit{Panel A: Humans (Dulleck et al.\ 2011, Table 4)}} \\
\midrule
Trade on consumer side & 0.730 & 0.850 & 0.880 & 0.930 \\
Avg \# consumers (given seller has $\geq 1$) & 1.540 & 1.540 & 1.630 & 1.600 \\
Trade on seller side & 0.470 & 0.550 & 0.540 & 0.580 \\
Efficiency  & 0.130 & 0.140 & 0.340 & 0.460 \\
Undertreatment (realized) & 0.730 & 0.640 & 0.530 & 0.360 \\
Overtreatment (realized) & 0.080 & 0.100 & 0.130 & 0.190 \\
Overcharging (realized) & 0.790 & 0.620 & --- & --- \\
$p^l$ with trade & 3.190 & 3.310 & 4.060 & 3.990 \\
$p^l$ without trade & 4.160 & 4.650 & 4.840 & 4.840 \\
$p^h$ with trade & 5.730 & 6.330 & 6.410 & 6.970 \\
$p^h$ without trade & 7.550 & 7.600 & 7.720 & 7.790 \\
Actually paid price & 5.350 & 5.800 & 5.190 & 5.580 \\
Profits sellers (per seller-period) & 2.360 & 2.650 & 1.960 & 1.900 \\
Profits consumers (per consumer-period) & 1.210 & 0.940 & 2.180 & 2.590 \\
Mean total income per period (A+B) & 14.280 & 14.360 & 16.560 & 17.960 \\
\addlinespace

\multicolumn{5}{l}{\textit{Panel B: Simulations (No Objective)}} \\
\midrule
Trade on consumer side & 0.772 & 0.875 & 0.985 & 0.995 \\
Avg \# consumers (given seller has $\geq 1$) & 3.030 & 3.100 & 3.371 & 2.637 \\
Trade on seller side & 0.256 & 0.282 & 0.294 & 0.381 \\
Efficiency  & 0.466 & 0.783 & 0.094 & 0.497 \\
Undertreatment (realized) & 0.998 & 0.867 & 1.000 & 0.994 \\
Overtreatment (realized) & 0.000 & 0.000 & 0.000 & 0.000 \\
Overcharging (realized) & 0.615 & 1.000 & 0.000 & 0.000 \\
$p^l$ with trade & 1.419 & 2.332 & 2.728 & 2.913 \\
$p^l$ without trade & 2.135 & 3.023 & 3.322 & 3.408 \\
$p^h$ with trade & 5.147 & 4.730 & 5.488 & 5.248 \\
$p^h$ without trade & 7.529 & 6.260 & 6.699 & 6.322 \\
Actually paid price & 4.656 & 4.627 & 2.695 & 2.892 \\
Profits sellers (per seller-period) & 2.041 & 2.174 & 0.685 & 0.874 \\
Profits consumers (per consumer-period) & 1.610 & 2.870 & 1.325 & 2.914 \\
Mean total income per period (A+B) & 14.602 & 20.175 & 8.043 & 15.150 \\
\addlinespace

\multicolumn{5}{l}{\textit{Panel C: Simulations (Self-Interested)}} \\
\midrule
Trade on consumer side & 0.666 & 0.777 & 0.886 & 0.997 \\
Avg \# consumers (given seller has $\geq 1$) & 3.187 & 2.697 & 3.179 & 2.523 \\
Trade on seller side & 0.211 & 0.298 & 0.279 & 0.397 \\
Efficiency & 0.502 & 0.190 & 0.197 & 0.351 \\
Undertreatment (realized) & 1.000 & 1.000 & 0.970 & 1.000 \\
Overtreatment (realized) & 0.000 & 0.000 & 0.000 & 0.002 \\
Overcharging (realized & 0.994 & 0.954 & 0.000 & 0.000 \\
$p^l$ with trade & 1.124 & 1.711 & 2.711 & 2.933 \\
$p^l$ without trade & 2.918 & 3.694 & 3.893 & 3.604 \\
$p^h$ with trade & 3.941 & 4.399 & 4.792 & 4.825 \\
$p^h$ without trade & 6.285 & 6.628 & 6.395 & 6.462 \\
Actually paid price & 3.948 & 4.106 & 2.668 & 2.858 \\
Profits sellers (per seller-period) & 1.309 & 1.571 & 0.537 & 0.851 \\
Profits consumers (per consumer-period) & 2.499 & 0.867 & 1.927 & 2.291 \\
Mean total income per period (A+B) & 15.232 & 9.752 & 9.860 & 12.570 \\
\addlinespace

\multicolumn{5}{l}{\textit{Panel D: Simulations (Inequity-Averse)}} \\
\midrule
Trade on consumer side & 1.000 & 0.988 & 0.942 & 0.998 \\
Avg \# consumers (given seller has $\geq 1$) & 3.895 & 2.901 & 4.000 & 3.200 \\
Trade on seller side & 0.257 & 0.347 & 0.235 & 0.315 \\
Efficiency & 0.891 & 0.852 & 0.748 & 0.830 \\
Undertreatment (realized) & 0.017 & 0.046 & 0.002 & 0.023 \\
Overtreatment (realized) & 0.259 & 0.248 & 0.392 & 0.307 \\
Overcharging (realized & 0.038 & 0.206 & 0.000 & 0.000 \\
$p^l$ with trade & 3.963 & 4.256 & 4.205 & 3.992 \\
$p^l$ without trade & 4.705 & 4.987 & 4.869 & 4.678 \\
$p^h$ with trade & 5.005 & 6.486 & 6.449 & 6.718 \\
$p^h$ without trade & 6.237 & 7.270 & 7.182 & 7.319 \\
Actually paid price & 4.365 & 5.323 & 5.800 & 5.779 \\
Profits sellers (per seller-period) & -0.010 & 0.929 & 0.852 & 1.150 \\
Profits consumers (per consumer-period) & 5.552 & 4.424 & 4.039 & 4.101 \\
Mean total income per period (A+B) & 22.167 & 21.413 & 19.565 & 21.005 \\
\addlinespace

\multicolumn{5}{l}{\textit{Panel E: Simulations (Efficiency-Loving)}} \\
\midrule
Trade on consumer side & 1.000 & 1.000 & 1.000 & 0.943 \\
Avg \# consumers (given seller has $\geq 1$) & 3.475 & 3.135 & 3.789 & 2.897 \\
Trade on seller side & 0.292 & 0.324 & 0.265 & 0.327 \\
Efficiency (consumer outside option only; baseline=6.4, max=24.0) & 0.866 & 0.849 & 0.950 & 0.843 \\
Undertreatment (realized) & 0.064 & 0.034 & 0.033 & 0.125 \\
Overtreatment (realized) & 0.089 & 0.319 & 0.065 & 0.067 \\
Overcharging (realized) & 0.071 & 0.070 & 0.000 & 0.000 \\
$p^l$ with trade & 2.855 & 2.871 & 2.539 & 3.016 \\
$p^l$ without trade & 3.361 & 3.259 & 3.157 & 3.487 \\
$p^h$ with trade & 5.157 & 4.834 & 6.106 & 5.982 \\
$p^h$ without trade & 5.810 & 6.316 & 6.715 & 6.868 \\
Actually paid price & 3.427 & 3.420 & 4.265 & 4.244 \\
Profits sellers (per seller-period) & -0.815 & -1.064 & 0.306 & 0.367 \\
Profits consumers (per consumer-period) & 6.229 & 6.424 & 5.558 & 4.944 \\
Mean total income per period (A+B) & 21.658 & 21.442 & 23.458 & 21.242 \\
\end{longtable}

\vspace{-0.5em}
\noindent\textit{Notes:}
C/N = competition, no reputation, no verifiability; CR/N = competition, reputation, no verifiability;
C/V = competition, no reputation, verifiability; CR/V = competition, reputation, verifiability.
Human values are taken from Dulleck et al.\ (2011), Table 4, restricted to these four treatments.
Simulation entries are cell means (15 simulations per cell, 16 rounds each). Note that efficiency in Dulleck et al. (2011) is derived from a setting in which both experts and consumers have outside options.
\par
\endgroup

\end{document}